\documentclass[twocolumn,aps,prx,showpacs,amsmath,superscriptaddress,10pt]{revtex4-2}

\usepackage{hyperref, soul}
\usepackage{ulem}
\hypersetup{linktocpage,,colorlinks=true}
\usepackage{threeparttable}
\usepackage{graphicx}
\usepackage{dcolumn}
\usepackage{bm}
\usepackage{color}
\usepackage{comment}
\usepackage{natbib}
\usepackage{amssymb}
\usepackage{subfigure}
\usepackage{amsfonts}
\usepackage{mathtools}
\usepackage{physics}
\usepackage{amsmath}
\usepackage{mathrsfs}
\usepackage{enumitem}
\usepackage{lmodern}
\usepackage{anyfontsize}
\usepackage{adjustbox}
\usepackage[inkscapearea=page]{svg}
\usepackage{tensor}
\usepackage[utf8]{inputenc}
\usepackage{titlesec}
\usepackage{indentfirst}
\usepackage{graphicx}

\numberwithin{equation}{section}

\begin{document}

\title{Towards a Topological Proof of the Strong Subadditivity}

\author{Chih-Yu Lo}
\email{chihyulo.jared@gmail.com}
\affiliation{Department of Physics, National Tsing Hua University, Hsinchu 30013, Taiwan}

\author{Po-Yao Chang}
\email{pychang@phys.nthu.edu.tw}
\affiliation{Department of Physics, National Tsing Hua University, Hsinchu 30013, Taiwan}

\date{\today}


\begin{abstract}

Topological entanglement entropy (TEE) represents an intrinsic contribution to the entanglement entropy (EE) in topologically ordered systems. In quantum information theory, strong subadditivity (SSA) is a fundamental property of EE, reflecting the non-negativity of conditional mutual information. TEE was originally believed to be a universal correction to the area law of EE, suggesting that its SSA would directly follow from the SSA of EE. However, due to spurious contributions, the correction term is not universal; consequently, the value predicted by topological quantum field theory (TQFT) provides only a lower bound.
In this work, we present a topological analysis showing that the SSA for TEE is equivalent to a specific inequality within the TQFT framework. We further verify that this inequality holds for all known unitary modular tensor categories (UMTCs) up to rank 11, supporting the conjecture that SSA holds universally in the TQFT framework. Conversely, assuming the validity of the SSA condition, the inequality can be interpreted as a consistency criterion for candidate UMTCs.

\end{abstract}

\maketitle

\tableofcontents

\section{Introduction}

Topologically ordered systems~\cite{TO} are gapped quantum phases characterized by robust ground state degeneracies and long-range entanglement. These systems have been extensively studied in various contexts, including fractional quantum Hall states~\cite{FQH1,FQH2}, quantum spin liquids~\cite{SpinLiquid}, and $p_x + ip_y$ superconductors~\cite{px+ipy}.
A hallmark of topological order is the topological entanglement entropy (TEE)~\cite{TEE,TEE2}, which serves as a universal diagnostic of long-range entanglement. In such systems, the entanglement entropy (EE) between a region $A$ and its complement typically follows the area law with a subleading correction:
\begin{equation}\label{Eq:Area}
    S_\text{EE}(A) = \alpha L_A - \gamma,
\end{equation}
where $\alpha$ is a constant, $L_A$ is the length of the boundary $\partial A$, and $\gamma \geq 0$ is a universal constant reflecting the system’s topological nature.

Since the low-energy physics of topologically ordered systems is effectively captured by topological quantum field theories (TQFTs), it was long believed that this universal correction $\gamma$ exactly corresponds to the TEE, denoted $S_\text{TEE}$. This quantity can be computed using the replica trick and surgery methods within TQFT, or alternatively via the Ishibashi state construction in the edge theory framework.~\cite{EdgeTEE,CARDY2006333,BCFT1,BCFT2}.
This expectation has been verified in numerous microscopic models~\cite{FQHTO1,FQHTO2,FQHTO3,SpinLiquidTO1,SpinLiquidTO2,ToricCodeTO,DimmerTO}.However, subsequent studies revealed that spurious contributions can contaminate the extraction of TEE from $\gamma$~\cite{Spurious1,Spurious2,Spurious3,Spurious4,Spurious5}.
That is, $-\gamma= S_\text{TEE}(A) + S_\text{spu.}(A)$ where  $ S_\text{spu.}(A)$ is a positive number coming from the spurious contributions.

As a result, it is now understood that the TQFT and edge theory calculations provide a universal lower bound on the topological correction:
\begin{equation}
\gamma \geq -S_\text{TEE}(A).
\end{equation}

A fundamental property in quantum information theory is strong subadditivity (SSA), which is satisfied by the von Neumann entropy~\cite{Nielsen}. Because the area term in Eq.(\ref{Eq:Area}) is extensive, it satisfies SSA as an equality. Consequently, the topological correction $\gamma$ is expected to obey SSA as well. However, due to the presence of spurious contributions, the SSA for $S_\text{TEE}$ cannot be directly implied, raising the question of whether a topological proof of SSA for $S_\text{TEE}$ is possible within the framework of TQFT.
At first glance, $S_\text{TEE}$ appears to be a von Neumann entropy, since it is derived from a reduced density matrix. However, in TQFTs, Hilbert spaces for subsystems are not well-defined, preventing a direct application of the standard SSA proof.

In this work, we aim to provide a topological proof of SSA by leveraging the TEE results from TQFT and edge theories. We show that the validity of SSA is equivalent to the following inequality:
\begin{equation}\label{Eq:Conjecture}
\sum_a |\psi_a|^2 (\ln d_a - \ln |\psi_a|) + |S\psi_a|^2 (\ln d_a - \ln |S\psi_a|) \geq 2 \ln \mathcal{D},
\end{equation}
where $|\psi_a|^2$ is the probability amplitude, $S$ is the modular $S$-matrix, $d_a$ are the quantum dimensions of the anyons, and $\mathcal{D}$ is the total quantum dimension. Thus, the SSA condition for TEE holds if and only if this inequality is satisfied.

We have numerically verified that Eq.~(\ref{Eq:Conjecture}) holds for unitary modular tensor categories (UMTCs) with rank less than $11$~\footnote{The numerical data for checking the validity of the inequality for UMTCs up to rank 11 is available at https://github.com/chihyulo/SSA-of-TEE-for-UMTC-up-to-rank-11}, including notable examples such as the Fibonacci anyons and the Semion model.
We conjecture that Eq.~(\ref{Eq:Conjecture}) is valid for all UMTCs, and that it may impose implicit constraints on the modular data.

The outline of this paper is as follows. In the remainder of this section, we briefly review the SSA condition.
In Sec.~\ref{Sec:bipartition}, we classify all possible bipartitions on a torus and discuss how to reduce them to canonical bipartitions.
In Sec.~\ref{Sec:TEE}, we review the TEE for the canonical bipartitions discussed in the previous literature (A general bulk derivation can also be found in Appendix.~\ref{Appendix:TEE}).
Readers familiar with these results may proceed directly to Sec.~\ref{Sec:SSA_iTEE}.
Sec.~\ref{Sec:SSA_iTEE} introduces the modified SSA for intrinsic TEE, supported by topological arguments.
Finally, in Sec.~\ref{Sec:SSA_TEE}, we will present the proof for the SSA of the TEE up to Eq.~(\ref{Eq:Conjecture}), along with several low-rank examples.

Let us begin by summarizing some key concepts that will be referenced throughout this paper.

\subsection{Conditional mutual information and the strong subadditivity}

In quantum information theory, the conditional mutual information (CMI) for the entanglement entropy (EE) measures the information that remains unrepresented in the Venn diagram: 
\begin{equation}
    \mathcal{I}_\text{EE}(A:B) = S_\text{EE}(A) + S_\text{EE}(B) - S_\text{EE}(A\cup B)- S_\text{EE}(A\cap B),
\end{equation}
where $S_\text{EE}$ is the EE between subregion and its complement.
The non-negativity of the CMI of EE:
\begin{equation}\label{Eq:SSA}
    \mathcal{I}_\text{EE}(A:B) \geq 0,
\end{equation}
is equivalent to the strong subadditivity (SSA) condition of EE.

Since the EE is invariant under complement, i.e., $S_\text{EE}(A) = S_\text{EE}(A^c)$, the conditional mutual information is also invariant under taking the simultaneous complement of both $A$ and $B$, i.e.,
\begin{equation}\label{Eq:ComplementInv}
    \mathcal{I}_\text{EE}(A:B) = \mathcal{I}_\text{EE}(A^c:B^c),
\end{equation}
That is, the SSA condition is also preserved under the simultaneous complement.

For convenience, we introduce the CMI for function $f$:
\begin{equation}\label{Eq:CMI}
    \mathcal{I}_{f}(A:B) = f(A) + f(B) - f(A\cup B) - f(A\cap B)
\end{equation}
which measures the additional information beyond the description of the Venn diagram.
If we take $f = S_\text{EE}$, then we recover the usual CMI for EE.
For $f = S_\text{TEE}$, we have the topological CMI, $\mathcal{I}_\text{TEE}(A:B)$. The SSA for TEE is therefore the non-negativity of the CMI for TEE.
In Sec.~\ref{Sec:TEE} and Sec.~\ref{Sec:SSA_iTEE}, we will take $f$ to be the intrinsic TEE $S_\text{iTEE}$ and the genus $g$, where we will show the modified SSA for the intrinsic TEE
\begin{equation}
    \mathcal{I}_\text{iTEE}(A:B) \geq 2 \mathcal{I}_{g}(A:B) \ln  \mathcal{D}.
\end{equation}

Furthermore, if a function $f$ satisfies $f(A) = f(A^c)$, then $\mathcal{I}_f(A:B)$ is invariant under simultaneous complement of $A$ and $B$,
\begin{equation}\label{Eq:SimCom}
    \mathcal{I}_f(A^c:B^c) = \mathcal{I}_f(A:B).
\end{equation}
For example, entropies such as the EE, TEE and iTEE all satisfy this property.

\section{Classification of bipartitions}\label{Sec:bipartition}

\subsection{Regularization of bipartitions}

To analyze the topology of our subregions more carefully, we need to pay close attention to whether a subregion includes its boundary points.

Typically, we think of subregions as open subsets. However, when considering a bipartition of a torus $T^2 = A \cup B$, the situation is more nuanced.
It is impossible for two open sets to disjointly cover a closed manifold like the torus.
Therefore, the appropriate condition for $A, B$ to form a bipartition of $T^2$ is that $\Bar{A} \cup B = T^2$, where $\Bar{A}$ is the closure of $A$, and their boundaries coincide, i.e. $\partial A = \partial B$, with $\partial$ denoting the boundary.
For convenience, we will describe such a bipartition by specifying the open region $A$; the region $B$ is then understood to be its complement, with a shared boundary.

In this context, constructing the reduced density matrix $\rho_A$ introduces an ambiguity: we must decide whether to glue the manifolds along $B$ or its closure $\Bar{B}$. 
For instance, consider the case where $B$ consists of two disjoint open disks whose boundaries intersect at a line segment.
In this situation, it is natural to treat these two disks as part of a single connected component, meaning we glue along their boundaries. This is illustrated as follows:
\begin{equation*}
    \adjustbox{valign = c}{\includegraphics[height = 1.3cm]{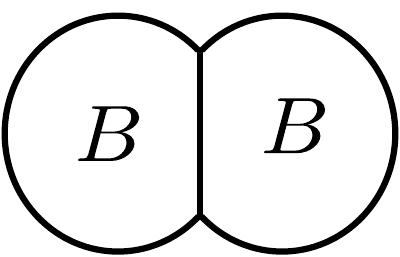}}\hspace{0.1cm} = 
    \adjustbox{valign = c}{\includegraphics[height = 1.3 cm]{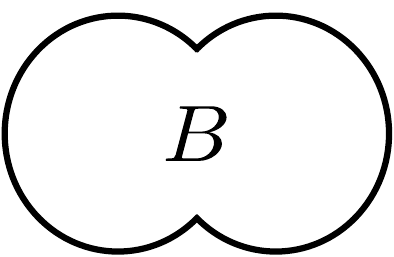}}\hspace{0.1cm}.
\end{equation*}
In this situation, the ambiguity can be resolved.

However, the situation becomes problematic when $A$ and $B$ intersect at a point rather than along a line segment, such as in the case of a tetrajunction. For example, consider the following bipartition on $S^2$
\begin{equation*}
    \adjustbox{valign = c}{\includegraphics[height = 1.4 cm]{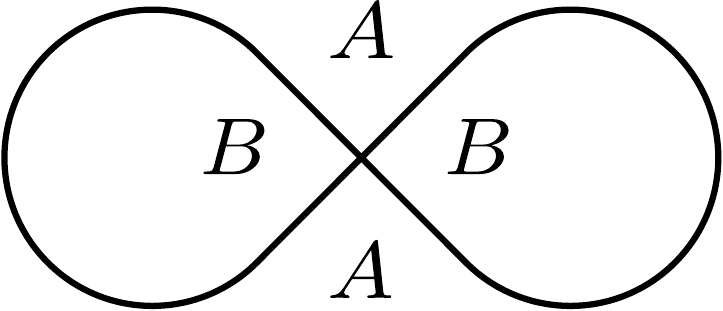}}\hspace{0.1cm}.
\end{equation*}
This configuration is the limiting case between two distinct situations:
\begin{equation*}
    \adjustbox{valign = c}{\includegraphics[height = 1.4 cm]{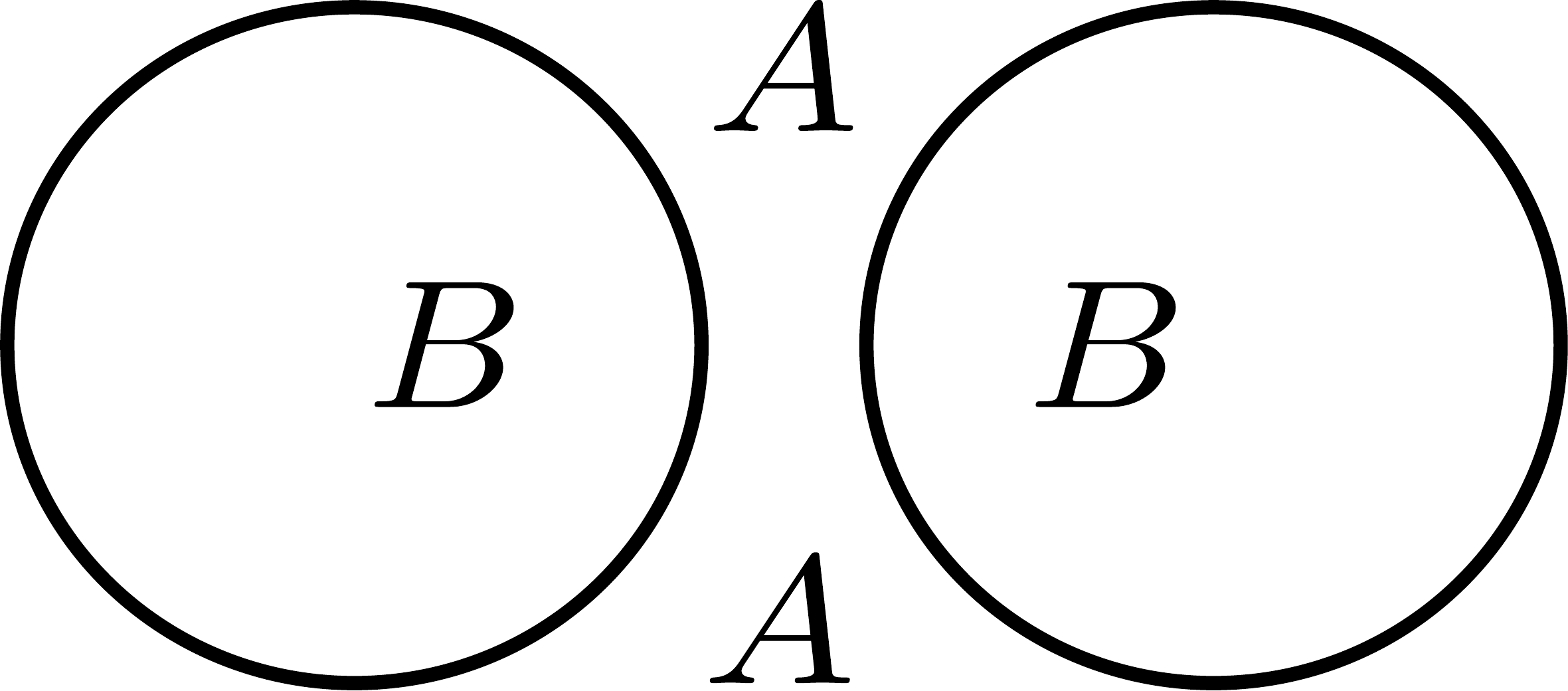}}
    \hspace{0.3cm} ,\hspace{0.3cm}
    \adjustbox{valign = c}{\includegraphics[height = 1.4 cm]{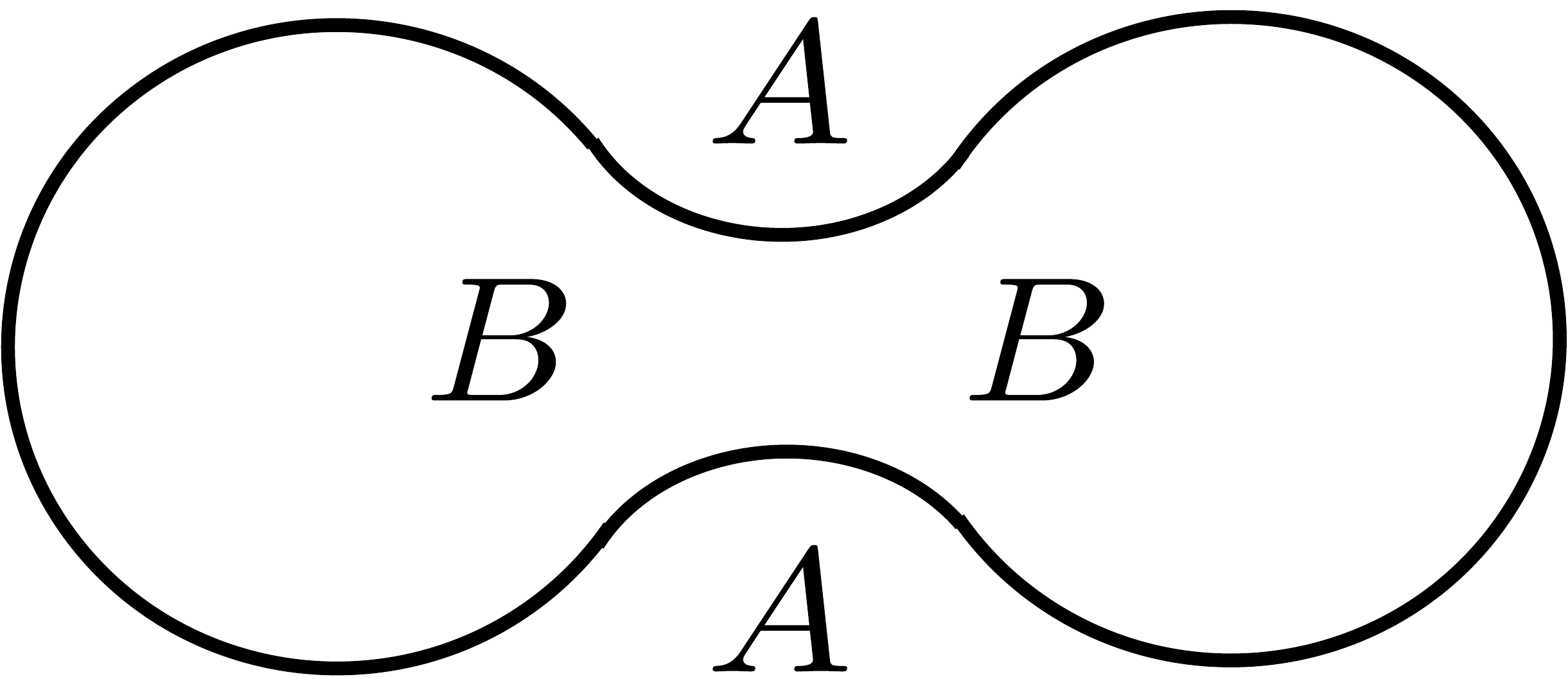}}.
\end{equation*}
The TEE for the left configuration is $-2\ln \mathcal{D}$, while for the right one it is $-\ln \mathcal{D}$.
Therefore, the limiting case with the tetrajunction is ill-defined in the context of TQFT.
Physically, as the length scale near the junction shrinks towards zero, it becomes smaller than the correlation length of the ultraviolet (UV) theory, meaning that the TQFT is no longer a valid approximation for the underlying UV theory.
We suspect that there may be a necessary corner contribution in such cases, which complicates the calculation of TEE.

To avoid these subtleties, we will focus on configurations where no such junctions occur.
More specifically, we require that the boundaries of different connected components remain disjoint.
With this physical consideration, the gluing of subregions becomes well-defined by gluing the closure of the subregions.
Consequently, we can consistently compute the TEE within the TQFT framework.

\subsection{Classification of bipartitions}

We now classify the bipartitions of a torus based on their entanglement interfaces.
These interfaces are composed of disjoint loops, which we categorize by the types of loops present.
On a torus, each loop is either a torus knot or a contractible loop.

Let $K(p,q)$ be the torus knot that winds $p\in \mathbb{Z}$ times around the meridian and $q\in \mathbb{Z}$ times around the longitude.
To prevent the knot from self-intersecting, we require that $p$ and $q$ be co-prime, meaning $\mathrm{gcd}(p,q) = 1$.
If the entanglement interface contains two different types of non-contractible torus knots (i.e., knots where at least one of $p$ or $q$ is different), these knots will necessarily intersect, violating our regularization assumption.
Thus, the entanglement interfaces we consider can only include one type of non-contractible torus knot, along with possibly some contractible loops.

Furthermore, any non-contractible torus knots must appear in pairs, as this is necessary for them to properly divide the torus into two distinct regions.
Thus, the general form of the entanglement interface can be expressed as follows:
\begin{equation}\label{Eq:GeneralBipartition}
    \partial A = \partial B =  (\amalg_{i=1}^n C_i) \amalg (\amalg_{j=1}^{2m} K_j(p,q)), 
\end{equation}
where the $C_i$'s represent contractible loops, $K_j(p,q) \simeq K(p,q)$ are non-intersecting torus knots of the same $p,q$, and $m,n \in \mathbb{Z}$ are integers.

\subsection{Canonical bipartitions on a torus}\label{Sec:CanonicalBipartition}

\begin{figure}[ht]
    \centering
    \includegraphics[height=2.5cm]{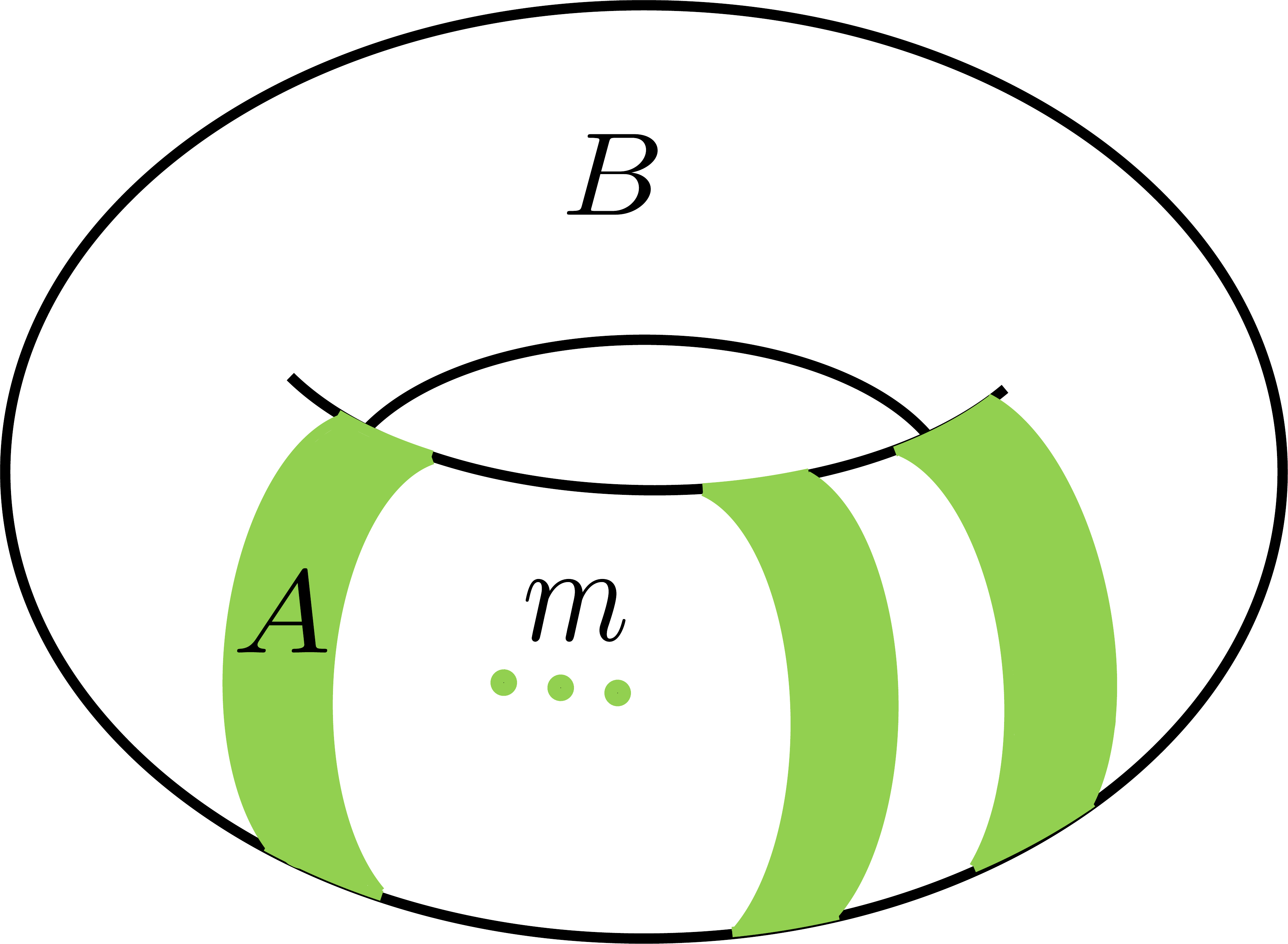}
    \caption{A canonical bipartition of $m$ rings.}
    \label{Fig:CanonicalBipartition1}
\end{figure}

\begin{figure}[ht]%
\centering
\subfigure[][]{%
\label{fig:ex1-a}%
\includegraphics[height = 2cm]{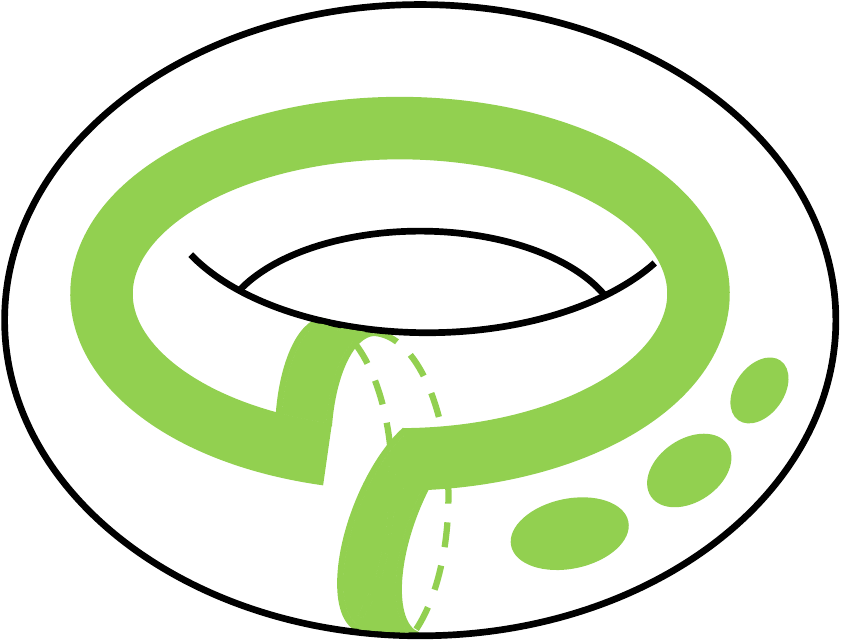}}%
\hspace{8pt}%
\subfigure[][]{%
\label{fig:ex1-b}%
\includegraphics[height = 2cm]{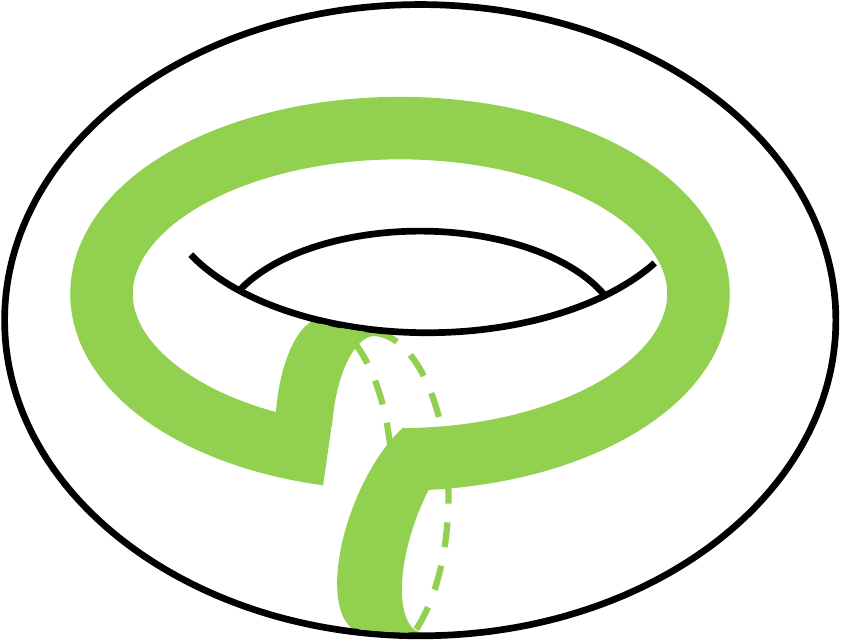}}%
\hspace{8pt}%
\subfigure[][]{%
\label{fig:ex1-c}%
\includegraphics[height = 2cm]{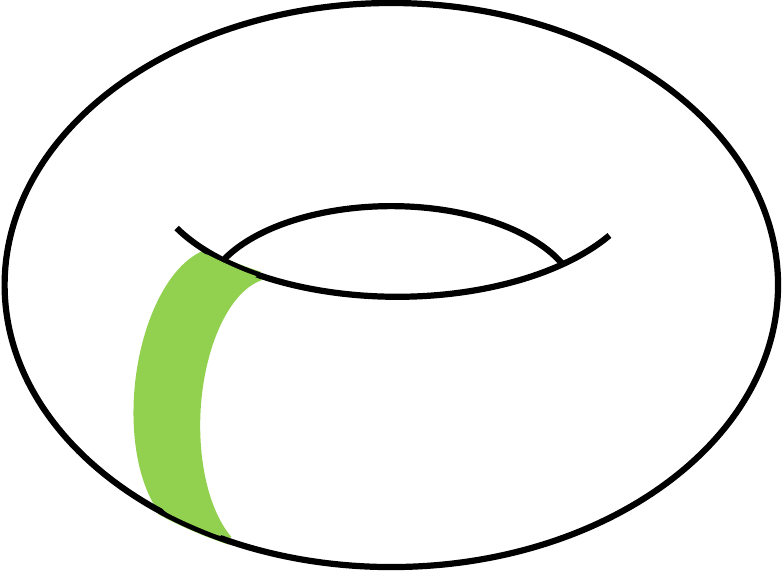}}%
\caption{An example of transforming a general bipartition into a canonical bipartition proceeds as follows: by removing the three contractible loops in \subref{fig:ex1-a}, we obtain the configuration shown in \subref{fig:ex1-b}, where the entanglement interface consists of two $K(1,1)$ knots. We then apply a coordinate transformation to this interface, converting the two $K(1,1)$ knots into two meridian loops. This transformation results in the canonical $R_1$ bipartition, as depicted in \subref{fig:ex1-c}.}%
\label{Fig:GeneralBipartition}
\end{figure}

We define a bipartition on the torus $T^2$ as a canonical bipartition $R_m$ if its entanglement interface consists solely of $2m$ meridians, i.e., 
\begin{equation}
    \partial A = \amalg_{j=1}^{2m} K_j(1,0),
\end{equation}
where we called $m$ the number of rings.
In this section, we will establish a connection between the TEE for an arbitrary bipartition and that of a canonical bipartition.
This connection will be made by systematically removing contractible loops (referred to as "bubbles") and applying appropriate coordinate transformations.
An illustrative example of this process can be found in Fig.~\ref{Fig:GeneralBipartition}.

\begin{figure}[ht]
    \centering
    \includegraphics[height=2cm]{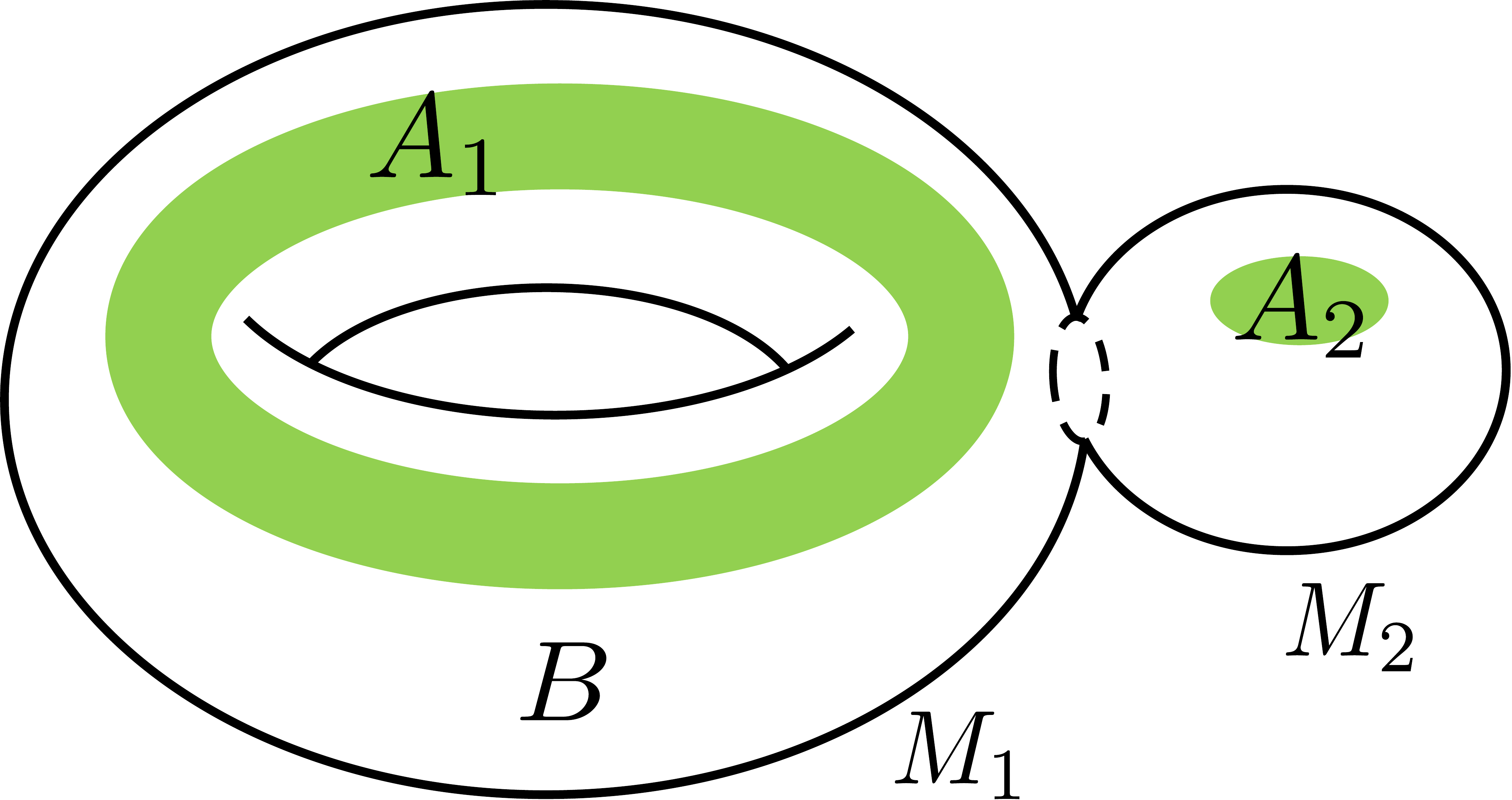}
    \caption{In this diagram, $A_1$ is represented by a non-contractible annulus, while $A_2$ corresponds to a disk. The dashed line indicates the $S^1$ boundary, and the disk bounded by the $S^1$ separates the solid torus into the left part $M_1$ and the right part $M_2$.}
    \label{Fig:Split1}
\end{figure}

Consider a vacuum state on the torus $T^2$, generated by a solid torus~\footnote{For simplicity, we will discuss the case without Wilson line insertions.
However, the argument remains valid in the presence of Wilson lines.} with the bipartition $T^2 = A \cup B$.
The subregion $A$ can be divided into two disjoint parts, $A = A_1 \amalg A_2$. Suppose that $A_2$ is contractible on $T^2$, there exist a circle $S^1 \subset T^2$ that encloses $A_2$, separating $B$ into $B_1$ and $B_2$.
We now extend such $S^1$ interface into a disk $D^2$ inside the bulk~\footnote{Any Wilson lines should be avoided in this consideration.}, which separates the solid torus into two components: $M_1 \simeq D^2\times S^1$ and $M_2 \simeq D^3$ (See Fig.~\ref{Fig:Split1}). In this setup, the original solid torus $M\simeq D^2 \times S^1$ can be viewed as the gluing of $M_1$ and $M_2$ along the common boundary $D^2$, i.e. $M = M_1 \cup_{D^2} M_2$. 

Next, we can construct the reduced density matrix by gluing the manifold with its orientation-reversed counterpart.
The two boundaries $D^2$ are glued to becomes a $S^2$ separating the glued manifold into two parts: $\rho_A = \Tilde{M_1} \cup_{S^2} \Tilde{M_2}$, where $\Tilde{M_i} = M_i \cup_{B_i} \Bar{M_i}$ with $i=1,2$. That is, the reduced density matrix becomes a connected sum of $\Tilde{M_1}$ and $\Tilde{M_2}$. One can apply the surgery method before performing the replica method
\begin{equation}
    \rho_A(M) = \frac{\rho_{A_1}(M_1) \otimes \rho_{A_2}(M_2)}{Z(S^3)}.
\end{equation}
Tracing the $n$-th power of the reduced density matrix, we obtain
\begin{equation}
    \mathrm{Tr} \rho_A(M)^n = \frac{ \mathrm{Tr}(\rho_{A_1}(M_1))^n \mathrm{Tr}(\rho_{A_2}(M_2))^n}{Z(S^3)^n}.
\end{equation}
In particular, if $A_2 \simeq D^2$, then $\mathrm{Tr}(\rho(M_2,A_2))^n = Z(S^3)$ and hence the TEE is given by
\begin{equation}
\begin{aligned}
    S(M,A) =& \lim_{n \rightarrow 1} \frac{1}{1-n}   \ln (\rho(M_1,A_1))^n \\
    &+ \lim_{n \rightarrow 1} \frac{1}{1-n} \ln Z(S^3)^{1-n}  \\
    =& S(M_1,A_1) - \ln \mathcal{D} 
\end{aligned}
\end{equation}
Thus, when computing the TEE, , we can effectively remove an isolated $D^2$ subregion—referred to as a "bubble"—by subtracting $\ln  \mathcal{D}$ from the final result. For example,
\begin{align}
   & S_{\text{TEE}} \Biggl( \adjustbox{valign = c}{\includegraphics[height = 2cm]{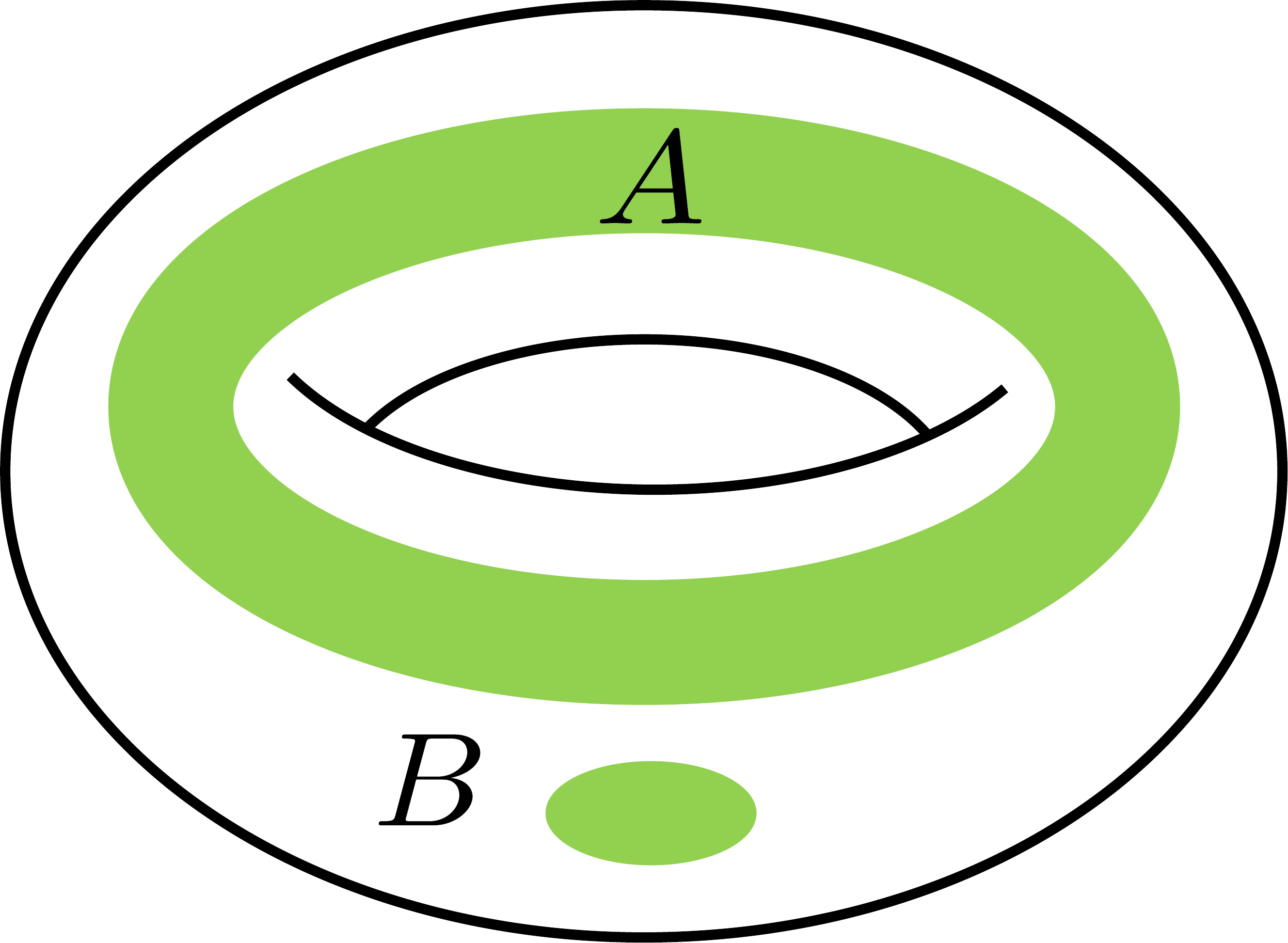}} \Biggl)  \notag\\
    = &S_{\text{TEE}} \Biggl( \adjustbox{valign = c}{\includegraphics[height = 2cm]{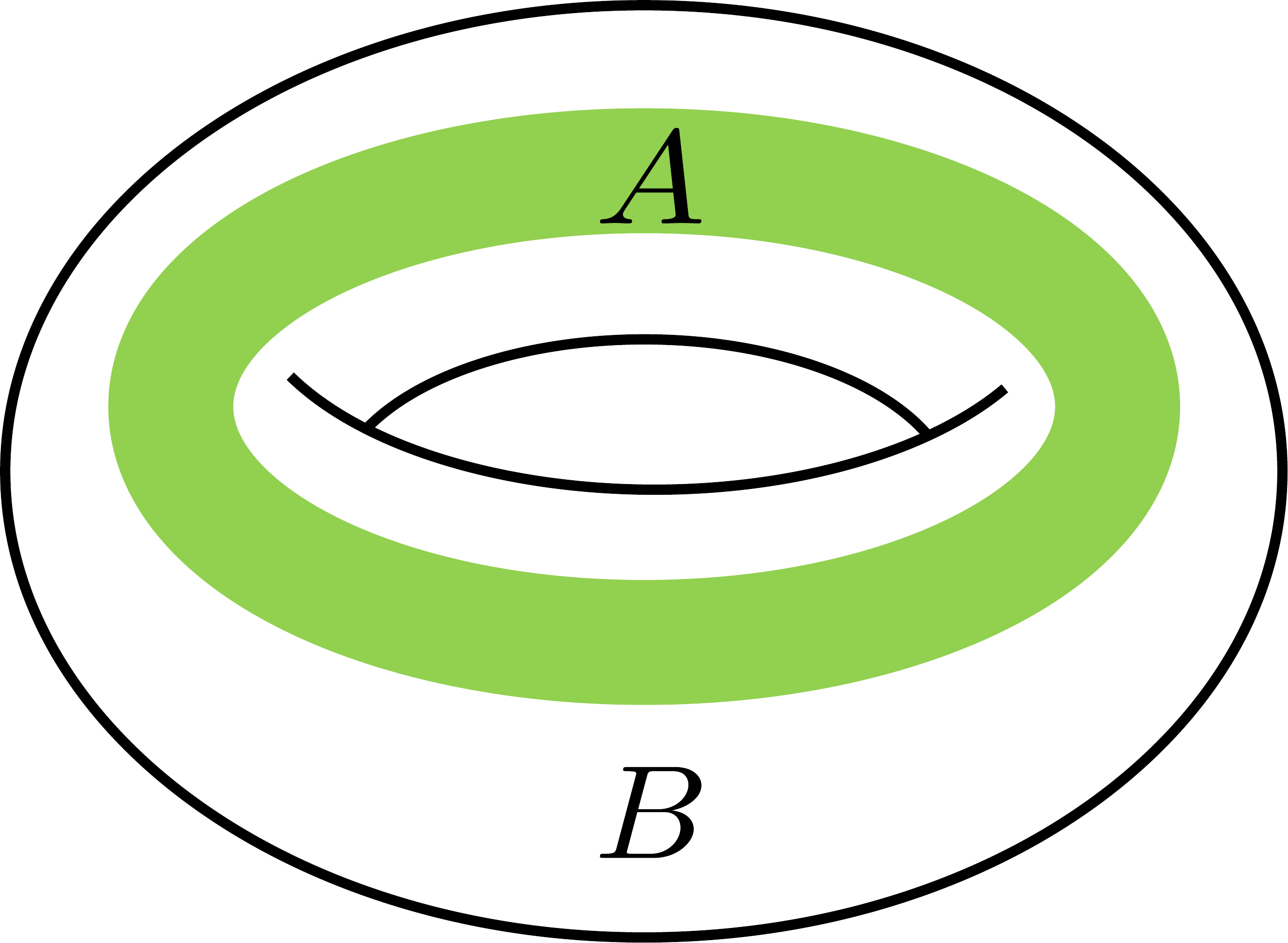}} \Biggl) - \ln  \mathcal{D}.
\end{align}
In general, if $A_2$ is a contractible region on the torus and $A_1 \cap A_2 = \emptyset$, then we have
\begin{equation}\label{Eq:Cut}
    S_{\text{TEE}}(T^2, A) = S_{\text{TEE}}(T^2, A_1) + S_{\text{TEE}}(S^3, A_2).
\end{equation}

There is also another type of bubble where the isolated $D^2$ subregion is located within $A$ rather than $B$.
In this scenario, we can leverage the fact that the TEE is identical for a subregion $A$ and its complement $B$.
By interchanging the roles of $A$ and $B$, the second type of bubble effectively transforms into the first type of bubble. Consequently, we can remove this bubble by adding $-\ln \mathcal{D}$ to the TEE.
For example, 
\begin{align}
    &S_{\text{TEE}} \Biggl( \adjustbox{valign = c}{\includegraphics[height = 2cm]{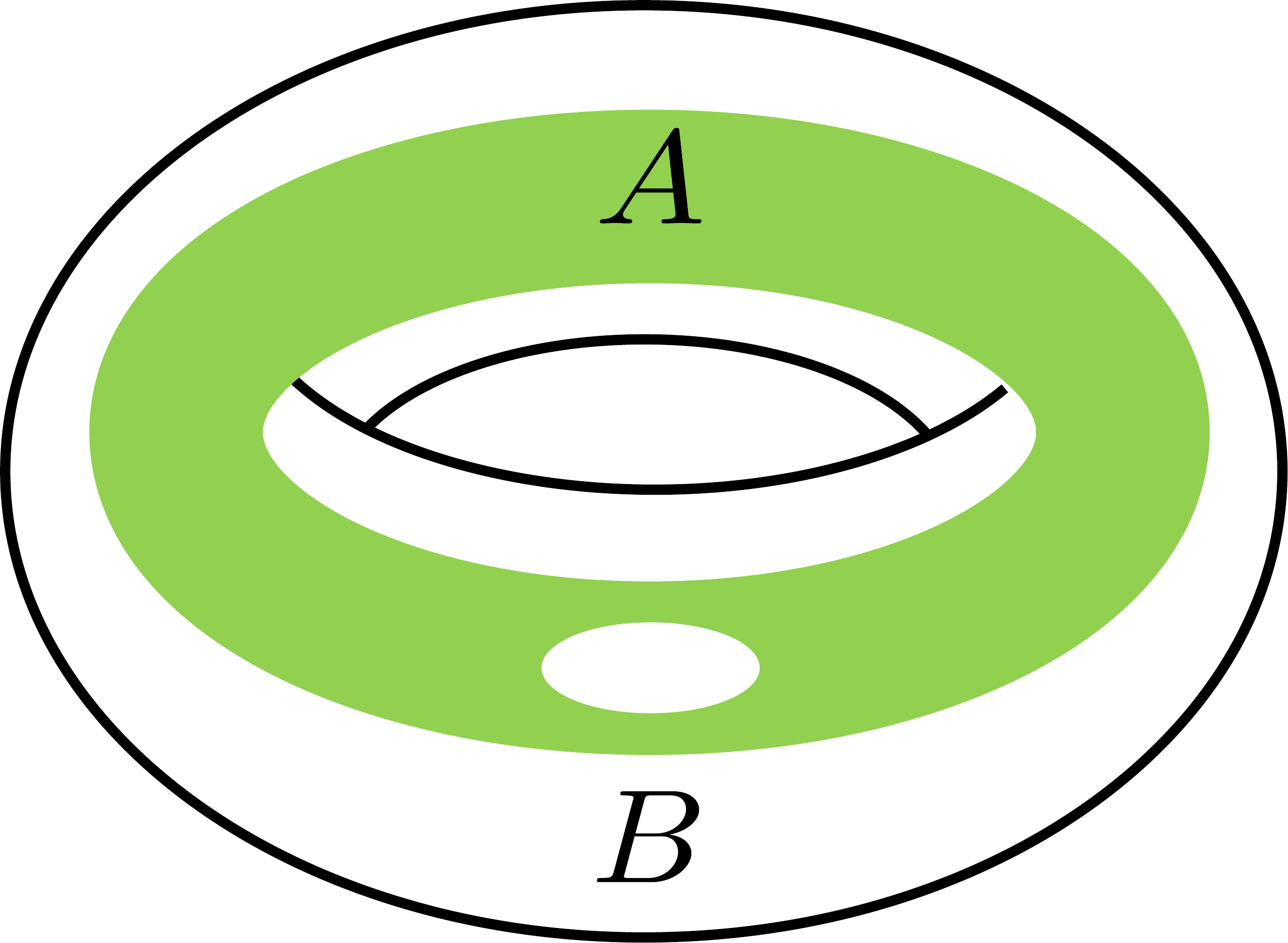}} \Biggl)  \notag \\
    = &S_{\text{TEE}} \Biggl( \adjustbox{valign = c}{\includegraphics[height = 2cm]{Figure/AnnulusBipartition.pdf}} \Biggl) - \ln  \mathcal{D}.
\end{align}

Finally, by induction, we can remove all contractible interfaces between the subregions by adding an integer multiple of $-\ln \mathcal{D}$. More explicitly, for a general bipartition represented by Eq.~(\ref{Eq:GeneralBipartition}), we have
\begin{align}\label{Eq:RemoveBubble}
    &S_{\text{TEE}}( (\amalg_{i=1}^n C_{i}) \amalg (\amalg_{j=1}^{2m} K_j(p,q)) )  \notag\\
    = &S_{\text{TEE}}( \amalg_{j=1}^{2m} K_{j}(p,q) ) - n \ln  \mathcal{D}.
\end{align}
Therefore, it remains to compute the TEE for a bipartition whose interface consists of $2m$ copies of torus knots.

In Ref.~\cite{KnotTEE}, we demonstrated that a state with a torus knot bipartition can be mapped to an effective state with a meridian bipartition.
In fact, the same mapping can be applied to a bipartition consisting of multiple torus knots.
Let $\ket{\psi} = \sum_a \psi_a \ket{a}$ represent the original state, and let $\mathcal{O}$ be the transformation that maps the torus knots back to the meridian. Then, we have:
\begin{equation} \label{Eq:EffectiveState}
    S(\ket{\psi}, \amalg_{j=1}^{2m} K_j(p,q)) = S(\ket{\mathcal{O}\psi}, R_m),
\end{equation}
where $\ket{\mathcal{O}\psi} = \sum_{a,b} \mathcal{O}_{ab} \psi_b \ket{a}$.
Therefore, we can study the TEE for any bipartition by examining the TEE for canonical bipartitions with Wilson lines inserted in the bulk.

\section{TEE for the canonical bipartitions}\label{Sec:TEE}

As discussed in the previous section, the TEE for a generic bipartition can be reduced to that of canonical bipartitions by removing contractible loops (bubbles) and untwisting torus knots. Therefore, it suffices to compute the TEE for canonical bipartitions with Wilson line insertions. The TEE for such multi-component canonical bipartitions has been explored in the literature~\cite{Reflect}, though prior work has primarily focused on general derivations using edge theory, with bulk computations performed only in specific cases. To address this gap, Appendix~\ref{Appendix:TEE} provides a systematic bulk derivation of the TEE for multi-component canonical bipartitions in the presence of Wilson lines, and we find our results to be in agreement with those obtained via the edge theory approach.

In summary, for a general bipartition, the TEE is given by
$S_\text{TEE}(\ket{\psi},A) = S_\text{iTEE}(A) + S_\text{cl}(\ket{\psi},A) + S_\text{Wil}(\ket{\psi},A)$.
The first term, $S_\text{iTEE}$, is a state-independent quantity that we refer to as the intrinsic TEE (iTEE):
\begin{equation}\label{Eq:iTEE}
    S_\text{iTEE}(A) = - \pi_{\partial A} \ln \mathcal{D},
\end{equation}
where $\pi_A$ denotes the number of entanglement interfaces, including both contractible and non-contractible components.
The second term is a classical contribution,
\begin{equation}
    S_\text{cl}(\ket{\psi},A) = - (1 - \delta_{m0}) \sum_a p_a \ln p_a,
\end{equation}
where $p_a = |\psi_a|^2$ is the classical probability associated with each topological sector, and $1 - \delta_{m0}$ ensures that this contribution appears only when there are non-contractible loops (i.e., $m\neq 0$).
The third term, $S_\text{Wil}$, captures the contribution from Wilson lines:
\begin{equation}
    S_\text{Wil}(\ket{\psi},A) = \sum_a p_a m\ln d_a,
\end{equation}
representing a weighted sum over Wilson line contributions, with $d_a$ denoting the quantum dimension of the anyon type $a$.

We define the ground state TEE to be the state dependent part
\begin{equation}
    S_\text{gs}(\ket{\psi},A) = S_\text{cl}(\ket{\psi},A) + S_\text{Wil}(\ket{\psi},A).
\end{equation}
One can find that if the subsystem $A$ is a union of several contractible loops  corresponding to $m=0$, the $S_\text{gs}(\ket{\psi},A)=0$ and $S_\text{TEE}(A)=S_\text{iTEE}(A) = - \pi_{\partial A} \ln \mathcal{D}$.
Since the ground state TEE is always non-negative, the iTEE serves as a lower bound for the TEE.
Furthermore, the ground state TEE vanishes for the effective vacuum state after untwisting, i.e., $\ket{\mathcal{O\psi}} = \ket{0}$.
Therefore, the iTEE is exactly the minimum of the TEE by varying the ground state
\begin{equation}
    S_{\text{iTEE}}(A) = \min_{\ket{\psi} \in \mathcal{H}} S_{\text{TEE}}(A, \ket{\psi}).
\end{equation}

\section{Modified SSA for the iTEE}\label{Sec:SSA_iTEE}

In this section, we will use topological analysis to demonstrate that the iTEE satisfies a modified version of the strong subadditivity (SSA):
\begin{equation}\label{Eq:WeakSubadditivity}
    \mathcal{I}_\text{iTEE}(A:B) \geq -2 \mathcal{I}_\mathrm{g}(A:B) \ln  \mathcal{D},
\end{equation}
where the $I_f(A:B)$ are defined in Eq.~\ref{Eq:CMI}, with $g$ representing the genus of subregions as will be discussed later.

\subsection{Properties of subregions}
We have previously discussed that the boundary of any subregion $A$ can be expressed as in Eq. (\ref{Eq:GeneralBipartition}).
The subregion $A$ can be decomposed into its connected components, $A = \amalg_{i=1}^{\pi_A} A_i$, where each $A_i$ is a connected component, and $\pi_A$ represents the number of these connected components.

These connected components are compact, connected surfaces, which can be obtained by adding $\pi_{\partial A_i}$ punctures to a compact surface without boundary of genus $g_{A_i}$.
The Euler characteristic of such a surface is
\begin{equation}
     \chi_{A_i} = 2 - 2g_{A_i} - \pi_{\partial A_i},
\end{equation}
where $\pi_{\partial A_i}$ denotes the number of connected components of the boundary $\partial A_i$.

\begin{figure}[ht]%
\centering
\subfigure[][]{%
\label{fig:ex3-a}%
\includegraphics[height = 2 cm]{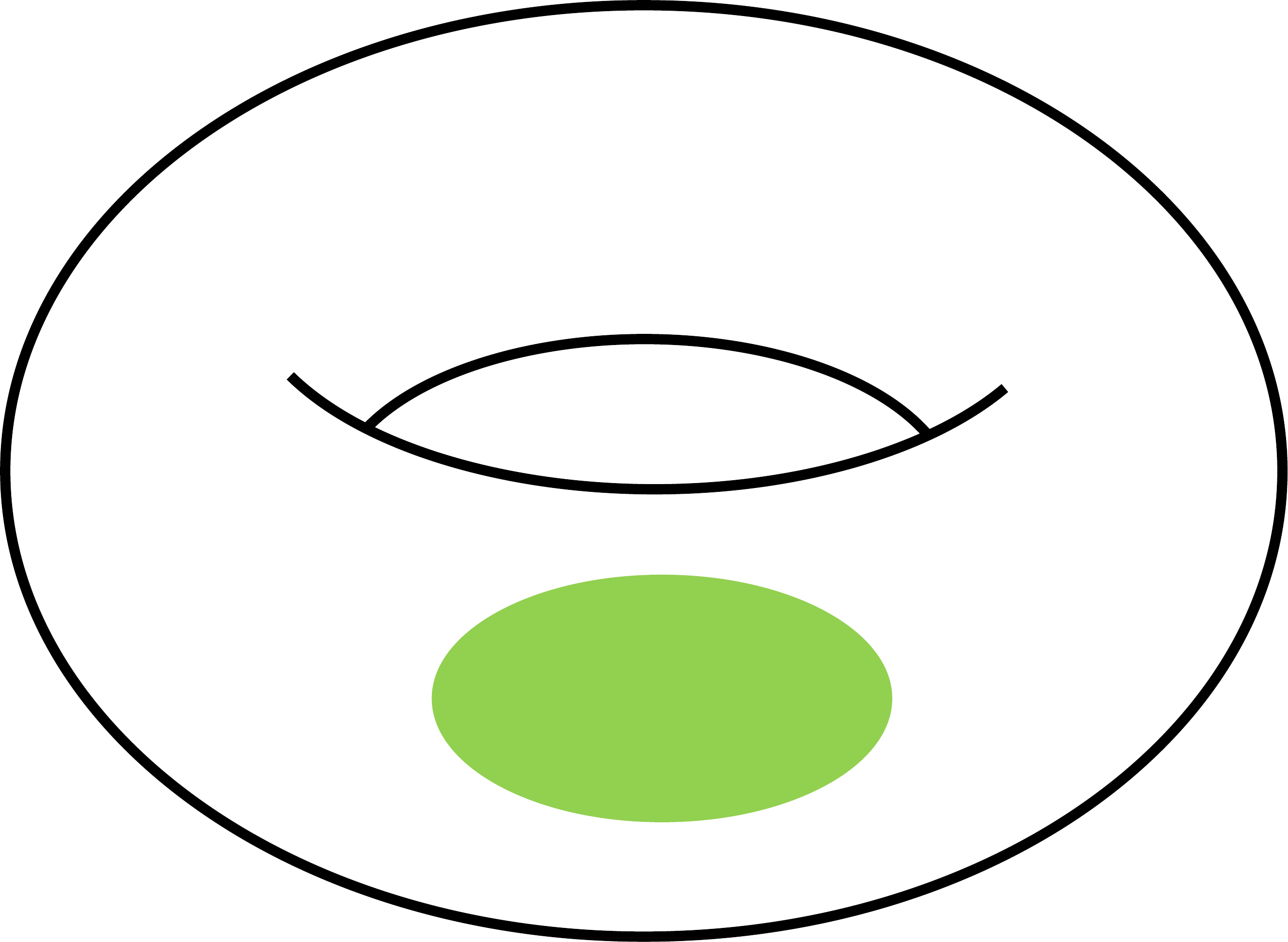}}%
\hspace{8pt}%
\subfigure[][]{%
\label{fig:ex3-b}%
\includegraphics[height = 2 cm]{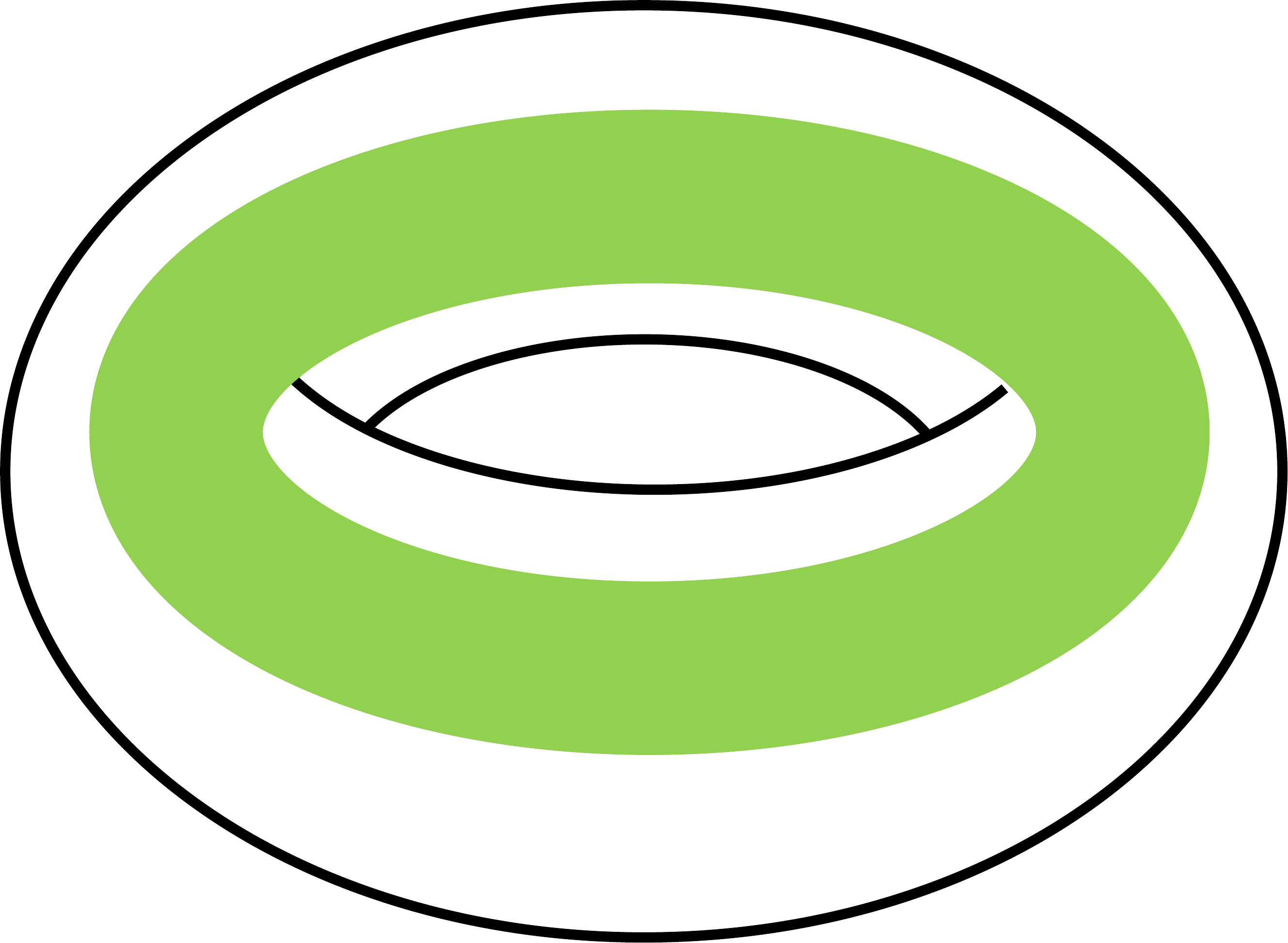}}
\subfigure[][]{%
\label{fig:ex3-c}%
\includegraphics[height = 2 cm]{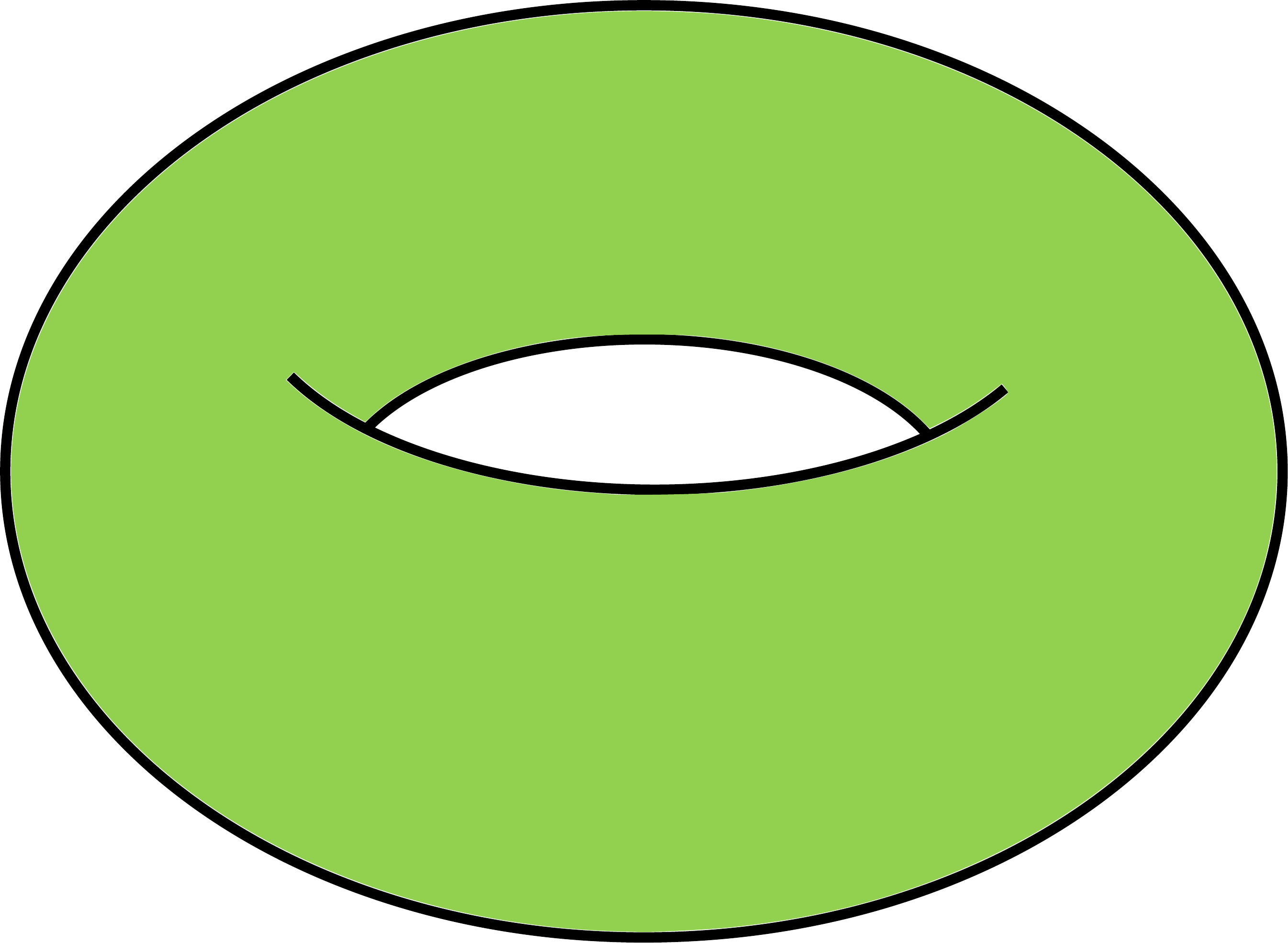}}%
\hspace{8pt}%
\caption[The three possible connected components.]{The green region $A_i$ on each torus illustrates the three possible connected components:
\subref{fig:ex3-a} Disk $(g_{A_i} = 0)$, \subref{fig:ex3-b} Ribbon $(g_{A_i} = 0)$, and \subref{fig:ex3-c} Torus $(g_{A_i} = 1)$. }%
\label{Fig:CntRegion}%
\end{figure}

Next, we classify the connected components of a subregion on the torus. There are three distinct types distinguished by the number of nontrivial torus cycles they encompass (See also Fig.~\ref{Fig:CntRegion})\footnote{In general, one can also consider extra punctures in the connected components. However, as we will show in later sections, these punctures will not affect the SSA properties.}

\begin{enumerate}
    \item \textbf{Disk}: This is the case in which the connected component contains no torus cycles. It can be obtained by puncturing a sphere $S^2$, giving it a genus $g_{A_i} = 0$.
    \item \textbf{Ribbon}: This connected component contains a single type of non-contractible torus knot $K(p,q)$. Its boundary consists of two such $K(p,q)$ knots. This surface also has a genus $g_{A_i} = 0$, as it can be derived from puncturing a sphere.
    \item \textbf{Torus}: This connected component contains all torus cycles, representing a full torus. This gives the surface a genus $g_{A_i} = 1$, corresponding to the genus of $T^2$.
\end{enumerate}

Now that we have determined the genus for each connected component $A_i$, the genus for the entire subregion $A$ is simply the sum of the $g_{A_i}$'s
\begin{equation}\label{Eq:Genus}
    2 g_A = 2 \sum_i g_{A_i}= 2\pi_A - \pi_{\partial A} - \chi_{A},
\end{equation}
where $\pi_A$ is the number of connected components, $\pi_{\partial A}$ is the number of boundary components, and $\chi_A$ is the Euler characteristic of the subregion $A$.

Moreover, since a torus component must intersects either with a ribbon or another torus, they cannot coexist within the same subregion $A$. 
As a result, subregions can be classified into only two types based on the genus: $g_A=0$ and $g_A=1$.

Furthermore, suppose $A \subset B$ and $g_A = 1$.
In this case, $A$ contains a torus component, which implies that there must be a component of $B$ that also contains this torus component.
Therefore, we have $g_B = 1 \geq g_A$.
This leads to the following monotonicity property:
\begin{equation}\label{Eq:Monotone}
    g_A \leq g_B \hspace{0.2cm} \text{if} \hspace{0.2cm} A \subset B.
\end{equation}

Since $A\subset A\cup B$ and $A\cap B \subset B$, by Eq.~(\ref{Eq:Monotone}) we have $g_{A} \leq g_{A\cup B}$ and $g_{A\cap B} \leq g_{B}$.
Therefore, we obtain the following bounds
\begin{equation}\label{Eq:IgBound}
    -1 \leq \mathcal{I}_g(A:B) \leq 1.
\end{equation}
After proving Eq.~(\ref{Eq:WeakSubadditivity}), this implies that the modification of the SSA for the intrinsic TEE is at most $2\ln \mathcal{D}$.

Now that we have expressed the number of boundary components in terms of the number of connected components and the Euler characteristic, we can leverage our understanding of topology to explore relationships related to the SSA.

\subsection{Relationship graph between subregions}\label{Sec:RelationGraph}
\begin{figure}[ht]
    \centering
    \includegraphics[height = 4 cm]{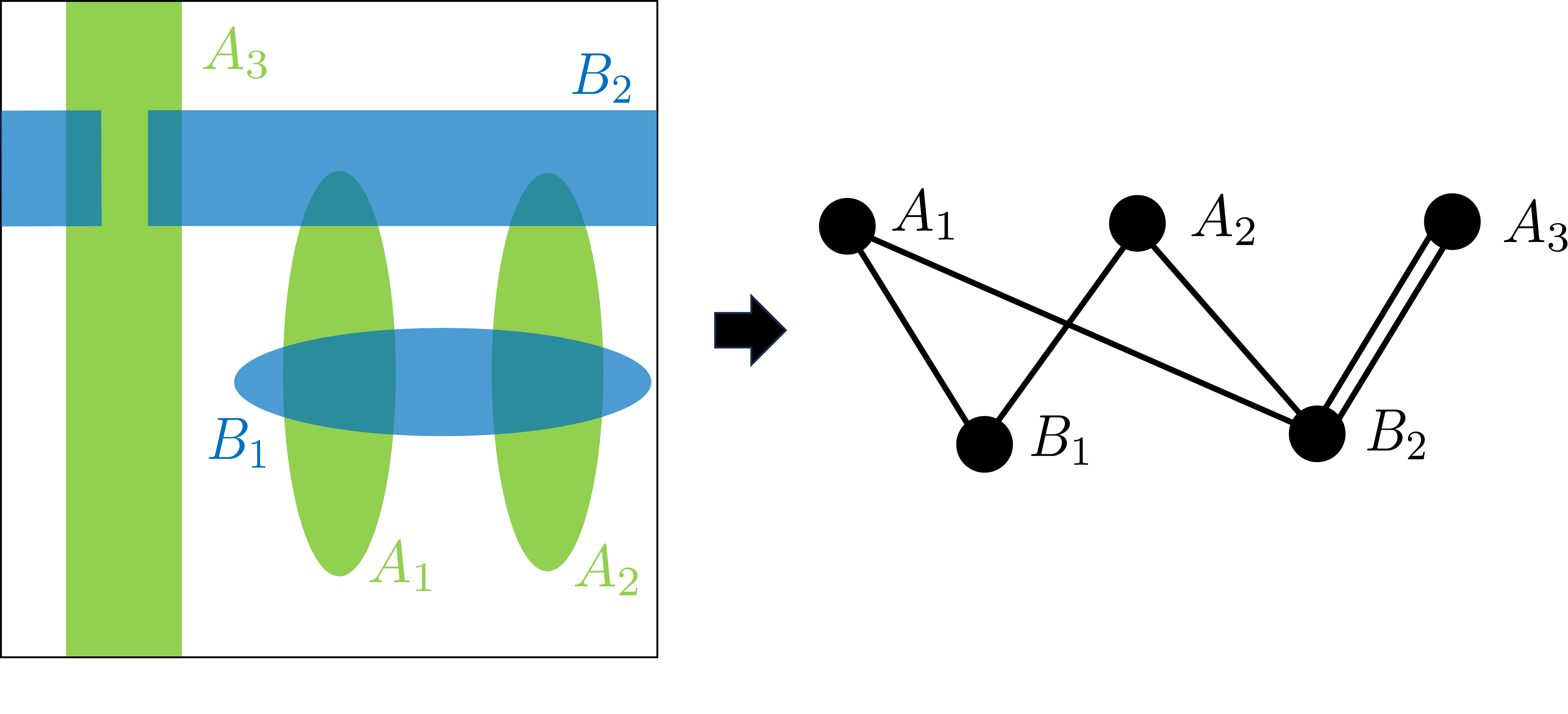}
    \caption{On the right is a graph showing the relationship graph of the configuration depicted in the figure on the left. In this case, there are $5$ vertices, $6$ edges, $1$ component and $2$ cycles.}
    \label{Fig:Graph}
\end{figure}

Since the ambient space is compact, we can write $A = \amalg_{i=1}^{\pi_A} A_i$, $B = \amalg_{j=1}^{\pi_B} B_j$, where $A_i, B_j$ represent the connected components of $A$ and $B$. We can construct a graph where the nodes correspond to the $A_i$ and $B_j$ components. We draw $n$ edges between $A_i$ and $B_j$ if the intersection $A_i \cap B_j$ contains $n$ connected components. An example is shown in Fig.~\ref{Fig:Graph}.

In graph theory, a well-known result states that for any connected graph, the number of fundamental cycles is given by:
\begin{equation}\label{Eq:Graph}
    \text{Cycles} = \text{Edges} - \text{Vertices} + \text{Components}.
\end{equation}
In our case, the graph consists of $\pi_{A\cup B}$ connected components, the number of vertices can be identified as $\pi_A + \pi_B$, and the number of edges is $\pi_{A\cap B}$.
Substituting these into the equation, we obtain: 
\begin{equation}\label{Eq:ConnectedComponent}
    \text{Cycles} = \pi_{A\cap B} + \pi_{A\cup B} - \pi_A - \pi_B.
\end{equation}
Using the notation introduced in Eq.~(\ref{Eq:CMI}), this relationship can be expressed compactly as: 
\begin{equation}\label{Eq:IPi}
    \mathcal{I}_\pi(A:B) = - \text{Cycles} \leq 0,
\end{equation}
where the number of fundamental cycles is a non-negative integer.

On the other hand, since our ambient space is compact, we can apply the inclusion-exclusion principle for the Euler characteristic, which states that:
\begin{equation}\label{Eq:EulerCharacteristic}
    \chi_A + \chi_B = \chi_{A\cup B} + \chi_{A\cap B},
\end{equation}
This means that the Euler characteristic is fully captured by the Venn diagram:
\begin{equation}\label{Eq:IChi}
    \mathcal{I}_\chi(A:B) = 0.
\end{equation}
Next, we use  Eq.~(\ref{Eq:iTEE}) and apply $\mathcal{I}$ to each term of Eq.~(\ref{Eq:Genus}), leading to the relation:
\begin{equation}\label{Eq:Inotation}
    \mathcal{I}_\text{iTEE} = 2 (\mathcal{I}_g - \mathcal{I}_\pi - \mathcal{I}_\chi )\ln  \mathcal{D}.
\end{equation}
With Eq.~(\ref{Eq:IPi}) and Eq.~(\ref{Eq:IChi}), we obtain the following inequality:
\begin{equation}\label{Eq:ModifiedSSA}
    \mathcal{I}_\text{iTEE}  = (\text{Cycles} +  2\mathcal{I}_g) \ln \mathcal{D} \geq  2\mathcal{I}_g \ln \mathcal{D} .
\end{equation}

In conclusion, the strong subadditivity of the intrinsic TEE receives a modification due to the presence of $\mathcal{I}_g$.
This modification is mild, as the absolute value of $\mathcal{I}_g$ is bounded by $1$.

\section{The SSA of \texorpdfstring{$S_{\text{TEE}}$}{S\_TEE}}\label{Sec:SSA_TEE}

For topologically ordered systems, the entanglement entropy receives a subleading correction to the area law~\cite{TEE,TEE2}:
\begin{equation*}
S_\text{EE}(A) = \alpha L_{A} - \gamma_A.
\end{equation*}
The area term, $\alpha L$, satisfies the additivity condition implied by the Venn diagram:
\begin{equation}
\mathcal{I}_L(A:B) = L_A + L_B - L_{A \cup B} - L_{A \cap B} = 0.
\end{equation}
As a result, the SSA condition for $-\gamma$ holds if and only if SSA is satisfied for $S_{\mathrm{EE}}$.

Since the SSA condition for $S_{\mathrm{EE}}$ is guaranteed by the properties of the von Neumann entropy, it follows that SSA must also hold for $-\gamma$. In particular, if the subleading correction corresponds exactly to the TEE computed via TQFT or edge theory ($S_\text{TEE} = -\gamma$), then the SSA condition applies directly to the TEE.

However, in practice, spurious (non-topological) contributions can obstruct a direct identification of the subleading term with the topological entanglement entropy. To address this, we adopt a topologically grounded perspective on SSA that remains independent of any specific microscopic model.

\begin{figure}[ht]%
\centering
\subfigure[][]{%
\label{fig:ex4-a}%
\includegraphics[height = 2.5 cm]{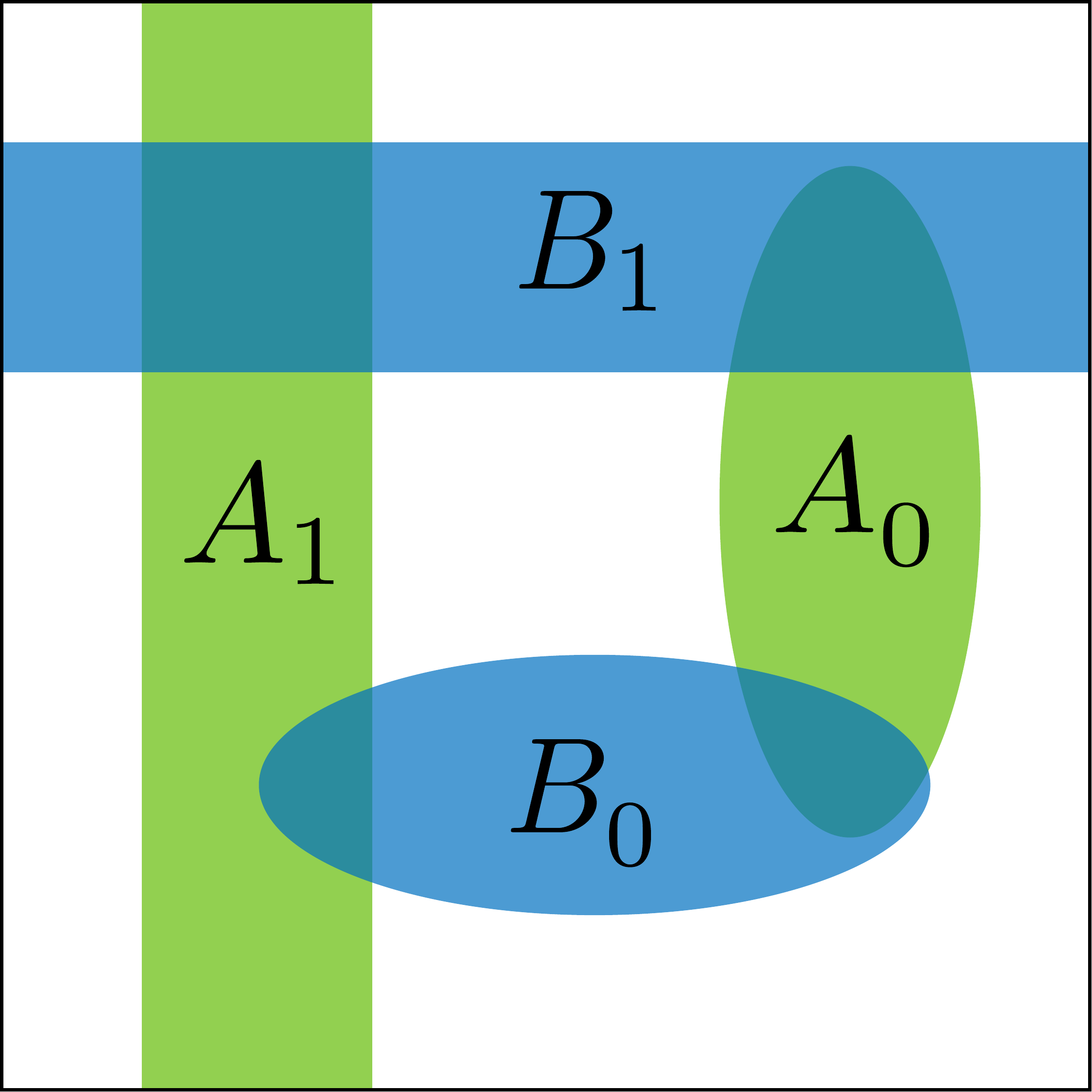}}%
\hspace{8pt}%
\subfigure[][]{%
\label{fig:ex4-b}%
\includegraphics[height = 2.5 cm]{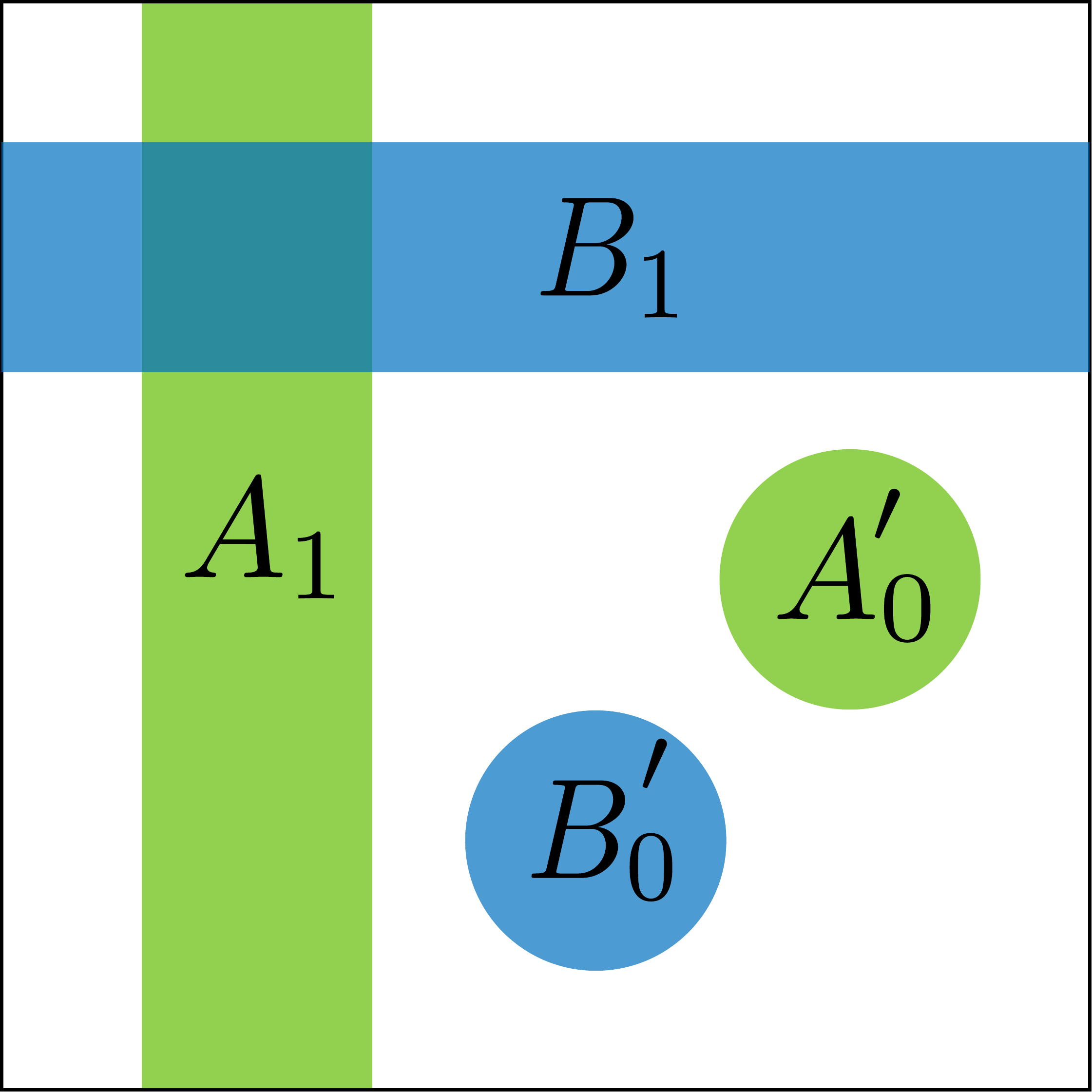}}%
\hspace{8pt}%
\subfigure[][]{%
\label{fig:ex4-c}%
\includegraphics[height = 2.5 cm]{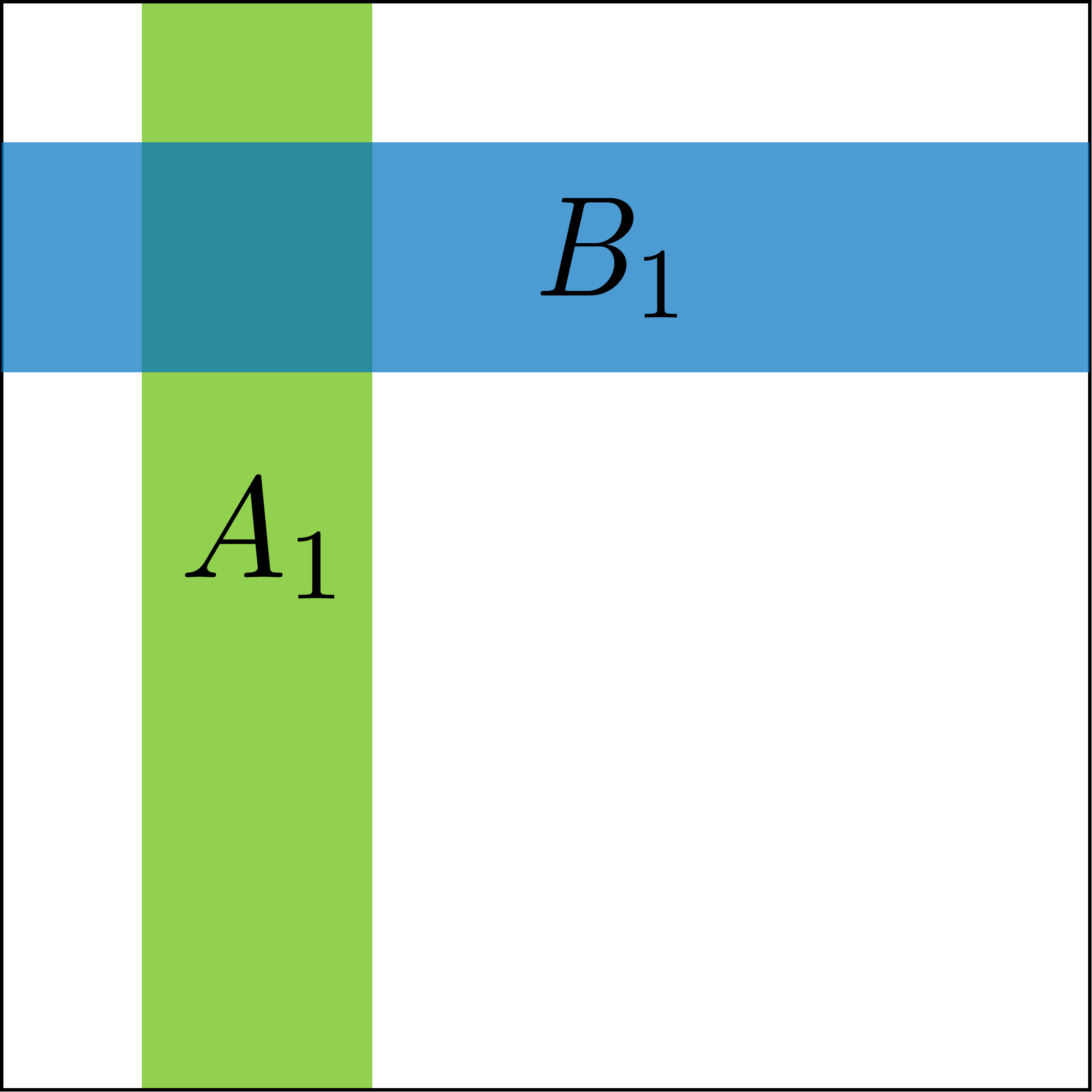}}%
\hspace{8pt}%
\caption[]{The schematic diagram for the three steps:
\subref{fig:ex3-a} which is the general case where $A_0, B_0$ represent the contractible subregions and $A_1, B_1$ represent the non-contractible ones,
\subref{fig:ex3-b} where all contractible regions are deformed such that they are isolated, and
\subref{fig:ex3-c} where $A, B$ become ribbons after the contractible isolated regions are removed.}%
\label{Fig:Conceptual}%
\end{figure}

More explicitly, we demonstrate that the SSA condition for $S_{\text{TEE}}$ holds if and only if Eq.(\ref{Eq:Conjecture}) is satisfied. Our argument proceeds in three conceptual steps, illustrated in Fig.\ref{Fig:Conceptual}.

First, We show that the topological CMI decreases when removable contractible intersections are removed.
Therefore, the SSA condition for general cases can be implied by those where all removable contractible intersections are removed.
In particular, all contractible regions can be isolated.

Next, we use the surgery method to demonstrate that the SSA condition is invariant under the removal of contractible, isolated regions. 
This further simplifies the problem, leaving us with cases where regions $A$ and $B$ consist solely of torus knot ribbons.

Finally, we classify these cases by the intersection number $i_{AB}$ between the torus knots of $A$ and $B$.
We show that the SSA condition for both the TEE and the iTEE holds for $i_{AB} \geq 2$ and $i_{AB} = 0$.
Moreover, we show that the SSA for $i_{AB} = 1$ holds if and only if Eq.~(\ref{Eq:Conjecture}) is satisfied.

Although we are unable to prove the general validity of Eq.~(\ref{Eq:Conjecture}), we have verified its correctness for all unitary modular categories up to rank $11$~\cite{Note1}. 
Thus, we conjecture that it holds for all unitary modular categories.

\subsection{Removing contractible intersections}\label{Subsec:Isolate}

In this subsection, we will give a proof that the topological CMI can only decrease when contractible intersections $A_0\cap B$ is homotopically removed, where $A_0$ is a connected component of $A$.~\footnote{Contractible is essential here. For the case where $A$ and $B$ are disjoint $R_1$ bipartitions, for which $\mathcal{I}_\text{TEE}(A:B) = S_\text{cl} \geq 0$. If we introduce an intersection between $A$ and $B$ at a $R_1$ region, we obtain $\mathcal{I}_\text{TEE}(A:B) = 0$, meaning that the topological CMI actually decreases when a non-contractible intersection is introduced.}.
Intuitively, shared region will correspond to shared topological CMI for such scenarios. 
Therefore, one can remove every contractible intersections between $A$ and $B$ by induction.

More explicitly, let $A = A_0 \amalg A_1$, where $A_0$ is a connected component of $A$ such that $A_0 \cap B$ consists of removable contractible disks.
If we were to homotopically deform $A_0$ to $A_0'$ such that $A'_0 \cap B =\emptyset$, then we will show that
\begin{equation}
    \Delta \mathcal{I}_\text{TEE} = \mathcal{I}_\text{TEE}(A': B) - \mathcal{I}_\text{TEE}(A: B) \leq 0
\end{equation}
We will first show in Sec.~\ref{Sec:IiTEE} that $\Delta \mathcal{I}_\text{iTEE} \leq 0$ and then demonstrate in Sec.~\ref{Sec:GsTEE} that $\Delta \mathcal{I}_\text{gs} > 0$ implies $\Delta \mathcal{I}_\text{TEE} \leq 0$.
Combining the results, we prove our claim.

\subsubsection{ \texorpdfstring{$\Delta \mathcal{I}_\text{iTEE} \leq 0$}{Delta I\_iTEE <=0}}\label{Sec:IiTEE}

In this subsection, we aim to show $\Delta \mathcal{I}_\text{iTEE} \leq 0$.
Using Eq.~(\ref{Eq:Inotation}) and Eq.~(\ref{Eq:IChi}), it suffice to show that
\begin{equation}
    \Delta \mathcal{I}_\pi  \geq \Delta \mathcal{I}_g.
\end{equation}

Since the deformation is homotopic, the genus of $A$, $B$ and $A \cap B$ remains unchanged, we have:
\begin{equation*}
    \Delta \mathcal{I}_g = - \Delta g_{A\cup B} \leq 1,
\end{equation*}
where we use the fact that $g_{A\cup B}$ can either be $0$ or $1$.

Since removing intersections correspond to removing edges to the relationship graph described in Sec.~\ref{Sec:RelationGraph}, the number of cycles can only decreases, i.e.,
\begin{equation*}
    \Delta \mathcal{I}_\pi = - \Delta \text{Cycles} \geq 0.   
\end{equation*}
Therefore, it remains to show the proof for $\Delta g_{A\cup B} = -1$.

\begin{figure}[ht]
    \centering
    \includegraphics[height = 3.2 cm]{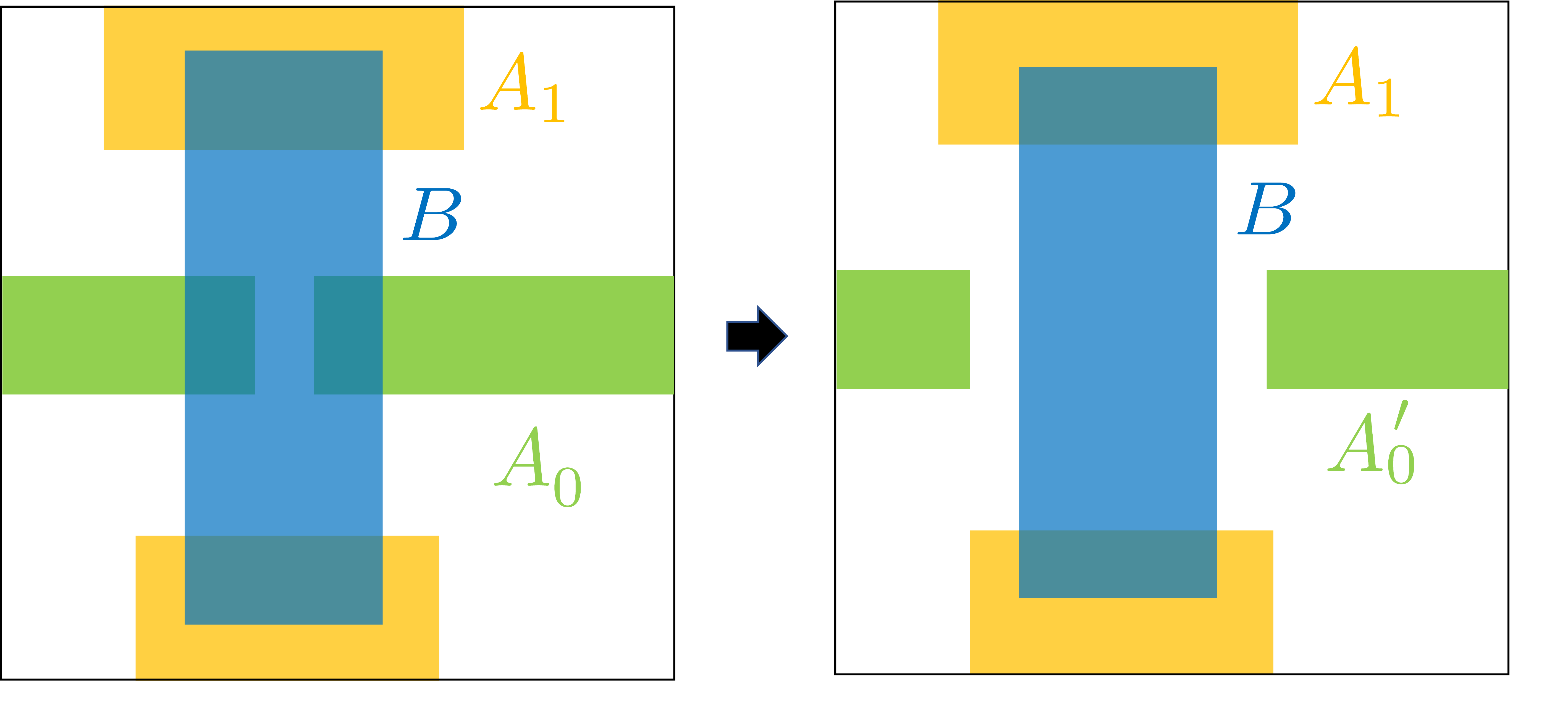}
    \caption{An example where $\Delta g_{A \cup B} = -1$ and $\Delta \text{Cycle} = -1$.}
    \label{Fig:NonTrivialIntersection1}
\end{figure}

In such a case, $A \cup B$ must form a torus component, which necessarily contains a torus cycle that is not present in $A' \cup B$.  
This torus cycle must be composed of parts from both $A$ and $B$, implying that it corresponds to a cycle in the relationship graph.  
Consequently, the cycle in the graph of $A \cup B$ that represents this fundamental torus cycle does not exist in the graph of $A' \cup B$ (see Fig.~\ref{Fig:NonTrivialIntersection1} for an illustration).
Consequently, we have:
\begin{equation}
    \Delta \mathcal{I}_\pi = - \Delta\text{Cycles} \geq 1 = \Delta \mathcal{I}_g,
\end{equation}
which finishes the proof.

\subsubsection{ \texorpdfstring{$\Delta \mathcal{I}_\text{gs} > 0$}{Delta I\_gs > 0} implies \texorpdfstring{$\Delta \mathcal{I}_\text{TEE} \leq 0$}{Delta I\_TEE <= 0}}\label{Sec:GsTEE}

Since the deformation is homotopic, $S_\text{gs}(A)$ and $S_\text{gs}(B)$ remain unchanged.
Moreover, the intersections removed from $A_0\cap B$ are contractible and therefore do not contribute to any change in $S_\text{gs}(A\cap B)$.
Thus, we have
\begin{equation*}
    \Delta\mathcal{I}_\text{gs}  = - \Delta S_\text{gs}(A\cup B).
\end{equation*}
Thus, to complete the argument, it suffices to show that
\begin{equation*}
    \Delta S_\text{gs}(A\cup B) < 0 \Rightarrow\Delta \mathcal{I}_\text{iTEE} - \Delta S_\text{gs}(A\cup B) \leq 0.
\end{equation*}

\begin{figure}[ht]
    \centering
    \includegraphics[height = 3.2 cm]{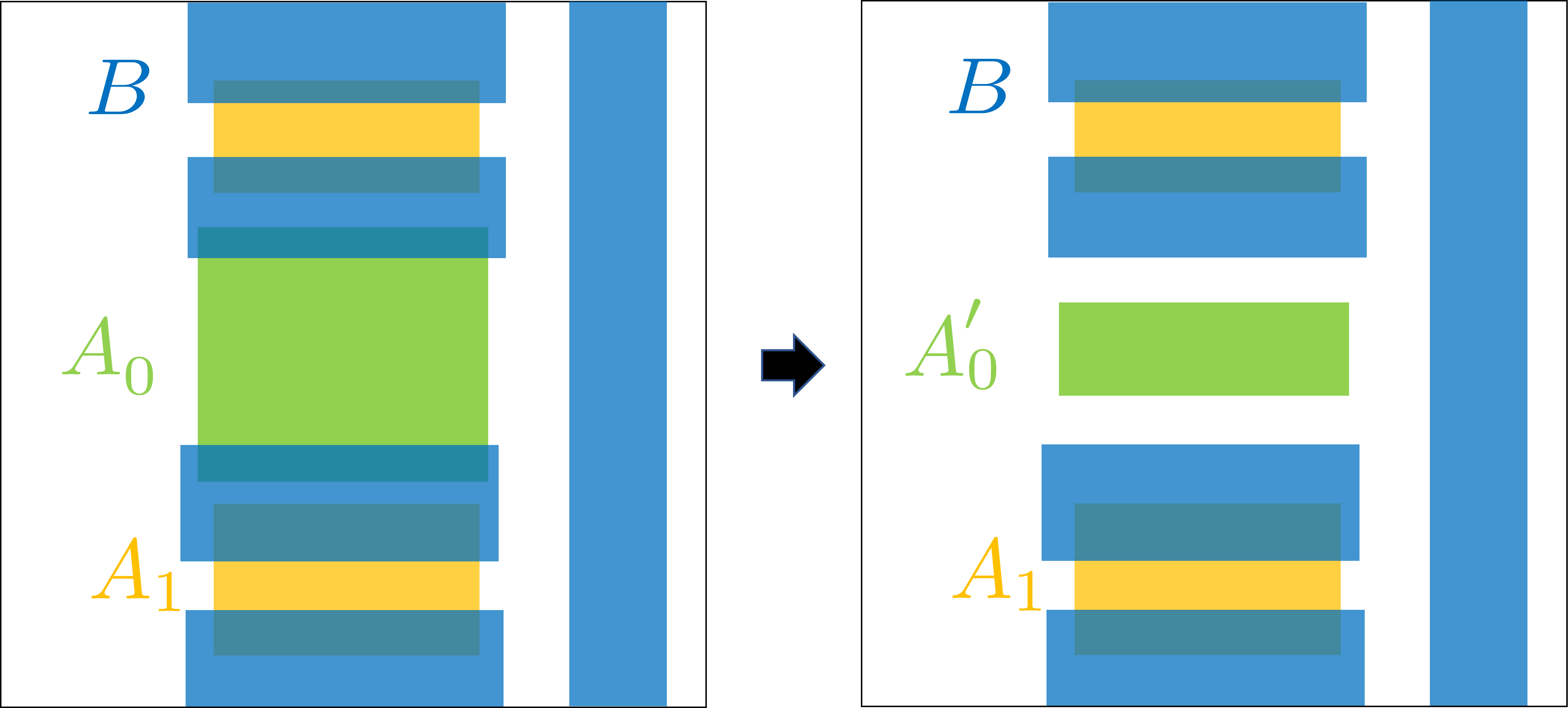}
    \caption{An example where $A\cup B$ lose a ribbon component. In this example, $\Delta S_\text{gs}(A\cup B) = -\sum_a 2|\psi_a|^2 \ln d_a$, and $\Delta S_\text{iTEE}(A\cup B) = 2\ln  \mathcal{D}$.}
    \label{Fig:NonTrivialIntersection2}
\end{figure}

For $\Delta S_\text{gs}(A\cup B) < 0$ to hold, the region $A'\cup B$ must lose at least one ribbon component as a result of the deformation (see Fig.~\ref{Fig:NonTrivialIntersection2}).
This ribbon necessarily includes parts from both $A$ and $B$, and therefore corresponds to a cycle in the relationship graph.
Consequently, $\Delta\text{Cycles} \leq -1$.
Furthermore, since both genus of $A\cup B$ remains unchanged after the deformation (i.e., $g_{A \cup B} = g_{A' \cup B}$), we have $\Delta \mathcal{I}_g = 0$.
Using Eq.~(\ref{Eq:ModifiedSSA}), we have
\begin{equation*}
    \Delta \mathcal{I}_\text{iTEE} \leq -2 \ln D.
\end{equation*}

On the other hand, from Eq.~(\ref{Eq:GSTEE}), we find
\begin{equation*}
    \Delta S_\text{gs}(A\cup B) = 
    \begin{cases}
        -\sum_a 2|\psi_a|^2 \ln \frac{d_a}{|\psi_a|},\; \text{if} \hspace{0.1cm} m_{A\cup B} = 0\\
        -\sum_a 2|\psi_a|^2 \ln d_a ,\; \text{if} \hspace{0.1cm} m_{A\cup B} > 0
    \end{cases},
\end{equation*}
where $m_{A\cup B}$ is the number of ribbon components in the original region $A\cup B$. 
In both cases, we find that
\begin{equation}
    \Delta S_\text{gs}(A\cup B) \geq -2 \ln \mathcal{D} \geq \Delta \mathcal{I}_\text{iTEE}.
\end{equation}
We conclude that $\Delta \mathcal{I}_\text{gs} < 0$ implies $\Delta \mathcal{I}_\text{TEE} \geq 0$, completing the proof.

\vspace{.5cm}
To sum up, in Sec.~\ref{Sec:IiTEE} and Sec.~\ref{Sec:GsTEE}, we proved that $\mathcal{I}_\text{TEE}$ decreases upon removing intersections. Therefore, the SSA condition for generic cases can be deduced by the cases where all removable contractible intersections are removed.

\subsection{Removing contractible isolated regions}\label{Subsec:RemovIso}

In the previous subsection, we established that the SSA condition for a general bipartition can be deduced by considering configurations in which all removable contractible intersections between $A, B$ are removed. In particular, any contractible component in $A, B$ can be isolated without loss of generality. In this subsection, we further demonstrate that the SSA condition remains invariant under the removal of such isolated contractible components.

Since $A_0$ is contractible, and $A_0 \cap A_1 = \emptyset$, we can apply Eq.~(\ref{Eq:Cut}) and obtain:
\begin{equation*}
    S_\text{TEE}(A) = S_\text{TEE}(A_0) + S_\text{TEE}(A_1).
\end{equation*}
Similarly, $A_0 \cap B = \emptyset$, we obtain:
\begin{equation*}
    S_\text{TEE}(A\cup B) = S_\text{TEE}(A_0) + S_\text{TEE}(A_1\cup B).
\end{equation*}
Additionally, because $A \cap B = A_1 \cap B$, it follows that:
\begin{equation*}
    S_\text{TEE}(A\cap B) = S_\text{TEE}(A_1\cap B).
\end{equation*}
By combining these results, we arrive at:
\begin{equation}
    \mathcal{I}_\text{TEE}(A:B) = \mathcal{I}_\text{TEE}(A_1:B).
\end{equation}
This demonstrates that the SSA condition for $A$ is equivalent to that for $A_1$, meaning that $A_0$ can be removed from the discussion of SSA.

Since any disk components are contractible and isolated, they can be removed when proving the SSA property.
Additionally, the complement of a torus component is a disk.
Together with Eq.~(\ref{Eq:ComplementInv}), which shows that the SSA condition is invariant under the simultaneous complement of $A$ and $B$, this allows us to remove these components as well when verifying the SSA condition.
Furthermore, punctures in connected components act as contractible components under complement operations, meaning they can also be removed.
Therefore, what remains is to establish the SSA for cases where both $A$ and $B$ consist solely of ribbon components.

\subsection{The SSA for \texorpdfstring{$S_\text{TEE}$}{S\_TEE} for ribbons}\label{Subsec:CanonicalSSA}

From Sec.~\ref{Subsec:Isolate} and Sec.~\ref{Subsec:RemovIso}, we have seen that showing the SSA for $S_\text{TEE}$ of arbitrary $A$ and $B$ reduces to showing the SSA for cases where $A$ and $B$ consist solely of ribbon components with minimal contractible intersections.

Let $A$ consist of $\pi_A$ copies of $K(p_A,q_A)$ ribbons and $B$ consist of $\pi_A$ copies of $K(p_A,q_A)$ ribbons. The number of intersections between them is given by $\pi_A \pi_B i_{AB}$, where $i_{AB} = |q_A p_B - q_B p_A|$ represents the intersection number between $K(p_A,q_A)$ and $K(p_B,q_B)$. In the following, we classify the cases based on $i_{AB}$.

\subsubsection{ \texorpdfstring{$i_{AB} \geq 2$}{i\_AB >= 2}}\label{Subsec:i=2}

First, we show that $\mathcal{I}_\text{iTEE} \geq 0$. By Eq.~(\ref{Eq:Inotation}), it suffice to show that $\mathcal{I}_\pi \leq \mathcal{I}_g$.
For $i_{AB} \neq 0$, the ribbons of $A$ and $B$ belong to different types of torus knots.
Therefore, $A\cup B$ is a torus with $\pi_{A\cup B} = 1$, while $A\cap B$ consists of $\pi_{A\cap B}$ disks, which can be written in terms of intersection number:
\begin{equation}
    \pi_{A\cap B} = \pi_A \pi_B i_{AB}.
\end{equation}
Since $g_{A\cup B}=1$ while $g_{A}=g_B = g_{A\cap B} = 0$, we have $\mathcal{I}_g = -1$.
Therefore,
\begin{equation}\label{Eq:Criteria}
    \mathcal{I}_\pi - \mathcal{I}_g = \pi_A + \pi_B - \pi_A \pi_B i_{AB}.
\end{equation}
Hence, for $i_{AB} \geq 2$, we have $\mathcal{I}_\text{iTEE} \geq 0$.
That is, the SSA condition for the iTEE holds.

Furthermore, $S_\text{gs}(A\cap B) = S_\text{gs}(A\cup B) = 0$, it follows that $\mathcal{I}_\text{gs} \geq 0$.
Therefore, $\mathcal{I}_\text{TEE} \geq \mathcal{I}_\text{iTEE} \geq 0$, which confirms that the SSA holds for both the iTEE and TEE.

\subsubsection{ \texorpdfstring{$i_{AB} = 0$}{i\_AB = 0}}

Next, we consider the case where $i_{AB} = 0$; that is, the torus knots associated with regions $A$ and $B$ are of the same type.  
As in the case with $i_{AB} \geq 2$, we first show that the SSA condition for the iTEE holds in this setting.

\begin{figure}[ht]
    \centering
    \includegraphics[height = 3.2 cm]{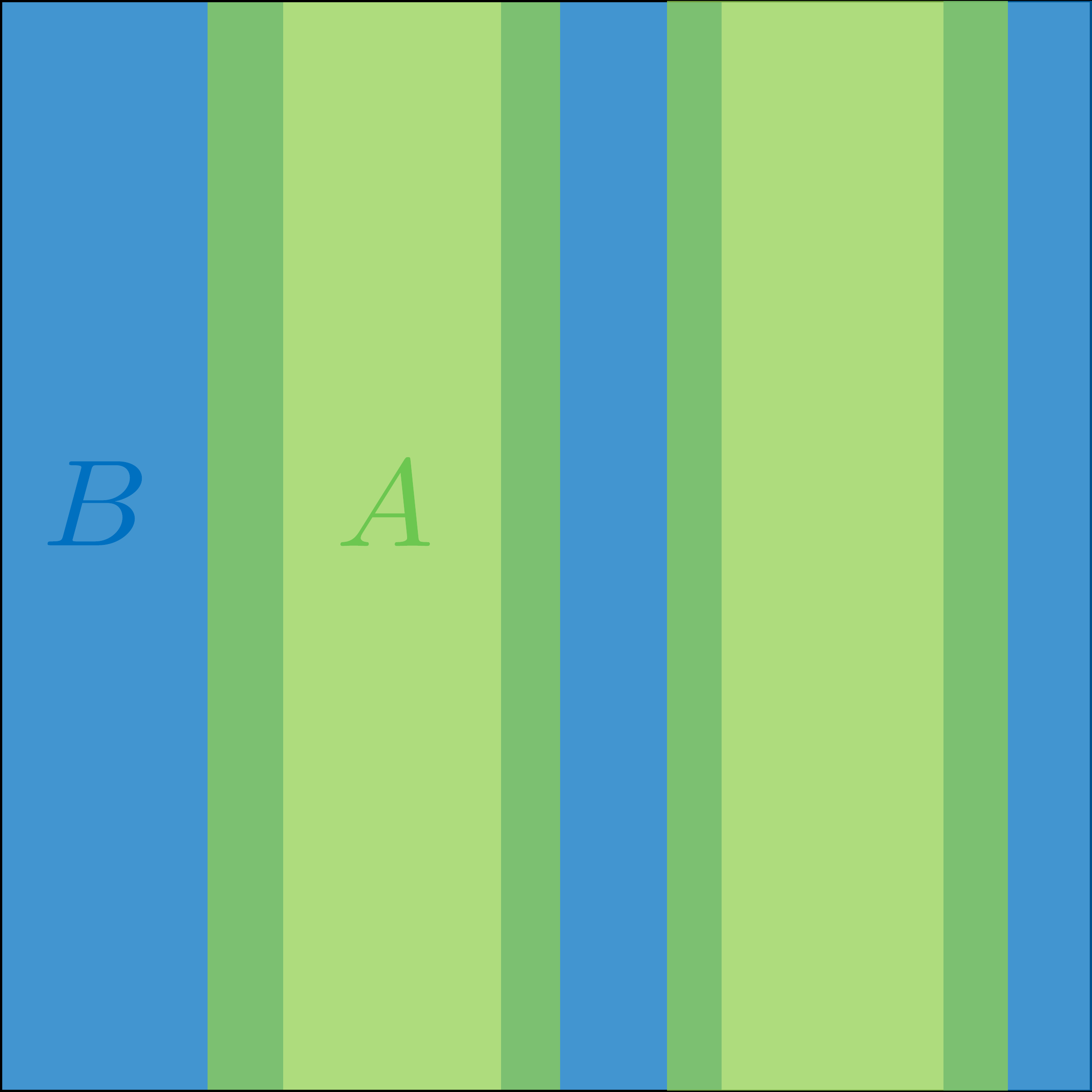}
    \caption{$A, B$ intersect at ribbon components to form a new torus cycle in $A\cup B$.}
    \label{Fig:RibbonToTorus}
\end{figure}

Since $\mathcal{I}_\pi = - \text{Cycle} \leq 0$, it suffices to show that $\mathcal{I}_\pi \leq \mathcal{I}_g$ in the case where $\mathcal{I}_g = -1$.
When $\mathcal{I}_g = -1$, the connected components of $A$ and $B$ must collectively form a loop that encloses a torus cycle distinct from those associated with either $A$ or $B$ (See Fig.~\ref{Fig:RibbonToTorus}). As a result, the relationship graph introduced in Sec.~\ref{Sec:SSA_iTEE} must contain at least one cycle. Therefore,
\begin{equation}
\mathcal{I}_\pi = -\text{Cycle} \leq -1 = \mathcal{I}_g,
\end{equation}
which completes the proof that the SSA condition holds for the iTEE in this case.

Furthermore, since all intersections must be ribbons, the boundaries of $A$ and $B$ cannot intersect.  
As a result, the total number of boundary components in $A$ and $B$ must equal the total number of boundary components in $A\cup B$ and $A\cap B$, that is,
\begin{equation}
m_A + m_B = m_{A \cup B} + m_{A \cap B}.
\end{equation}
Applying Eq.~(\ref{Eq:GeneralTEE}), we conclude that $\mathcal{I}_\text{gs} \geq 0$.  
Combining this with the SSA condition for the iTEE, we find that the SSA condition also holds for the full topological entanglement entropy $S_\text{TEE}$.

\subsubsection{ \texorpdfstring{$i_{AB} =1$}{i\_AB = 1}}

For $i_{AB} = 1$, suppose that $\pi_A > 1$ and $\pi_B > 1$. Then, by Eq.~(\ref{Eq:Criteria}), we have $\mathcal{I}_\text{iTEE} \geq 0$.  
Similarly, as in the case $i_{AB} \geq 2$, $A \cap B$ is composed entirely of disks, while $A \cup B$ constitutes a torus.
Consequently, $S_\text{gs}(A \cap B) = S_\text{gs}(A \cup B) = 0$, which implies $\mathcal{I}_\text{gs} \geq 0$.  
Therefore, we conclude that both the iTEE and the TEE satisfy the SSA condition.

\begin{figure}[ht]
    \centering
    \includegraphics[height = 2.5 cm]{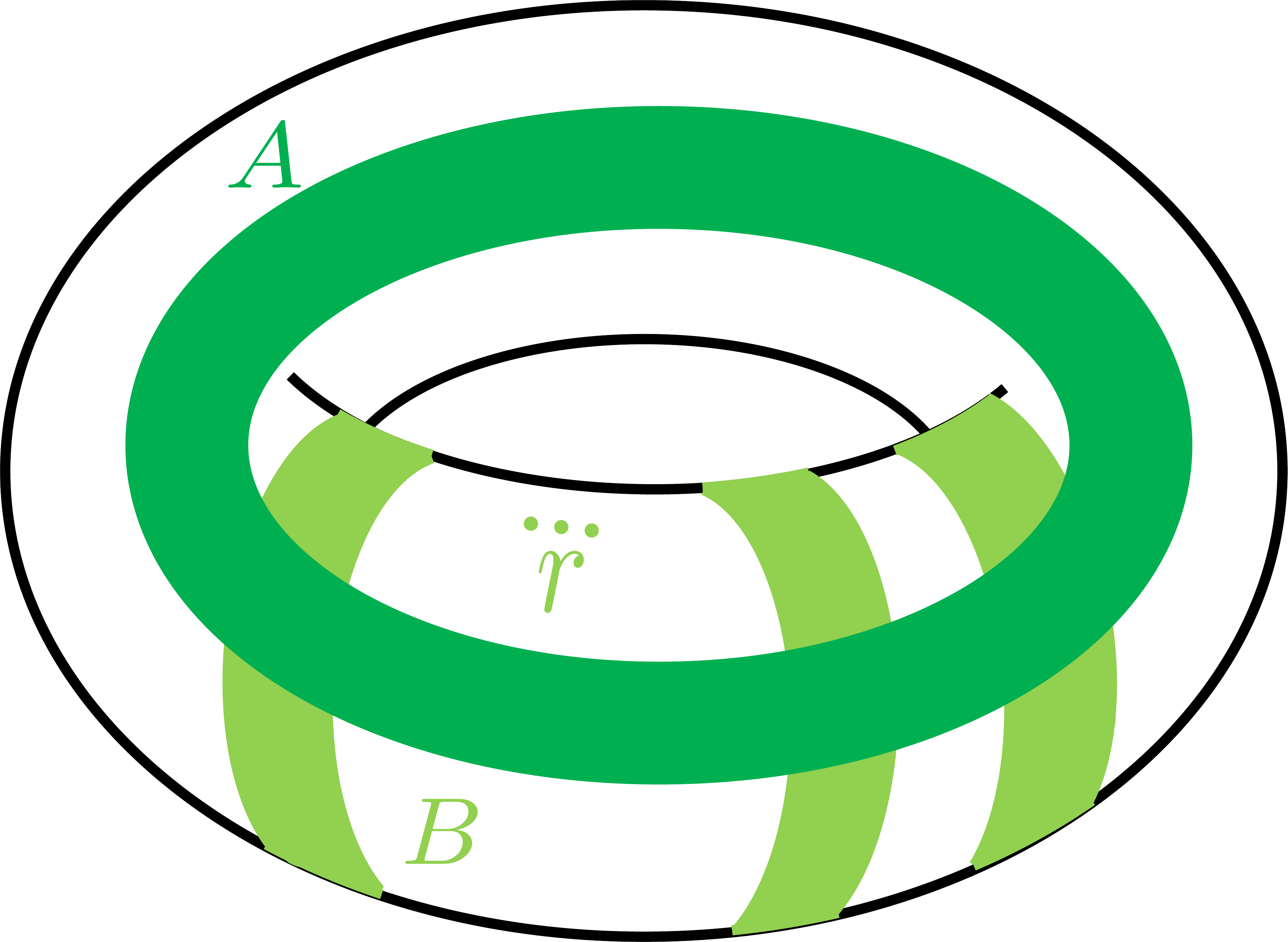}
    \caption{$A$ consists of a single ribbon and $B$ consists of $r$ ribbons. The intersection number between the $A$ ribbon and any one of the $B$ ribbon is $i_{AB} = 1$.}
    \label{Fig:Violation}
\end{figure}

For the remaining cases, without loss of generality, we can assume $(\pi_A, \pi_B)=(1,r), $ where $ r\in \mathbb{Z}_{>0}$ is a positive integer (See Fig.~\ref{Fig:Violation}).
Using Eq.~(\ref{Eq:Inotation}) and Eq.~(\ref{Eq:Criteria}) we get:
\begin{equation}
    \mathcal{I}_\text{iTEE} = -2 \ln  \mathcal{D}.
\end{equation}

As for the $\mathcal{I}_\text{gs}$, $A \cap B$ and $A \cup B$ does not constitute to the ground state TEE. Therefore, we have
\begin{align}
   & \mathcal{I}_\text{gs} = S_\text{gs}(A) + S_\text{gs}(B)  \notag\\
    = &\sum_a 2 \left[ |\psi_a|^2 (\ln d_a - \ln |\psi_a|) + |S\psi_a|^2 (r\ln d_a - \ln |S\psi_a|) \right].
\end{align}
Since the second term increases with $r$, it suffices to check the minimal case, $r = 1$.
Therefore, the SSA condition for the TEE, $\mathcal{I}_\text{gs} + \mathcal{I}_\text{iTEE} \geq 0$, holds if and only if the following inequality is satisfied:
\begin{equation}\label{Eq:Conjecture2}
    \sum_a |\psi_a|^2 (\ln d_a - \ln |\psi_a|) + |S\psi_a|^2 (\ln d_a - \ln |S\psi_a|) \geq  \ln  \mathcal{D}.
\end{equation}

Although we are unable to proof this inequality, we have verified~\cite{Note1} the condition for existing modular data classified by~\cite{ModularData} for unitary modular tensor categories (UMTCs) up to rank $11$.

For illustrative purposes, we present a few low-rank examples here.
At rank 2, we consider the Fibonacci anyon and the Semion whose modular S-matrices are given by: 
\begin{equation}
    S^\text{Fib}=\begin{pmatrix}
        1&\frac{1-\sqrt{5}}{2}\\
        \frac{1+\sqrt{5}}{2}&-1
    \end{pmatrix}
    ,\hspace{0.3cm}
    S^\text{Sem}=\begin{pmatrix}
        1&1\\
        1&-1
    \end{pmatrix}
\end{equation}
At rank $3$, we take the UMTC with modular $S$ matrix:
\begin{equation}
    S=\begin{pmatrix}
        1&-c^3_7&\xi^3_7\\
        -c^3_7&-\xi^3_7&1\\
        \xi^3_7&1&c^3_7
    \end{pmatrix},
\end{equation}
where $c^n_m = \zeta^n_m + \zeta^{-n}_m$, $\xi^n_m = \frac{\zeta^n_{2m} - \zeta^{-n}_{2m}}{\zeta^1_{2m} - \zeta^{-1}_{2m}}$, and $\zeta^n_m = e^{\frac{2 n\pi i}{m}}$ are the cyclotomic numbers.

\begin{figure}[ht]
    \centering
    \includegraphics[height = 5 cm]{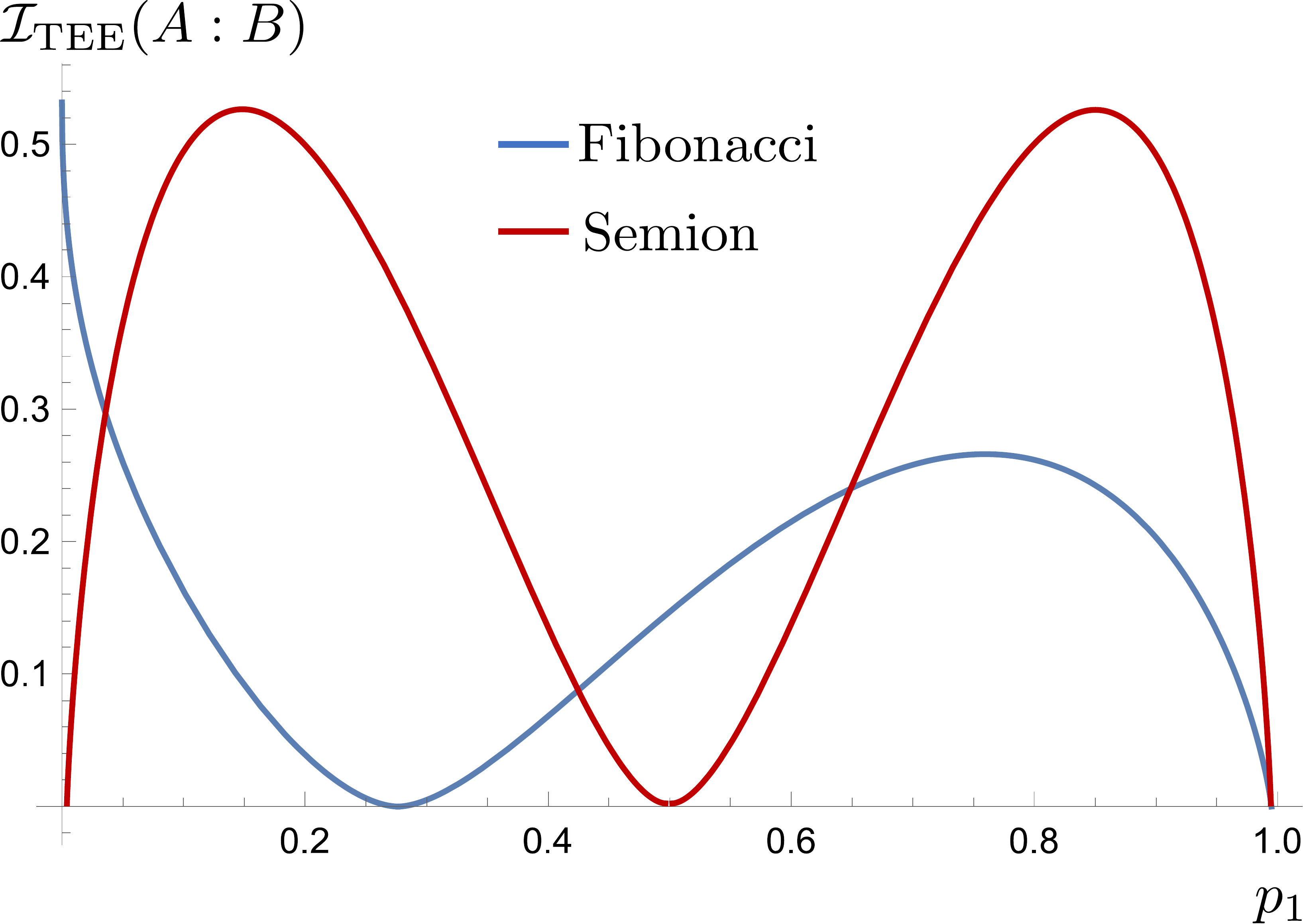}
    \caption{$\mathcal{I}_\text{TEE}(A:B)$ as a function of $p_1 = |\psi_1|^2 $ for Fibonacci anyon and Semion. Eq.~(\ref{Eq:Conjecture}) holds if the graph is above zero.}
    \label{Fig:Rank2}
\end{figure}

\begin{figure}[ht]
    \centering
    \includegraphics[height = 6 cm]{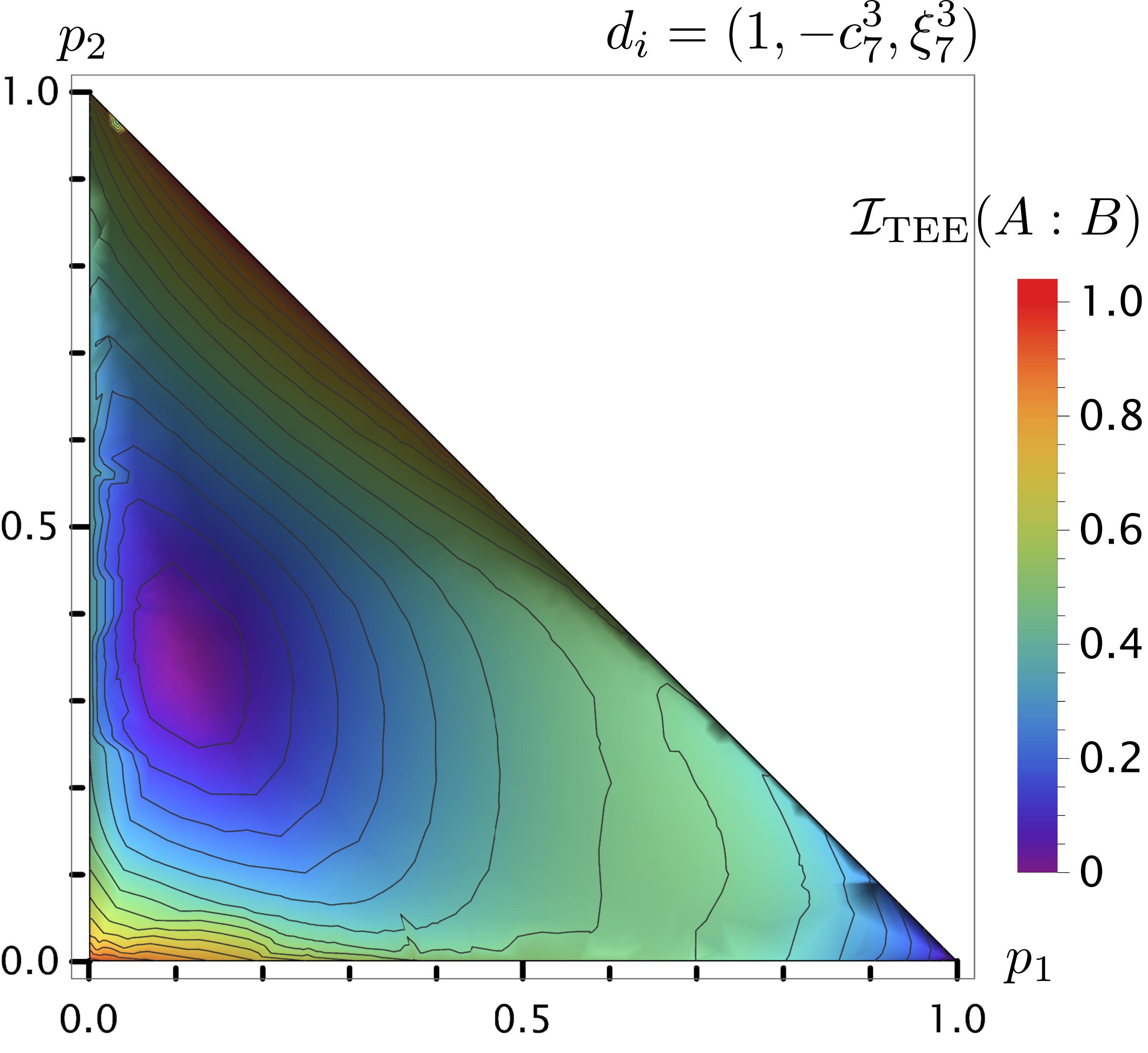}
    \caption{$\mathcal{I}_\text{TEE}(A:B)$ as a function of $p_1,p_2$ for a rank $3$ UMTC. Eq.~(\ref{Eq:Conjecture}) is valid if the height is above zero.}
    \label{Fig:Rank3}
\end{figure}

In both examples shown in Fig.~\ref{Fig:Rank2} and Fig.~\ref{Fig:Rank3}, we observe that the inequality holds for all possible ground states, and equality is achieved precisely when the ground state is the vacuum $\ket{0}$ or can be transformed into the vacuum by the $S$ matrix, i.e., $S\ket{0}$.

\section{Conclusion}
In this paper, we classify bipartitions of a torus based on their entanglement interfaces.
Each bipartition consists of an even number of torus knots of the same type, along with other contractible loops.
By removing bubbles and applying coordinate transformations, any bipartition can be reduced to a canonical form, whose topological entanglement entropy (TEE) is well studied.
The TEE consist of a state independent part which we term the intrinsic TEE and a ground state dependent part referred to as the ground state TEE.

Building on these results, we investigate the strong subadditivity (SSA) property for the TEE. We first show that the topological conditional mutual information decreases upon eliminating removable contractible intersections. 
This key observation enables us to restrict attention to configurations where the regions $A$ and $B$ consist solely of torus knot ribbons.
Within this simplified setting, we classify bipartitions by the intersection number $i_{AB}$ between the torus knots.
For $i_{AB} \neq 1$, we rigorously prove that the SSA inequality  is always satisfied.
In the special case $i_{AB} = 1$, we establish that SSA holds if and only if the following inequality is satisfied:
\begin{equation*}
    \sum_a |\psi_a|^2 (\ln d_a - \ln |\psi_a|) + |S\psi_a|^2 (\ln d_a - \ln |S\psi_a|) \geq  \ln  \mathcal{D}.
\end{equation*}

Numerical tests confirm that the inequality is upheld for all unitary modular tensor categories (UMTCs) of rank up to 11. This leads us to conjecture that the inequality may hold universally for all UMTCs. Conversely, assuming the validity of SSA for TEE implies that inequality imposes a nontrivial constraint on the modular data of permissible UMTCs.

In parallel, we derive a modified SSA inequality for the intrinsic topological entanglement entropy, purely in topological arguments:
\begin{equation*}
\mathcal{I}_\text{iTEE}(A:B) \geq 2\mathcal{I}_g(A:B) \ln \mathcal{D}.
\end{equation*}
Here, $\mathcal{I}_g(A:B)$ is a genus-dependent term characterizing the topology of the subregions, and and it satisfies $|\mathcal{I}_g(A:B)| \leq 1$ on the torus. This modified SSA deviates from the standard SSA only in the case $\mathcal{I}_g(A:B) = -1$.

\section{Acknowledgments}
We are grateful to Xueda Wen for useful discussions. P.-Y.C. acknowledges support from the National Science and Technology Council of Taiwan under Grants No. NSTC 113-2112-M-007-019. Both P.-Y.C and C.-Y. L. thank the National Center for Theoretical Sciences, Physics Division for its support.

\appendix

\section{Bulk approach derivation of the TEE for multi-component canonical bipartition}\label{Appendix:TEE}

\subsection{Vacuum state} \label{SubSec:Vacuum}

As a warm-up, we first demonstrate that by applying the replica and surgery method, the TEE for the vacuum state with the canonical bipartition of $m$ rings is given by:
\begin{equation}\label{Eq:CanonicalTEE}
    S_{\text{TEE}}(\ket{0}, R_m) = -2m \ln  \mathcal{D}.
\end{equation}

Consider a vacuum state with the $R_m$ bipartition, represented as follows:
\begin{equation}
    \ket{0} = \adjustbox{valign = c}{\includegraphics[height = 2.5 cm]{Figure/CanonicalBipartition1.pdf}}.
\end{equation}
For simplicity of visualization, let us consider the case where $m=2$. 
We also contract our solid torus into a circle, where each point represents a $D^2$
\begin{equation}
    \ket{0} = \adjustbox{valign = c}{\includegraphics[height = 1.5 cm]{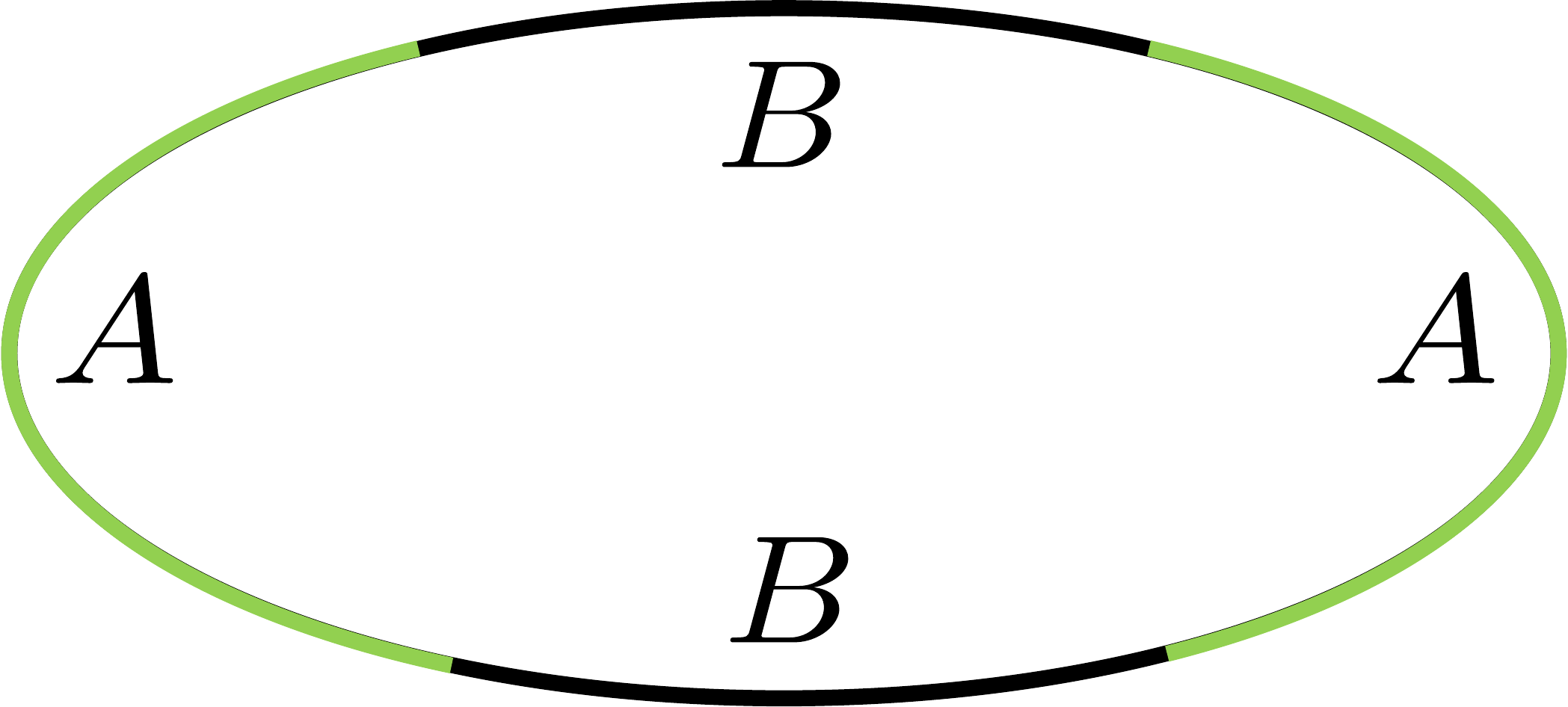}}.
\end{equation}
The reduced density matrix is obtained by adding an orientation reversed copy of $\ket{0}$ and gluing them together along the $B$ subregion:
\begin{equation}
    \rho_A = \mathrm{Tr}_B \ket{0} \bra{0} = \hspace{0.1cm} \adjustbox{valign = c}{\includegraphics[height = 2 cm]{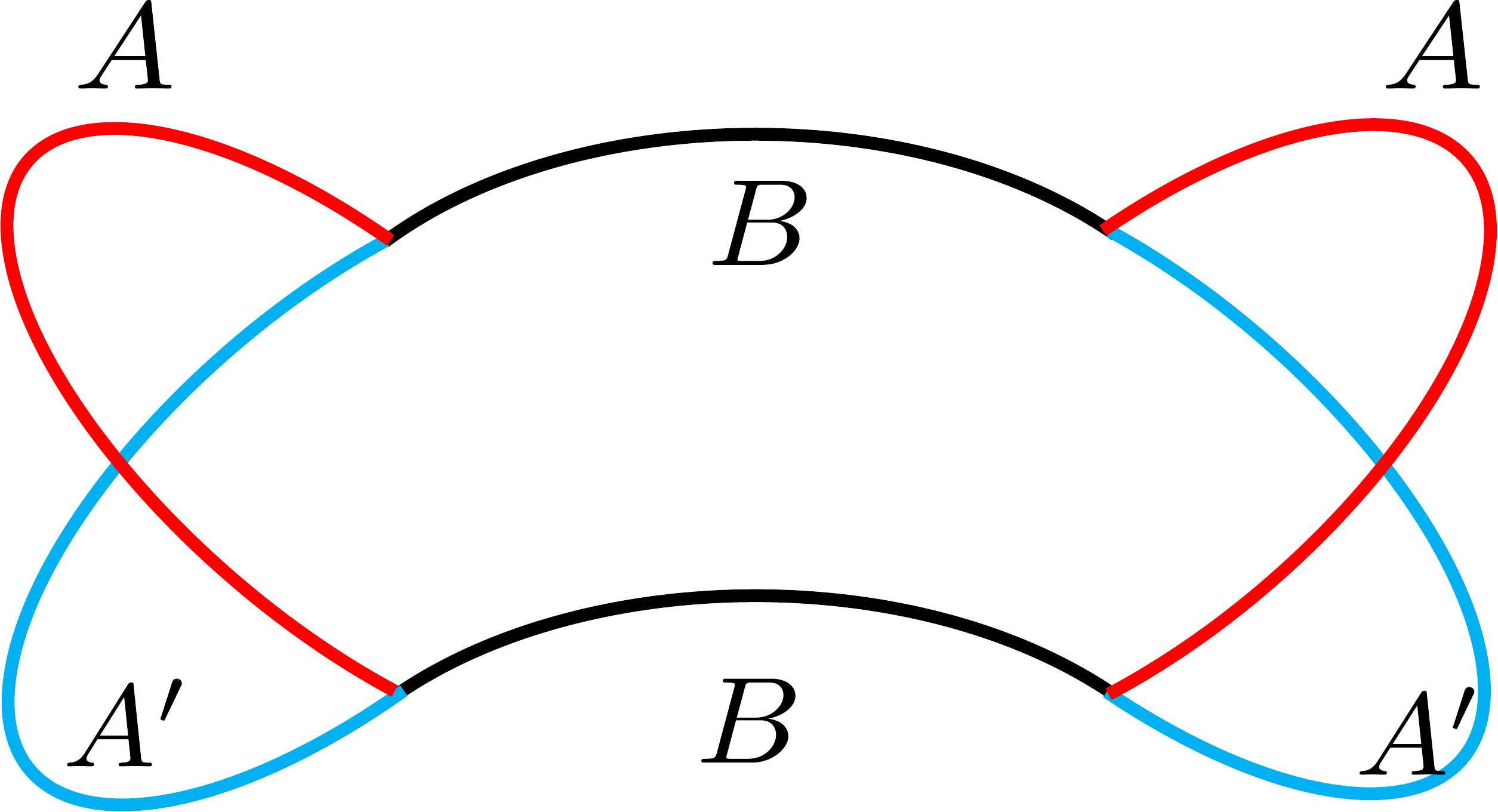}}.
\end{equation}
In the diagram, we use different colors to differentiate the subregions of the two copies, facilitating identification during the gluing process.
Each point on the $B$ regions (indicated by black lines) now represents an $S^2$, while each point in the $A$, $A'$ regions (indicated by red and blue lines) continues to represent a $D^2$.
To compute $\rho_A^2$, we simply add another copy of $\rho_A$ and glue the red region of the first copy to the blue region of the second copy:
\begin{equation}
    \rho_A^2 = \adjustbox{valign = c}{\includegraphics[height = 2 cm]{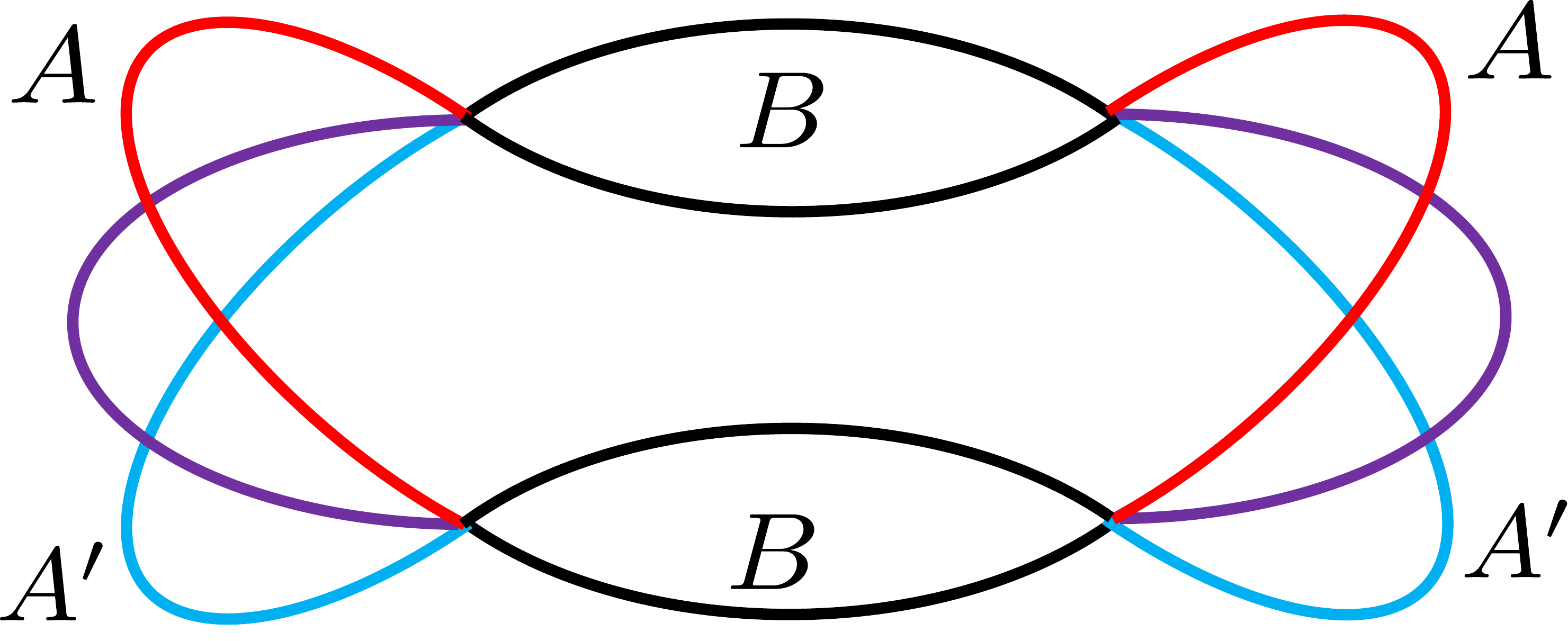}}.
\end{equation}
In this representation, the purple regions indicate where the red and blue regions are glued together.
Consequently, each point on the purple region represents an $S^2$.

Finally, taking trace of $\rho_A^2$ corresponds to gluing the remaining red and blue regions together:
\begin{align}
    \mathrm{Tr} (\rho_A)^2 &= \adjustbox{valign = c}{\includegraphics[height = 2 cm]{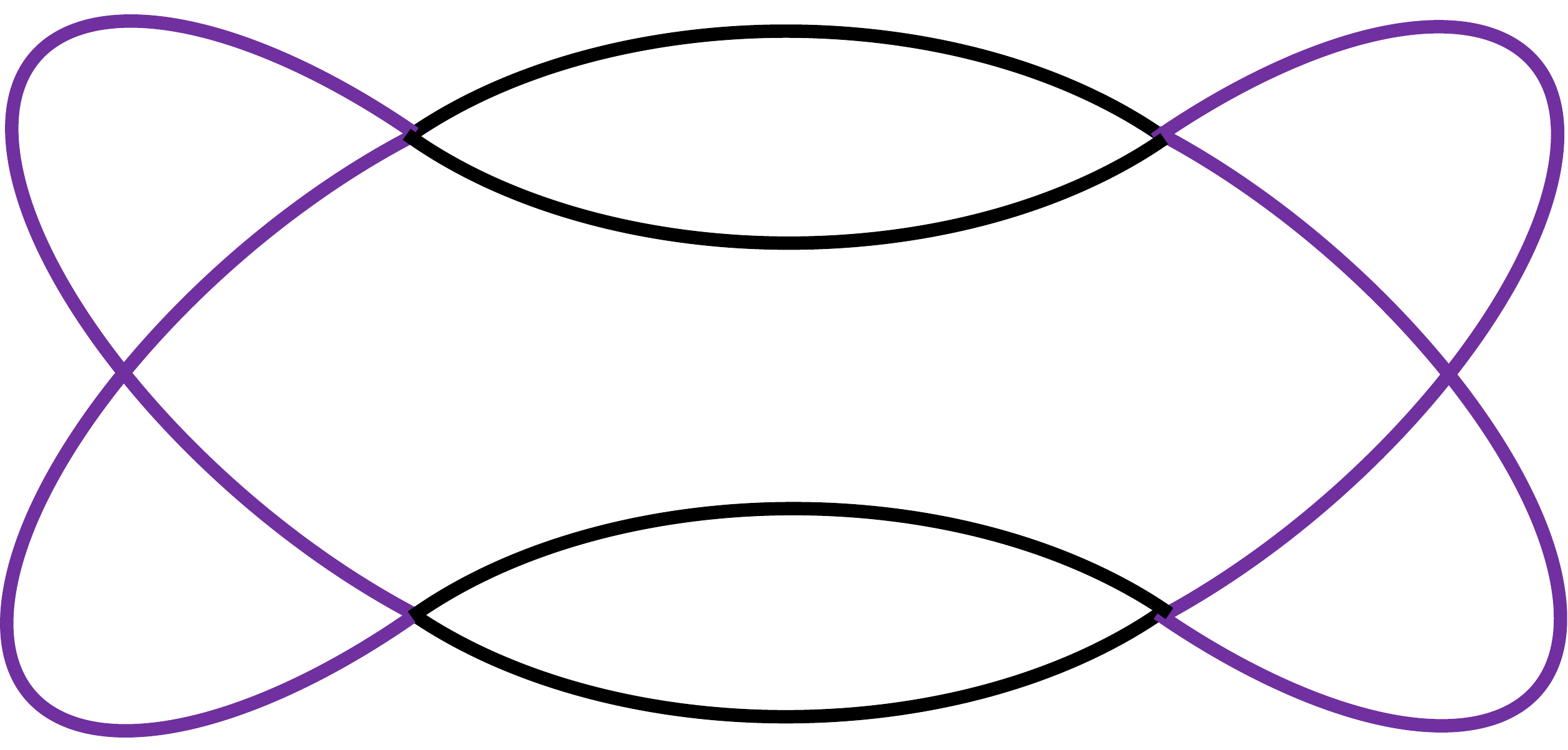}}  \notag\\
    &= \hspace{.4cm} \adjustbox{valign = c}{\includegraphics[height = 2.5 cm]{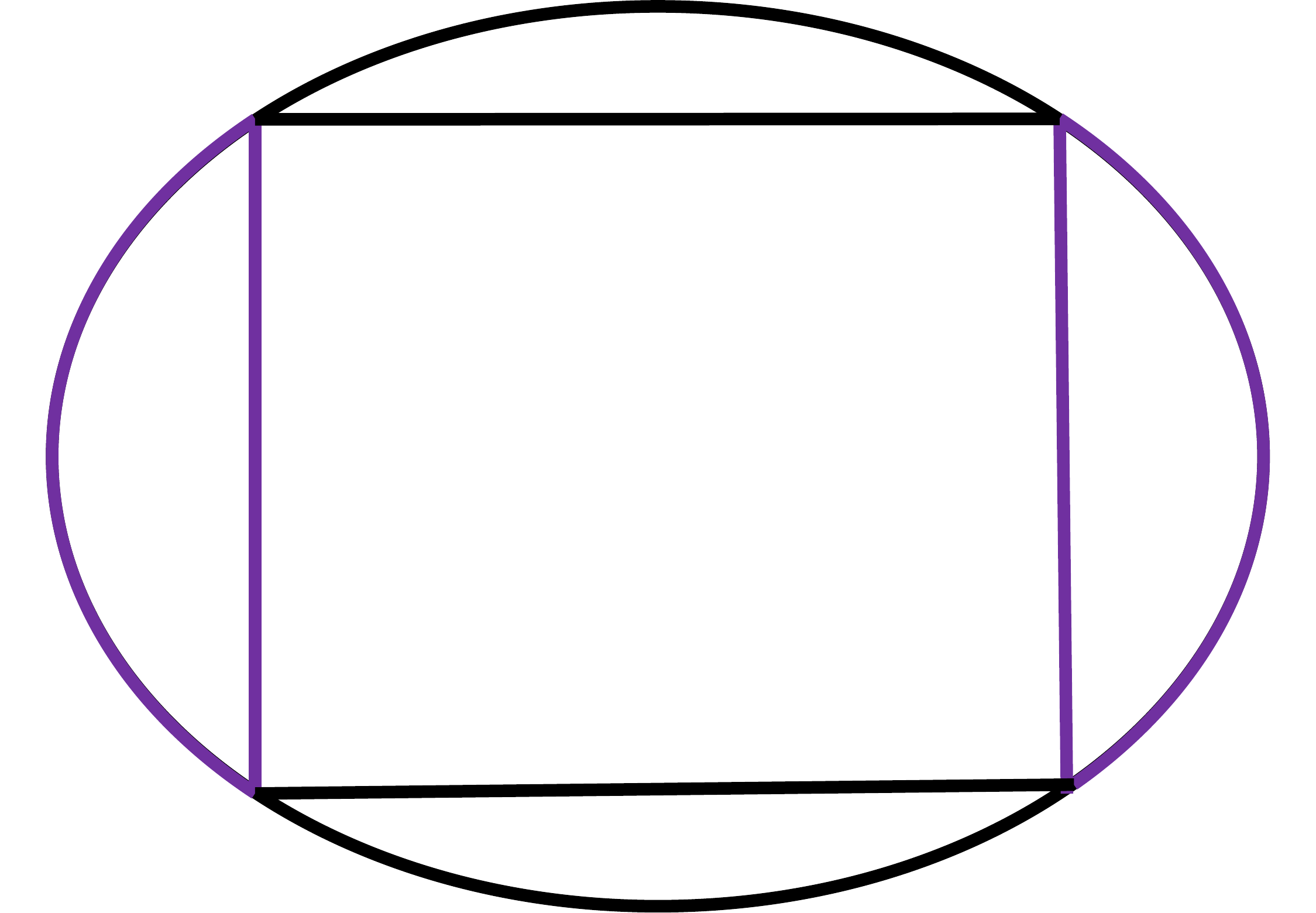}}.
\end{align}
In the second figure, we place everything onto a plane. Now that all the boundaries are properly glued, each point (both purple and black) in this configuration represents an $S^2$. Therefore, the configuration is a connected sum of five copies of $S^2\times S^1$'s.
To evaluate the partition function of such a configuration, we perform the surgery method by inserting four copies of $S^3$ to separate the $S^2 \times S^1$ components:
\begin{equation}
    \mathrm{Tr} (\rho_A)^2 = \frac{Z(S^2 \times S^1)^5}{Z(S^3)^4} =  \mathcal{D}^4,
\end{equation}
where we use the fact that $Z(S^2 \times S^1) = 1$ and $Z(S^3) =  \mathcal{D}^{-1}$ (see \cite{Witten1989}).

In general, for $\mathrm{Tr} (\rho_A)^n$, we will have $n$ lines connecting each neighboring pair of nodes:
\begin{equation}
    \mathrm{Tr} (\rho_A)^n = \hspace{0.2cm} \adjustbox{valign = c}{\includegraphics[height = 3 cm]{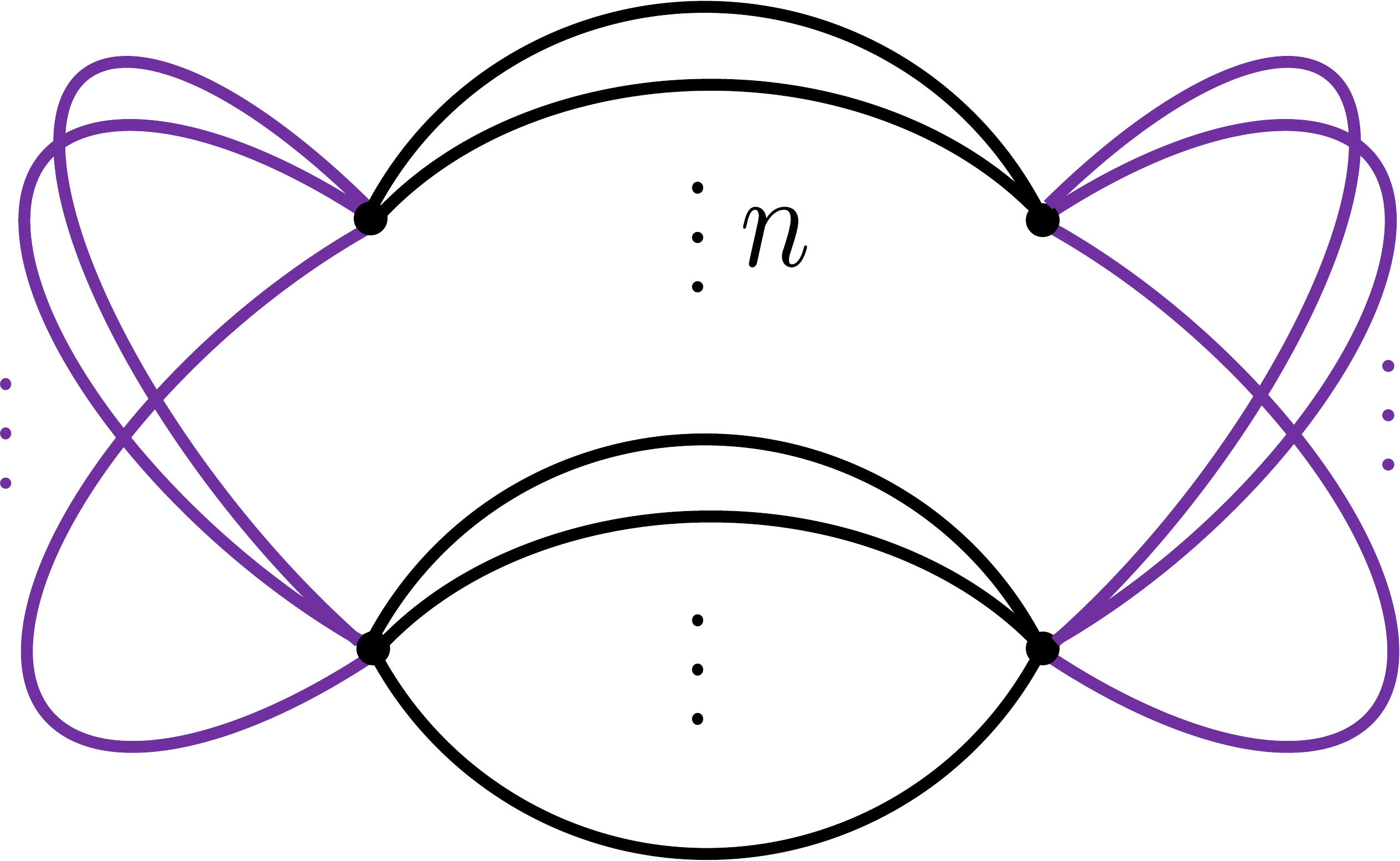}}.
\end{equation}
This figure consist of $4(n-1)+1$ holes\footnote{A simple way to compute the number of holes $F$ is to apply the Euler characteristic formula with $\chi = 1$, $E = 4n$ and $V=4$.}, which form a connected sum of $4(n-1)+1$ copies of $S^2 \times S^1$. Therefore, we can perform the surgery by inserting $4(n-1)$ copies of $S^3$ to separate the configuration:
\begin{equation}
    \mathrm{Tr} (\rho_A)^n = \frac{Z(S^2\times S^1)^{4(n-1)+1}}{Z(S^3)^{4(n-1)}} =  \mathcal{D}^{4(n-1)}.
\end{equation}
The TEE is then given by:
\begin{equation}
    S_{\text{TEE}}(\ket{0}, R_2) = \lim_{n\rightarrow 1} \frac{1}{1-n} \ln \mathrm{Tr} (\rho_A)^n = -4\ln  \mathcal{D}.
\end{equation}

For general $R_m$ bipartition, the configuration of $\mathrm{Tr} (\rho_A)^n$ consists of $2m$ nodes, with $n$ lines connecting neighboring nodes. As a result, the configuration forms a connected sum of $2m(n-1)+1$ copies of $S^2 \times S^1$. The partition function is evaluated by inserting $2m(n-1)$ copies of $S^3$ to separate the components. Thus, we have:
\begin{equation}
    \mathrm{Tr} (\rho_A)^n =  \mathcal{D}^{2m(n-1)}.
\end{equation}
The TEE for the $R_m$ bipartition is then given by:
\begin{equation}
    S_{\text{TEE}}(\ket{0}, R_m) = -2m\ln  \mathcal{D}.
\end{equation}
This completes the proof of Eq.~(\ref{Eq:CanonicalTEE}).

\subsection{Generic state on \texorpdfstring{$R_2$}{R\_2} bipartition}\label{Sec:R2Bipartition}

Next, we consider a generic state on a torus with a canonical $R_2$ bipartition:
\begin{equation}
    \ket{\psi} = \sum_a \psi_a \hspace{0.1cm} \adjustbox{valign = c}{\includegraphics[height = 2.5 cm]{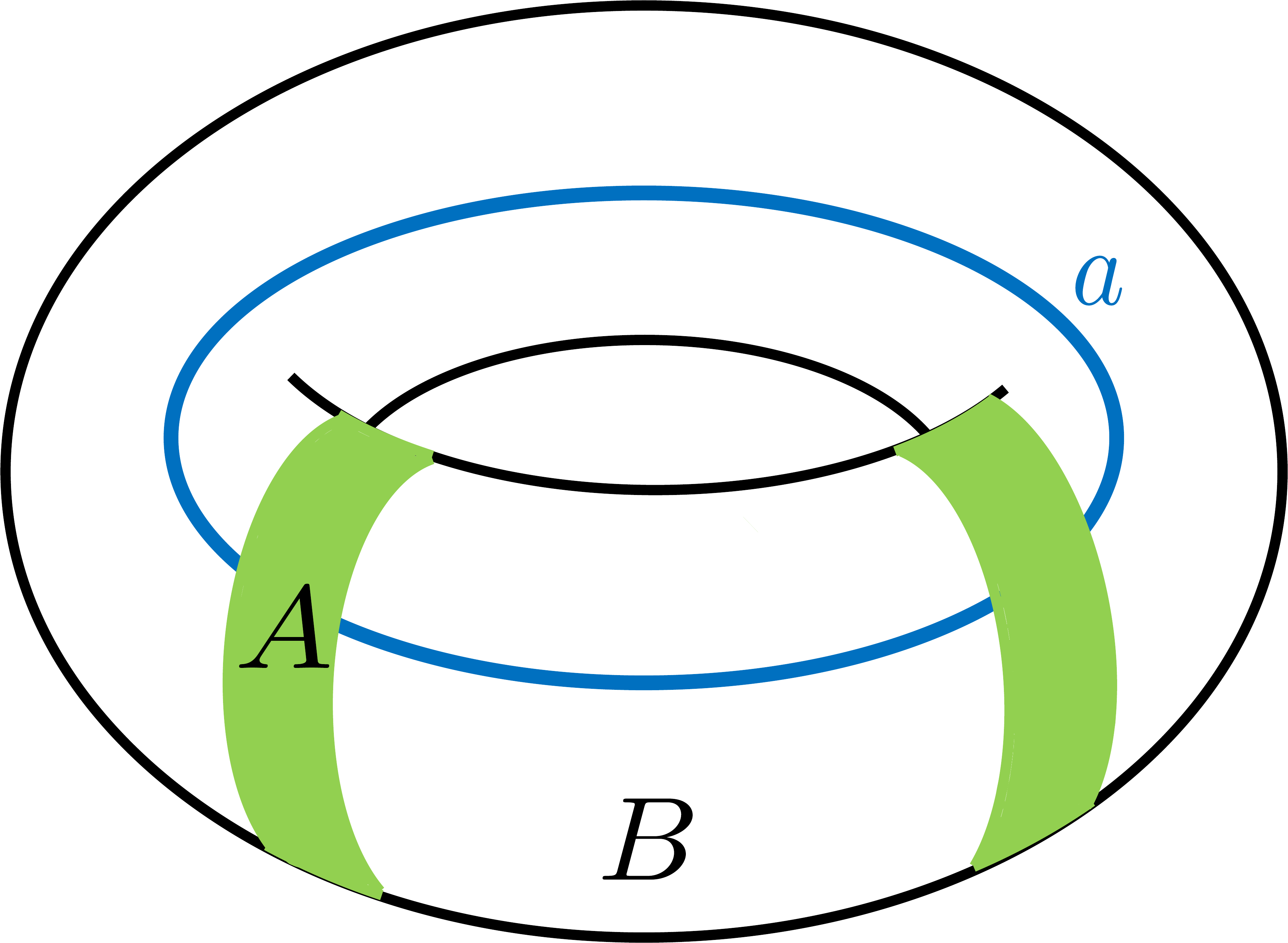}}.
\end{equation}
From the arguments in Sec.~\ref{SubSec:Vacuum}, we know that the bulk configuration of $\mathrm{Tr} (\rho_A)^2$ is a connected sum of five $S^2\times S^1$ components. The primary difference here is that we now need to keep track of the Wilson lines inserted in the bulk. 
Following the notation from Sec.~\ref{SubSec:Vacuum}, the reduced density matrix becomes:
\begin{equation}
    \rho_A = \sum_{a,b} \psi_a \Bar{\psi}_b \hspace{0.3cm} \adjustbox{valign = c}{\includegraphics[height = 2 cm]{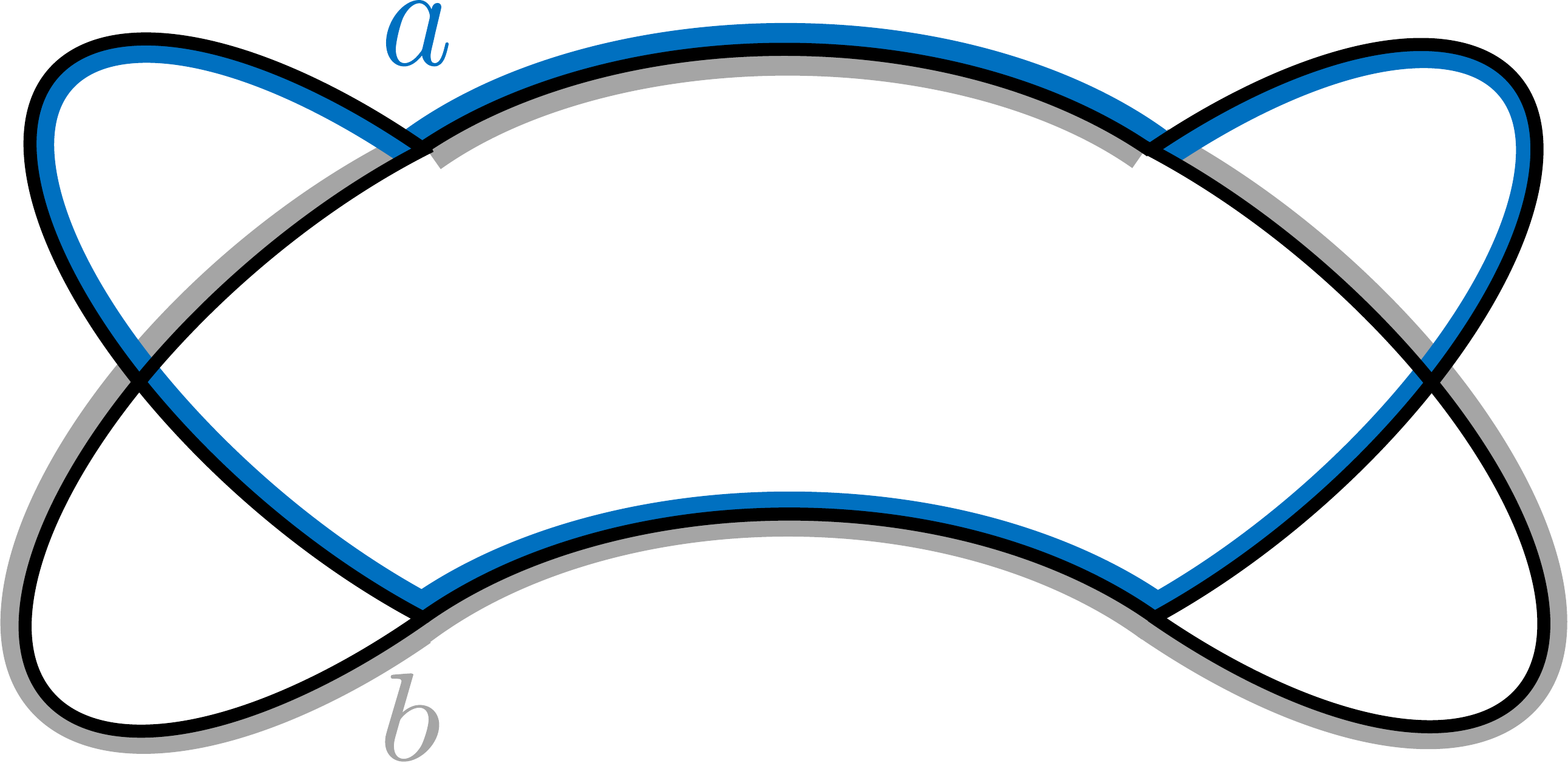}}.
\end{equation}
In this expression, we color-code the Wilson lines rather than the bulk configurations. Wilson lines are represented as the blue and gray loops in the diagram~\footnote{The gray line is actually continuous; the diagram is simplified for ease of visualization.}.
These loops represent the threading of the Wilson lines through the bulk, contributing to the overall entanglement structure. The color-coding helps distinguish between different sets of Wilson lines, making it easier to track their effects throughout the gluing process.

By adding another copy of $\rho_A$ and gluing the corresponding subregions together, we obtain:
\begin{align} 
    &\mathrm{Tr} (\rho_A)^2 \notag\\
     = & \sum_{a,b,c,d} \psi_a \Bar{\psi_b} \psi_c \Bar{\psi_d} \hspace{0.3cm} \adjustbox{valign = c}{\includegraphics[height = 2.2 cm]{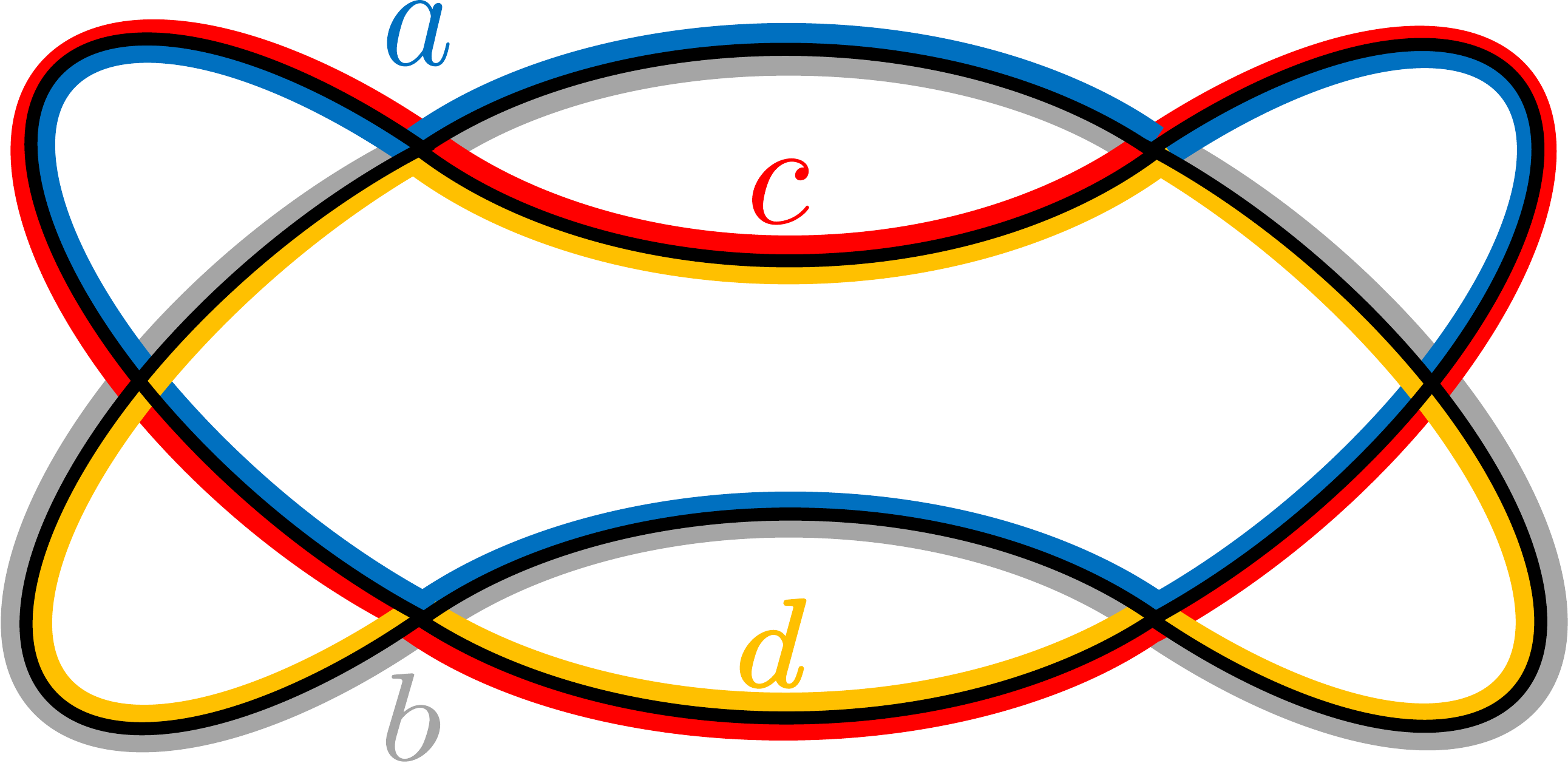}}.
\end{align}
In this configuration, there are four nodes arranged in a circle, and two lines (referred to as "up" and "down" leaves) connect each neighboring pair of nodes. We label each Wilson line according to the leaves they thread through. Thus, the Wilson line configurations correspond to the labels $(a,b,c,d) = (uuuu,dudu,udud,dddd)$. Notice that all the even labels of each Wilson line are identical, and similarly, all the odd labels are the same.
To improve visualization of this configuration, we flatten the 3D structure down to a plane and mark each hole of the bulk configuration using a cross.
Following a convention where we place all the "up" leaves inside and the "down" leaves outside, the configuration transforms as follows:
\begin{align}
   & \adjustbox{valign = c}{\includegraphics[height = 2.2 cm]{Figure/2RingABCD.pdf}} \hspace{0.3cm}  \notag\\
    = & \hspace{.7cm} \adjustbox{valign = c}{\includegraphics[height = 3 cm]{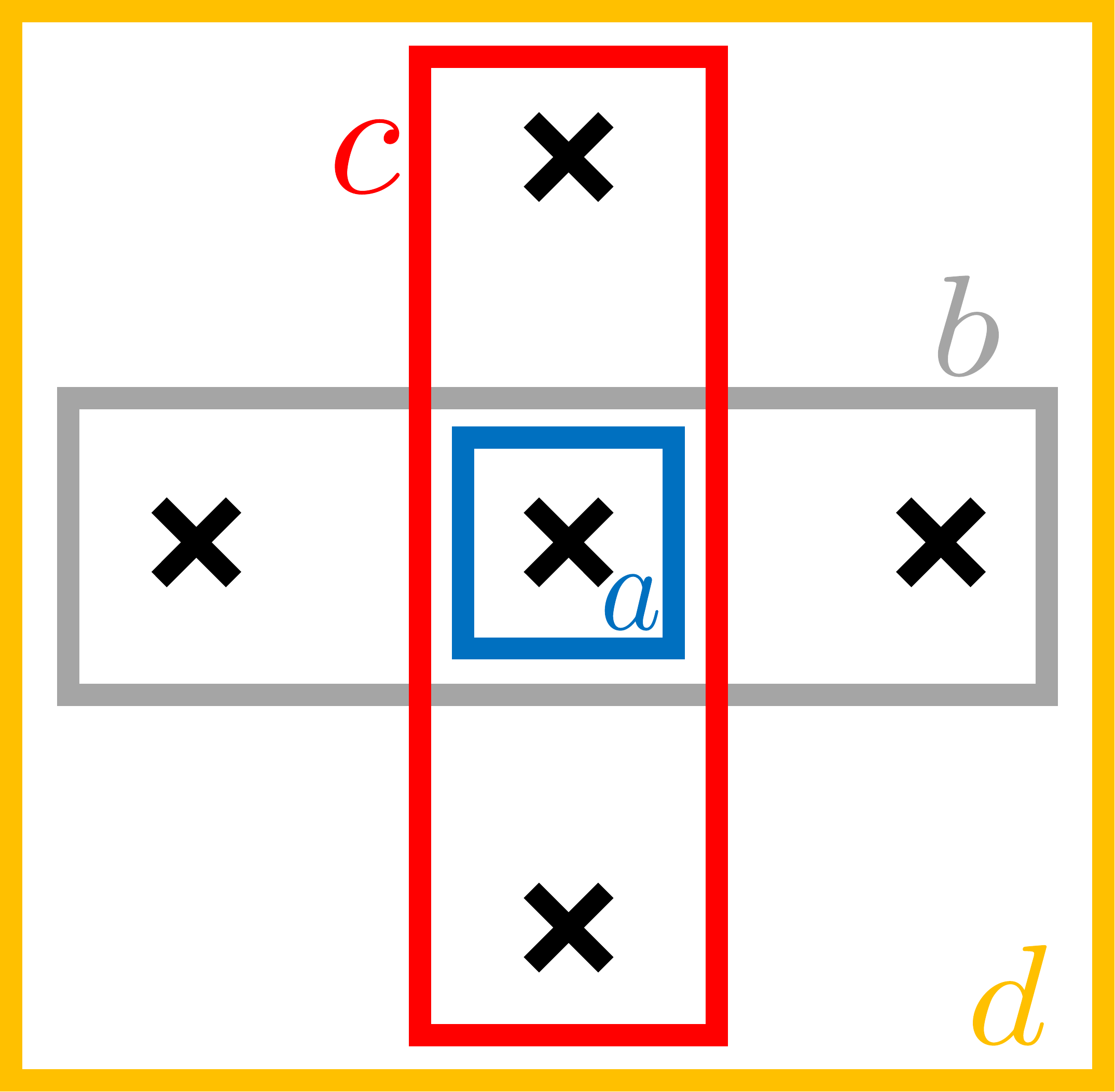}}.
\end{align}
In this flattened representation, each bulk region between the holes is threaded by exactly two Wilson lines. The bulk configuration remains a connected sum of five $S^2\times S^1$ components, as expected.

To evaluate this configuration, we apply the surgery method to separate the holes.
We begin by making a cut to isolate the right hole:
\begin{equation*}
    \adjustbox{valign = c}{\includegraphics[height = 4 cm]{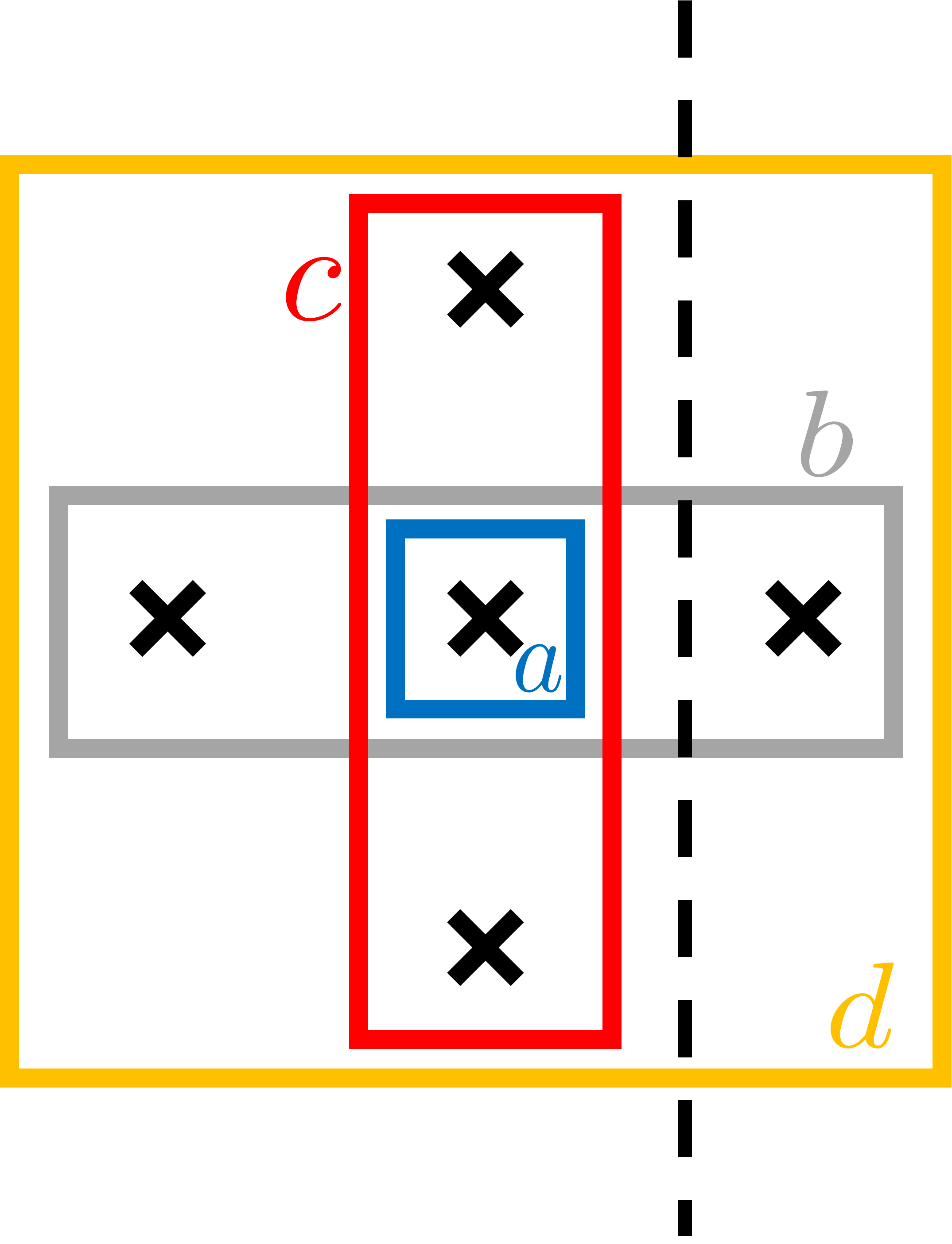}} \hspace{0.2cm} = \hspace{0.2cm} \bra{L} \ket{R},
\end{equation*}
where the right side of the dashed line is denoted by $\ket{R}$ and the left side by $\ket{L}$.
This cut creates an $S^2$ with four marked points: $b,\Bar{b}, d, \Bar{d}$. Thus, the states $\ket{L}$ and $\ket{R}$ belong to the Hilbert space $\mathcal{H}_{S^2, b,\Bar{b}, d, \Bar{d}} = V^{bd}_{bd}$.
By applying the surgery method, we insert a complete orthonormal basis in $V^{bd}_{bd}$:
\begin{equation}
    \ket{e,\mu,\nu} = \hspace{0.2cm} \adjustbox{valign = c}{\includegraphics[height = 3 cm]{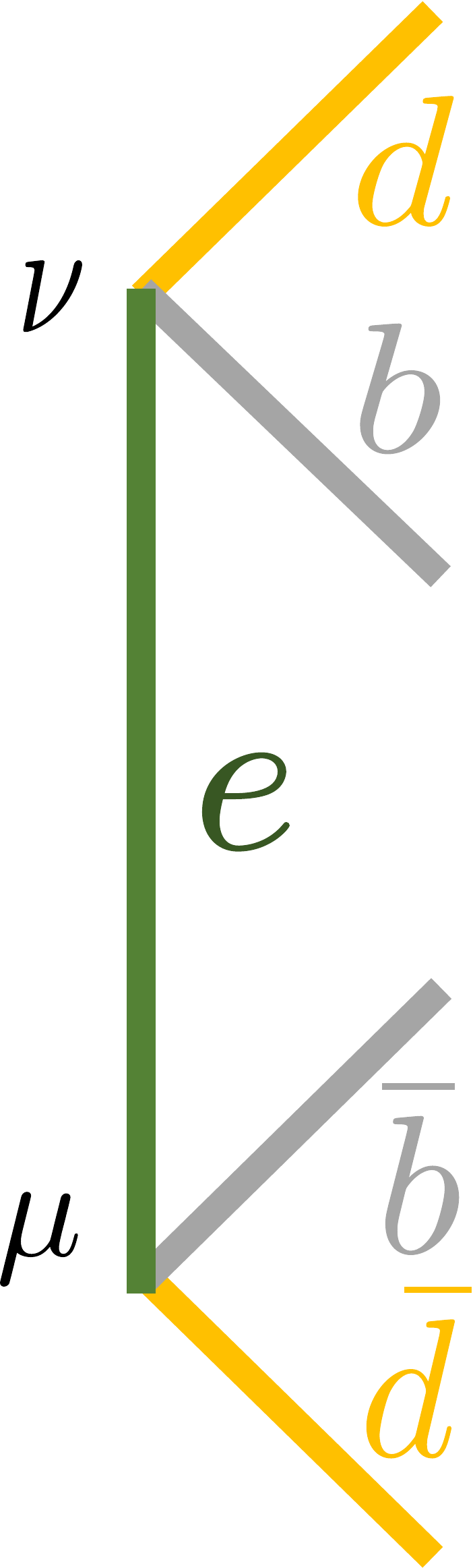}},
\end{equation}
where we represent the marked points on the $S^2$ using external legs, without depicting the sphere itself.
Using the bubble merging diagram (see, for instance, \cite{Bonderson} for a detailed review), one can verify that this basis is indeed orthogonal, with normalization:
\begin{equation}
    \braket{e,\mu,\nu}{f,\alpha,\beta} = \hspace{0.1cm} \adjustbox{valign = c}{\includegraphics[height = 3 cm]{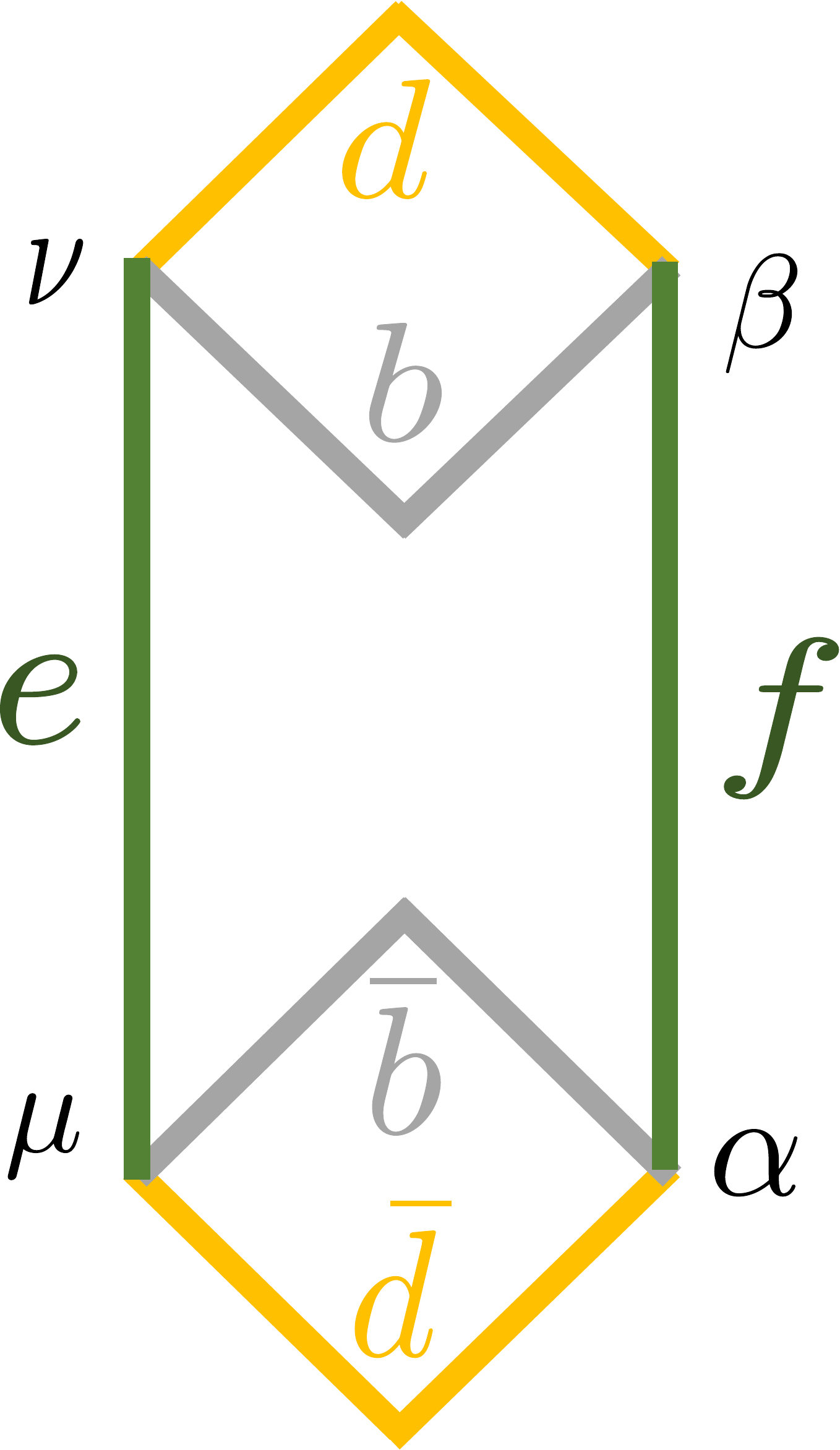}} \hspace{0.1cm} = \delta_{ef} \delta_{\mu \alpha} \delta_{\nu \beta} \frac{d_b d_d}{ \mathcal{D}}.
\end{equation}

We apply the surgery method by inserting the complete orthonormal basis:
\begin{equation}
    \bra{L}\ket{R} = \sum_{e,\mu,\nu} \frac{\braket{L}{e,\mu,\nu} \braket{e,\mu,\nu}{R}}{\braket{e,\mu,\nu}{e,\mu,\nu}}.
\end{equation}
Pictorially, this is represented by:
\begin{align} \label{Eq:SurgeryCross}
    & \adjustbox{valign = c}{\includegraphics[height = 4 cm]{Figure/Cut.pdf}} \hspace{0.1cm} \notag\\
    = &\hspace{0.1cm} \sum_{e,\mu,\nu}   \frac{ \mathcal{D}}{d_b d_d} \hspace{0.3cm}  \adjustbox{valign = c}{\includegraphics[height = 4 cm]{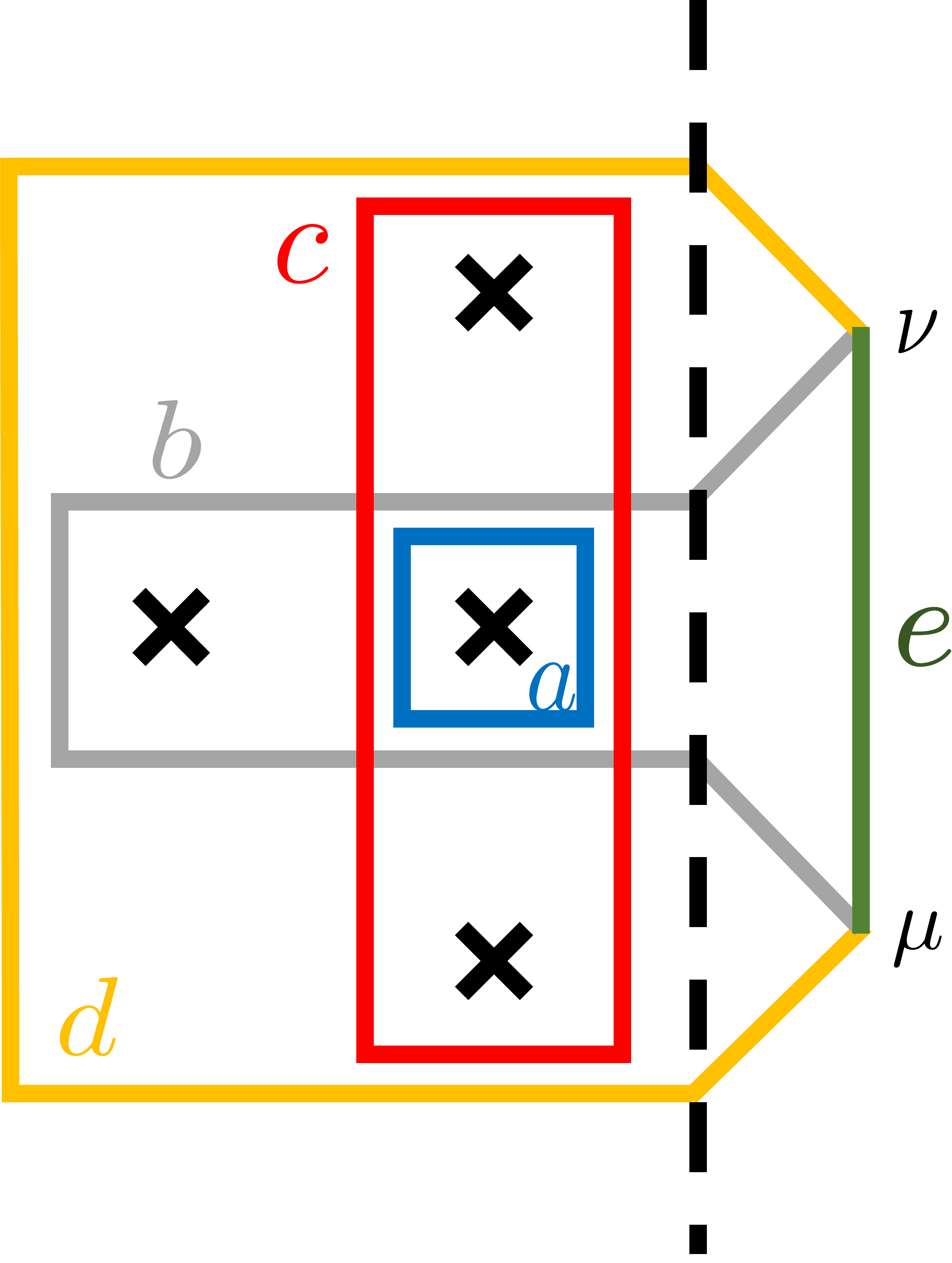}} \hspace{0.3cm} \adjustbox{valign = c}{\includegraphics[height = 4 cm]{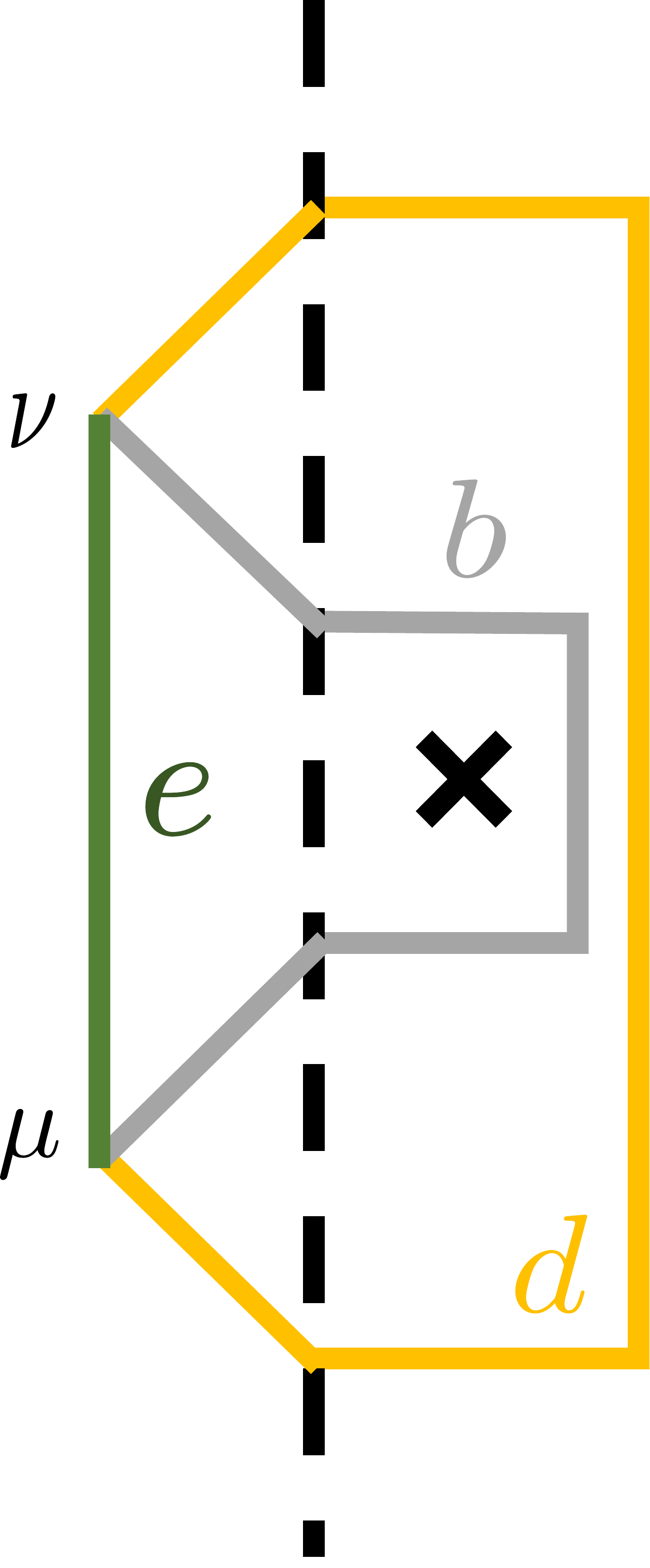}}. 
\end{align}
For the right diagram in the equation, we evaluate it using the bubble merging diagram.
This simplifies to:
\begin{equation}
    \braket{e,\mu,\nu}{R} = \adjustbox{valign = c}{\includegraphics[height = 4 cm]{Figure/RightAssemble.pdf}} = \sqrt{\frac{d_b d_d}{d_e}} \delta_{0e}  \delta_{\mu,\nu}.
\end{equation}
Thus, the sum over $e$ collapses due to $\delta_{0e}$, leaving us with:
\begin{align} \label{Eq:SurgeryLeft}
    \adjustbox{valign = c}{\includegraphics[height = 3 cm]{Figure/Cross.pdf}} \hspace{0.1cm}
    = \sum_{\mu,\nu} \hspace{0.1cm} \frac{ \mathcal{D}}{\sqrt{d_b d_d}} \hspace{0.1cm} \adjustbox{valign = c}{\includegraphics[height = 3 cm]{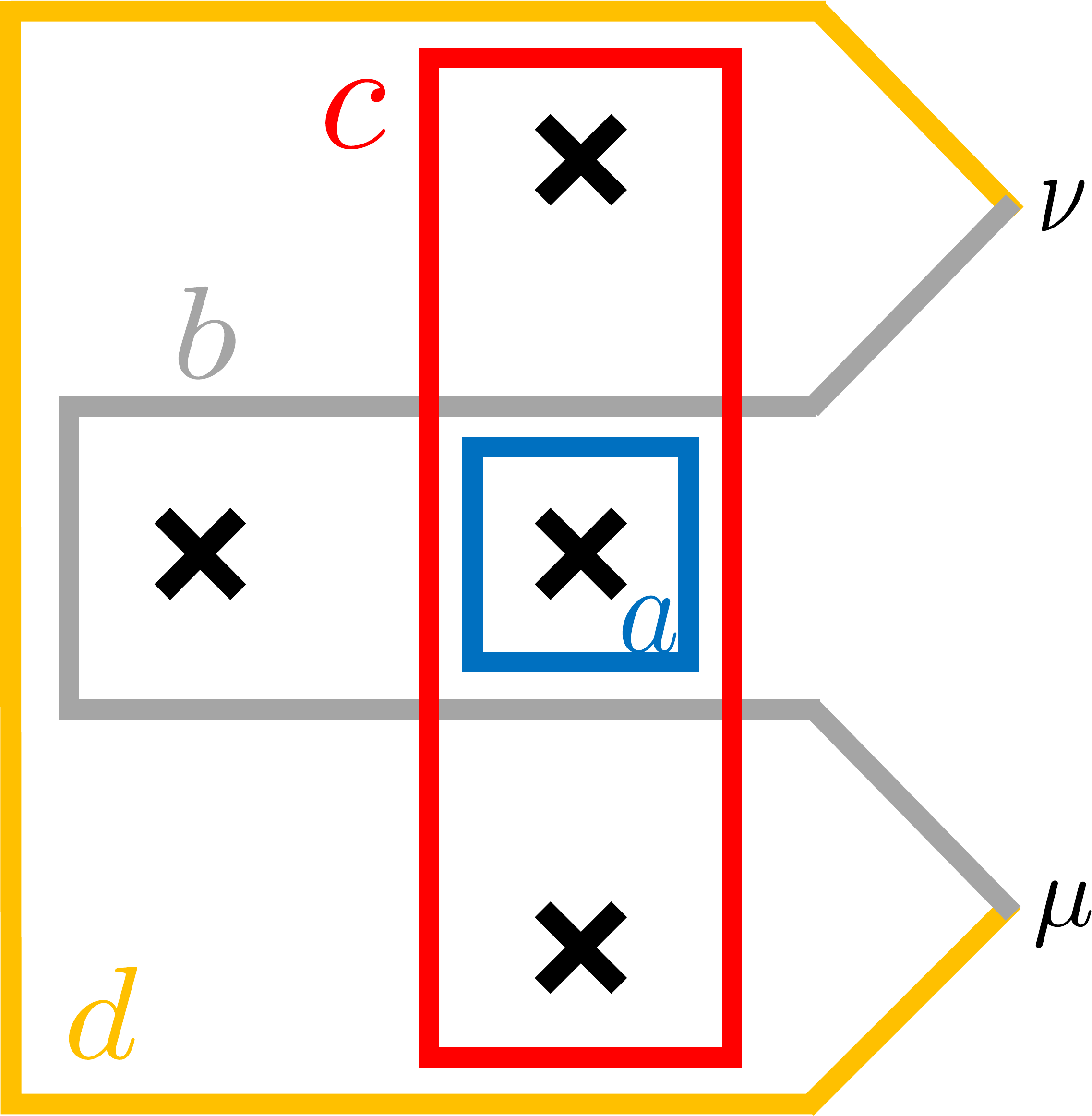}}.
\end{align}

Following the process, we next cut off the left hole in the diagram. Applying the same surgery method as before, we obtain:
\begin{equation} \label{Eq:SurgeryRight}
    \adjustbox{valign = c}{\includegraphics[height = 3 cm]{Figure/Cross2.pdf}} \hspace{0.1cm} = \sum_{\mu',\nu'}\hspace{0.1cm} \frac{ \mathcal{D}}{\sqrt{d_b d_d}} \hspace{0.1cm} \adjustbox{valign = c}{\includegraphics[height = 3 cm]{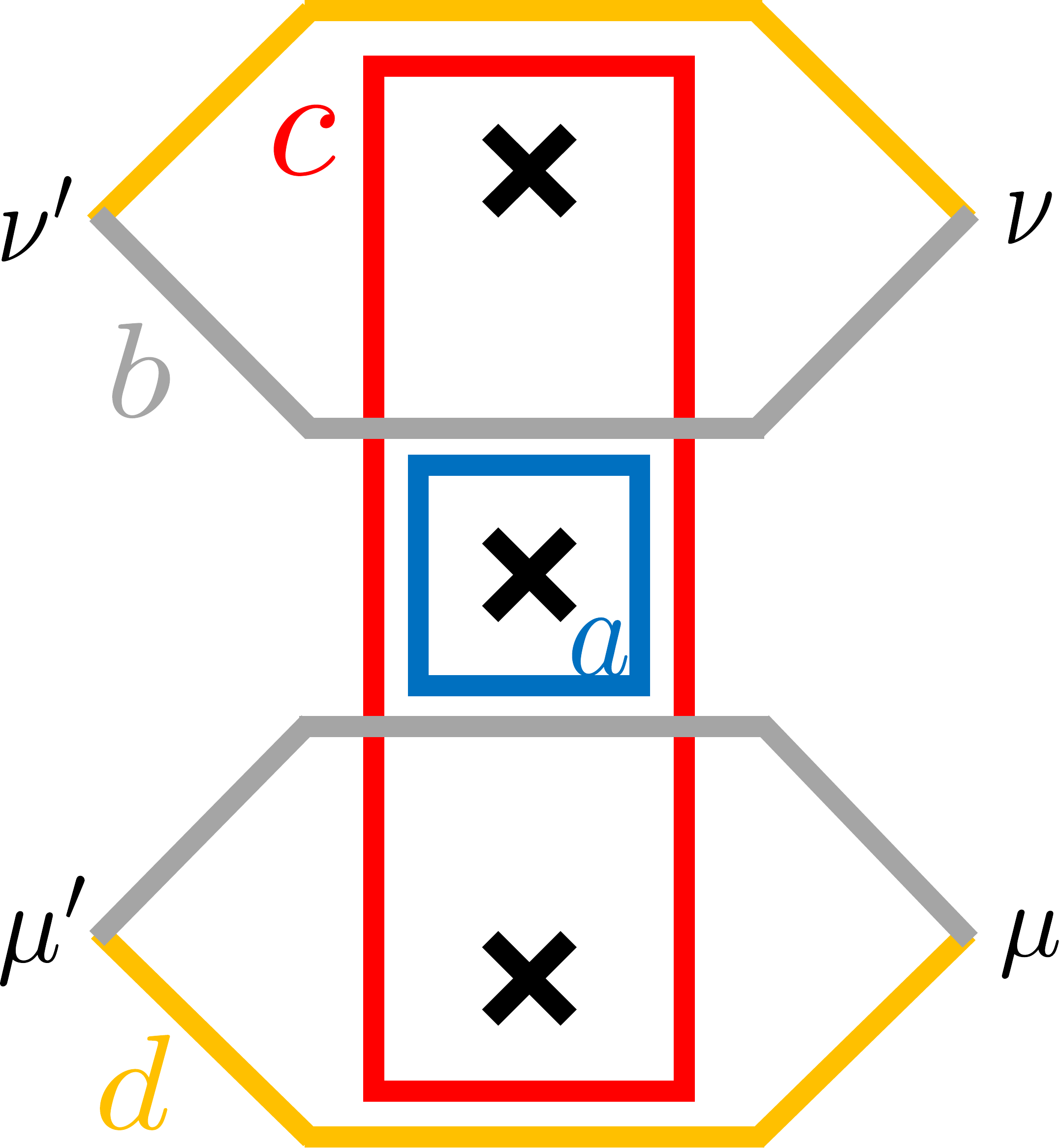}}.
\end{equation}
This step introduces an additional constraint: $b = \Bar{d}$. Therefore, the indices $\mu,\nu, \mu',\nu'$ can be neglected since there's a unique channel for an anyon to merge with its inverse. 

Finally, we apply the surgery method twice more to separate all three holes:
\begin{align} \label{Eq:SurgeryUD}
    \adjustbox{valign = c}{\includegraphics[height = 3 cm]{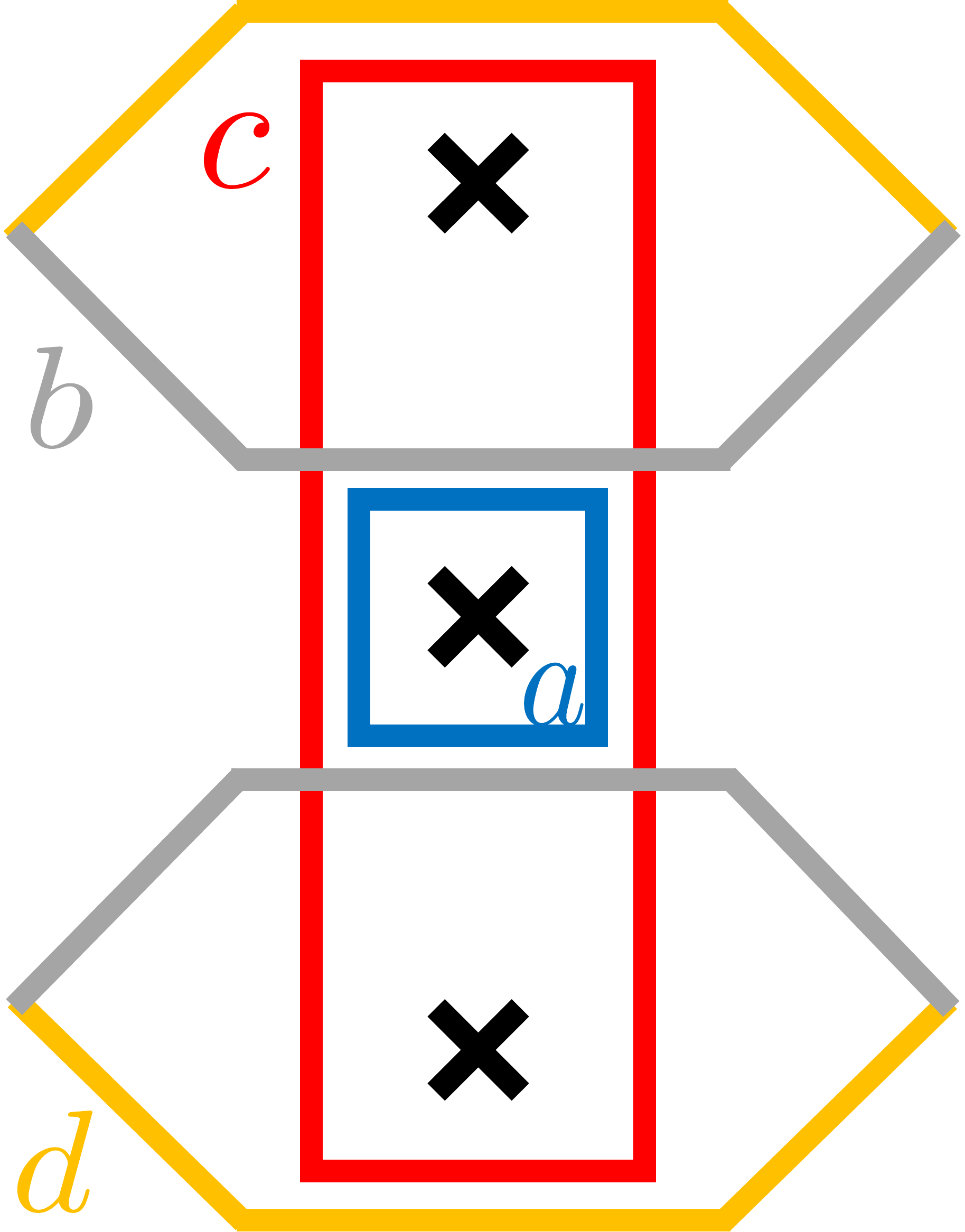}} &= \Big(\frac{ \mathcal{D}}{d_c} \Big)^2 \hspace{0.1cm} \Big( \adjustbox{valign = c}{\includegraphics[height = 1.2 cm]{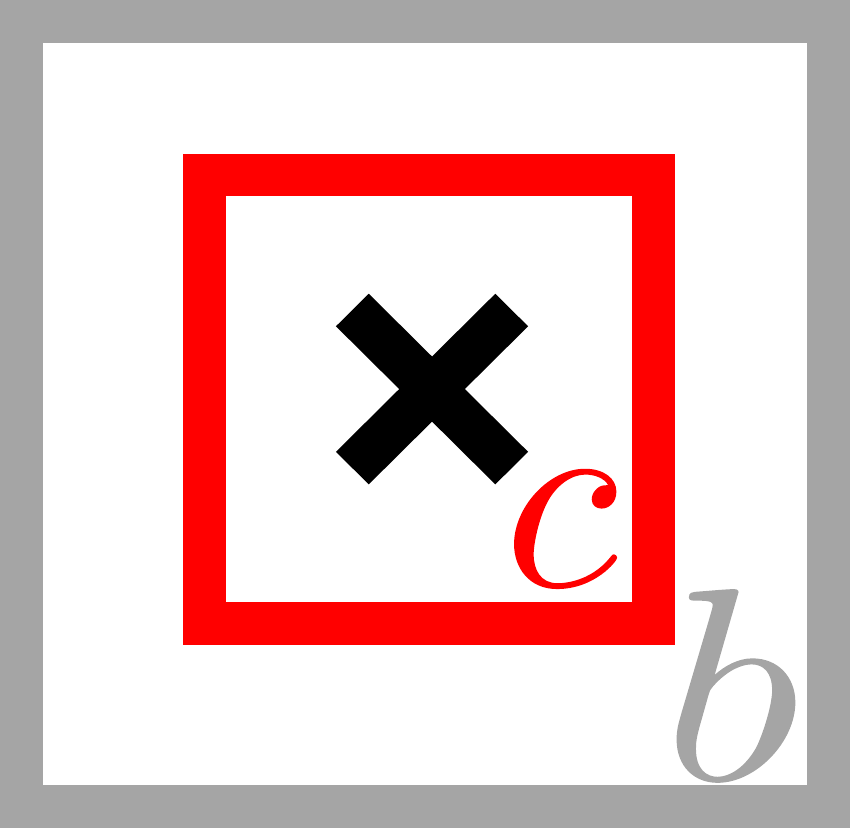}} \Big)^2 \hspace{0.1cm} \adjustbox{valign = c}{\includegraphics[height = 1.2 cm]{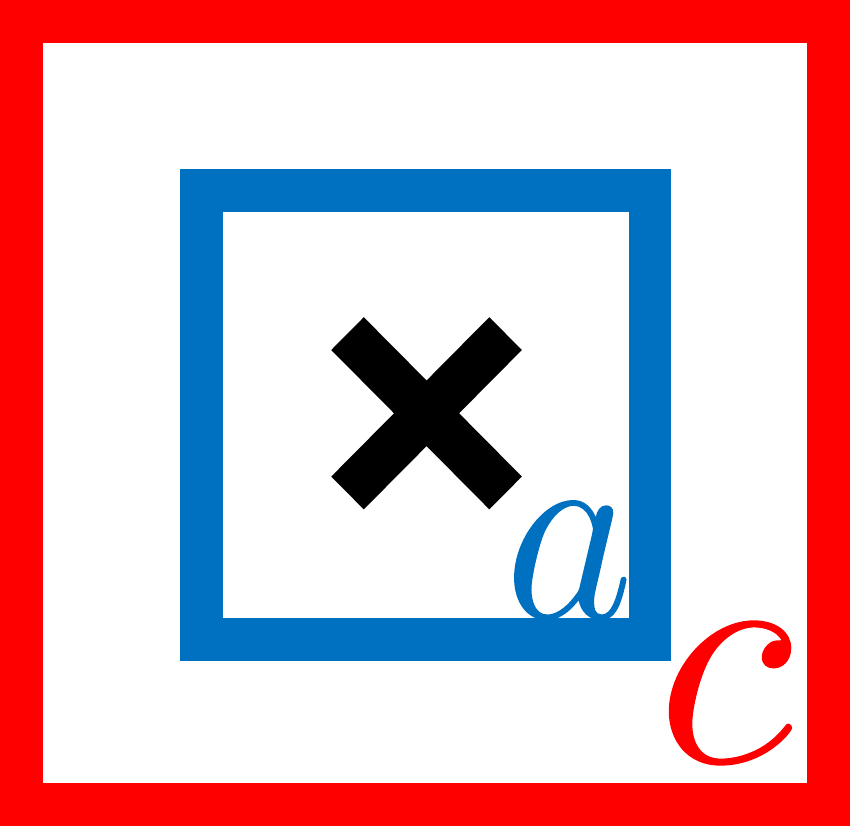}}  \notag\\
    &= \Big(\frac{ \mathcal{D}}{d_c} \Big)^2  \delta_{bc} \delta_{ac}.
\end{align}
Combining the above results, we have:
\begin{equation}
    \mathrm{Tr} (\rho_A)^2 = \sum_{a} |\psi_a|^4 \Big(\frac{ \mathcal{D}}{d_a} \Big)^4.
\end{equation}

Next, we consider the case of $\mathrm{Tr} (\rho_A)^3$. In this configuration, we have four nodes arranged in a circle, with each neighboring pair connected by three lines. If we label the leaves from inside to outside as $1,2,3$, the possible labels for the Wilson lines can be represented as: $1111,1212,2222,2323,3333,3131$.
Notably, because the even labels are identical to each other and the odd labels are also identical, we can simplify our labeling. For example, we can denote the Wilson lines by their even and odd labels rather than writing out each full label explicitly.

To visualize the configuration, we can represent it as follows:
\begin{align}
    \mathrm{Tr} (\rho_A)^3  
    = &\sum_{\substack{a_{11}, a_{12}, a_{22}, \\ a_{23}, a_{33}, a_{31}} } \psi_{11} \Bar{\psi}_{12} \psi_{22} \Bar{\psi}_{23} \psi_{33} \Bar{\psi}_{31} \notag\\
    & \times \hspace{0.2cm} \adjustbox{valign = c}{\includegraphics[height = 4. cm]{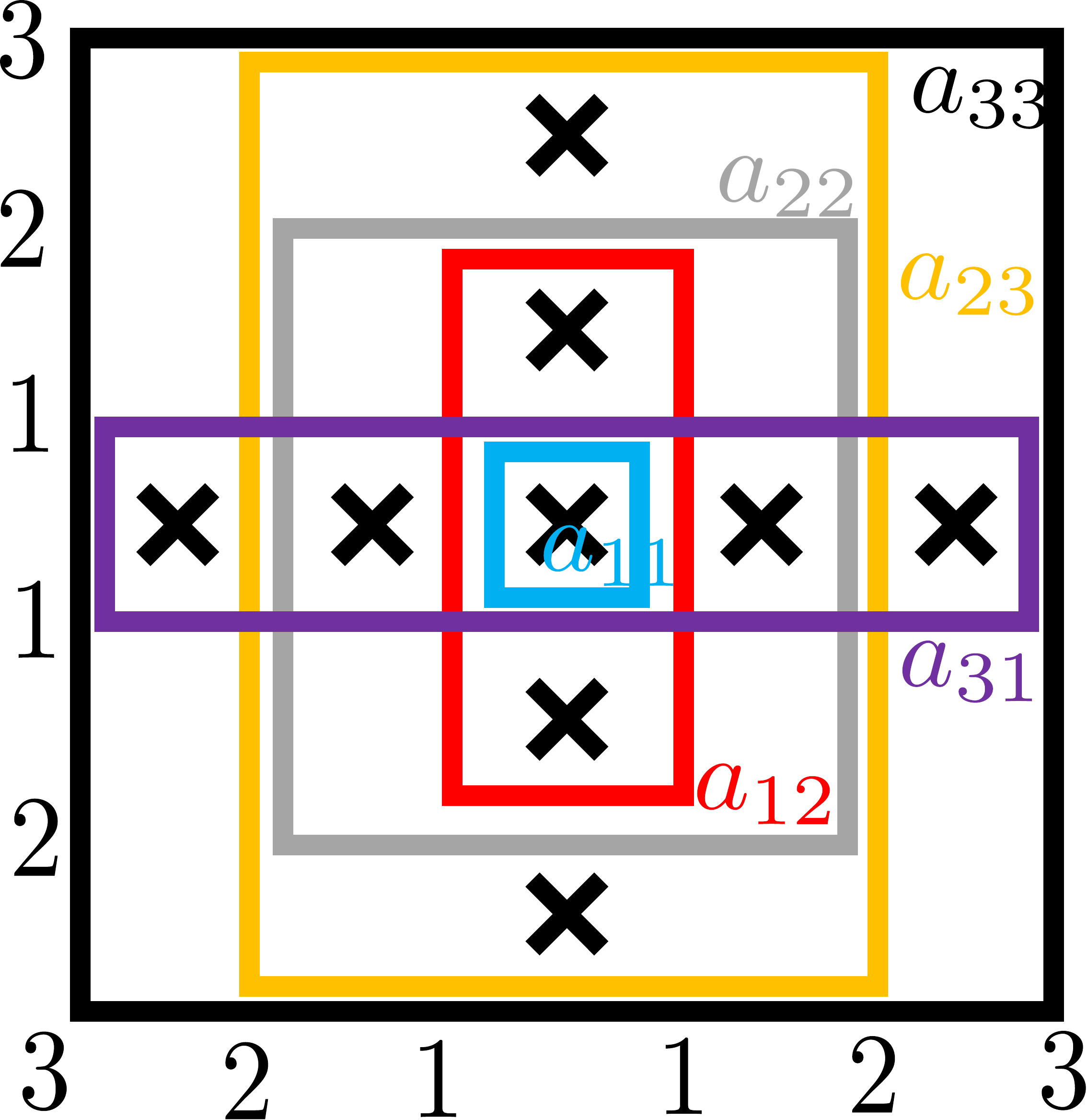}}.
\end{align}
To make the induction pattern clearer, we can exchange the rows of the diagram such that the internal rows are arranged in decreasing order from inside to outside, except for the outermost one:
\begin{equation}
    \adjustbox{valign = c}{\includegraphics[height = 4. cm]{Figure/SwitchRow.pdf}} \hspace{0.2cm} = \hspace{0.2cm} \adjustbox{valign = c}{\includegraphics[height = 4. cm]{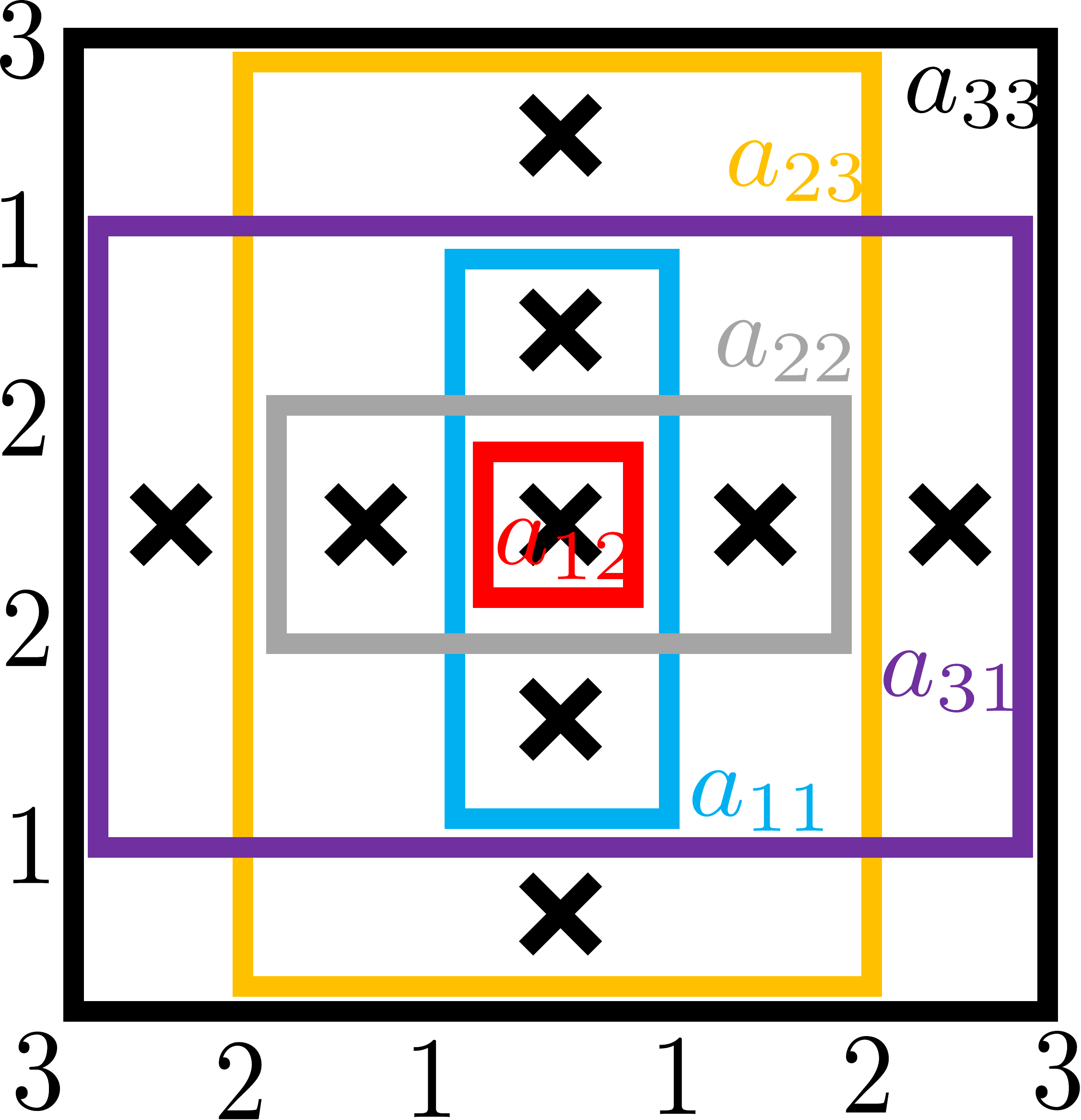}}.
\end{equation}
This adjustment helps to visualize the similarity between the configurations in each layer.

By applying the surgery method to the left and rightmost holes as detailed in Eq.~(\ref{Eq:SurgeryLeft}) and Eq.~(\ref{Eq:SurgeryRight}), we then obtain
\begin{equation}
    \adjustbox{valign = c}{\includegraphics[height = 4 cm]{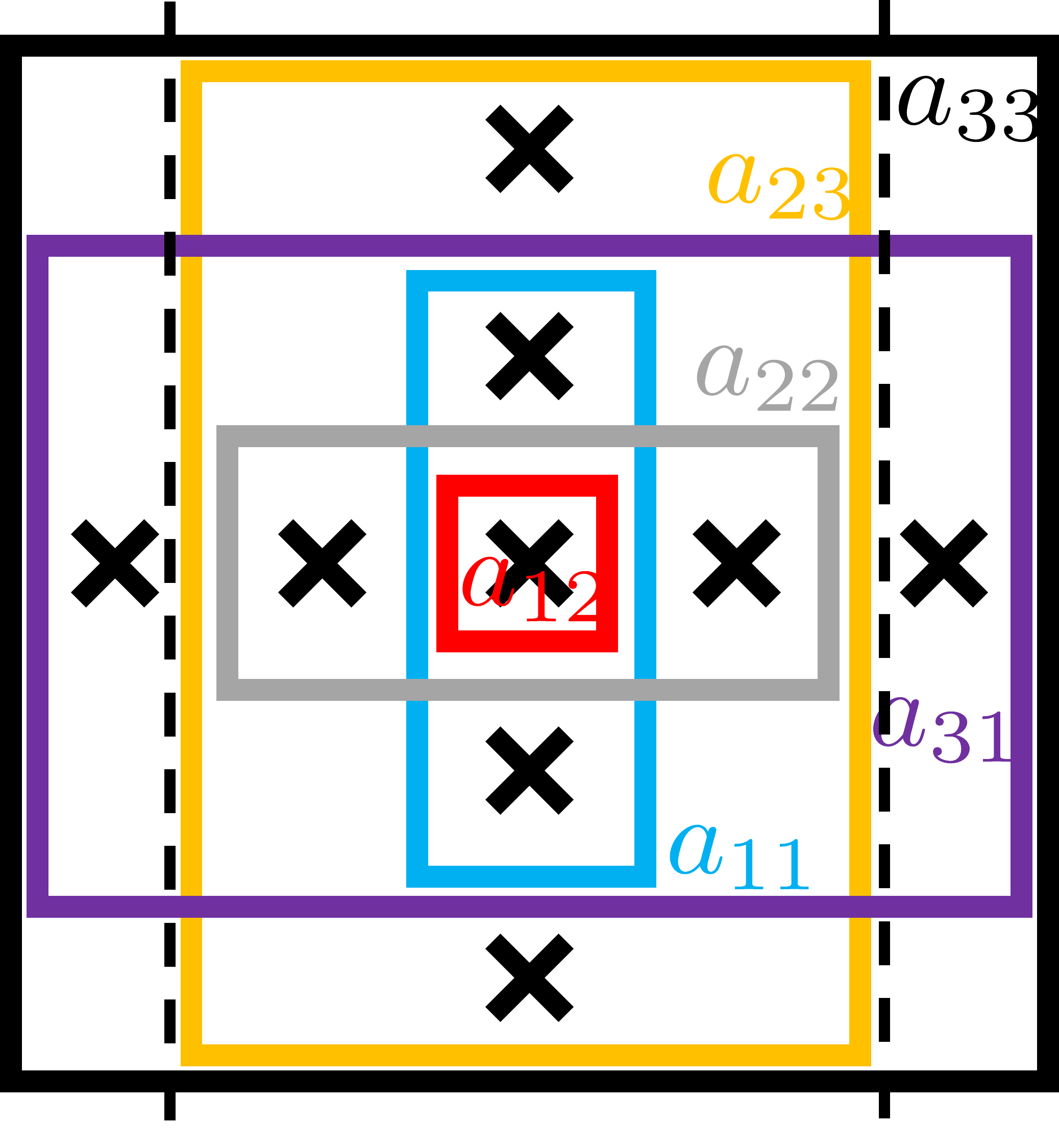}} \hspace{0.1cm} = \Big(\frac{ \mathcal{D}}{d_{33}} \Big)^2 \hspace{0.3cm}\adjustbox{valign = c}{\includegraphics[height = 4 cm]{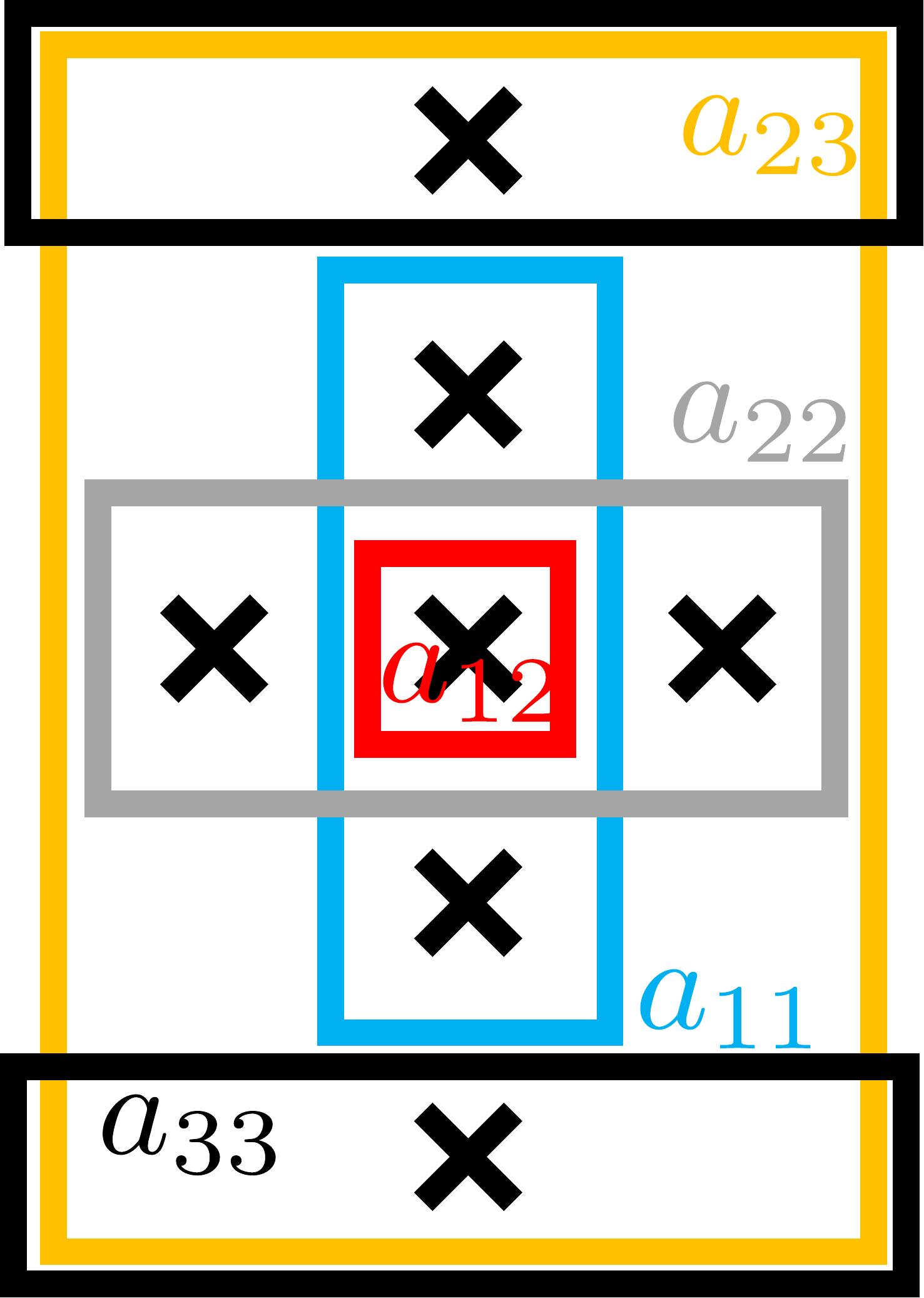}}.
\end{equation}
Next, we proceed to remove the uppermost and lowermost holes, resulting in:
\begin{equation}
    \adjustbox{valign = c}{\includegraphics[height = 4 cm]{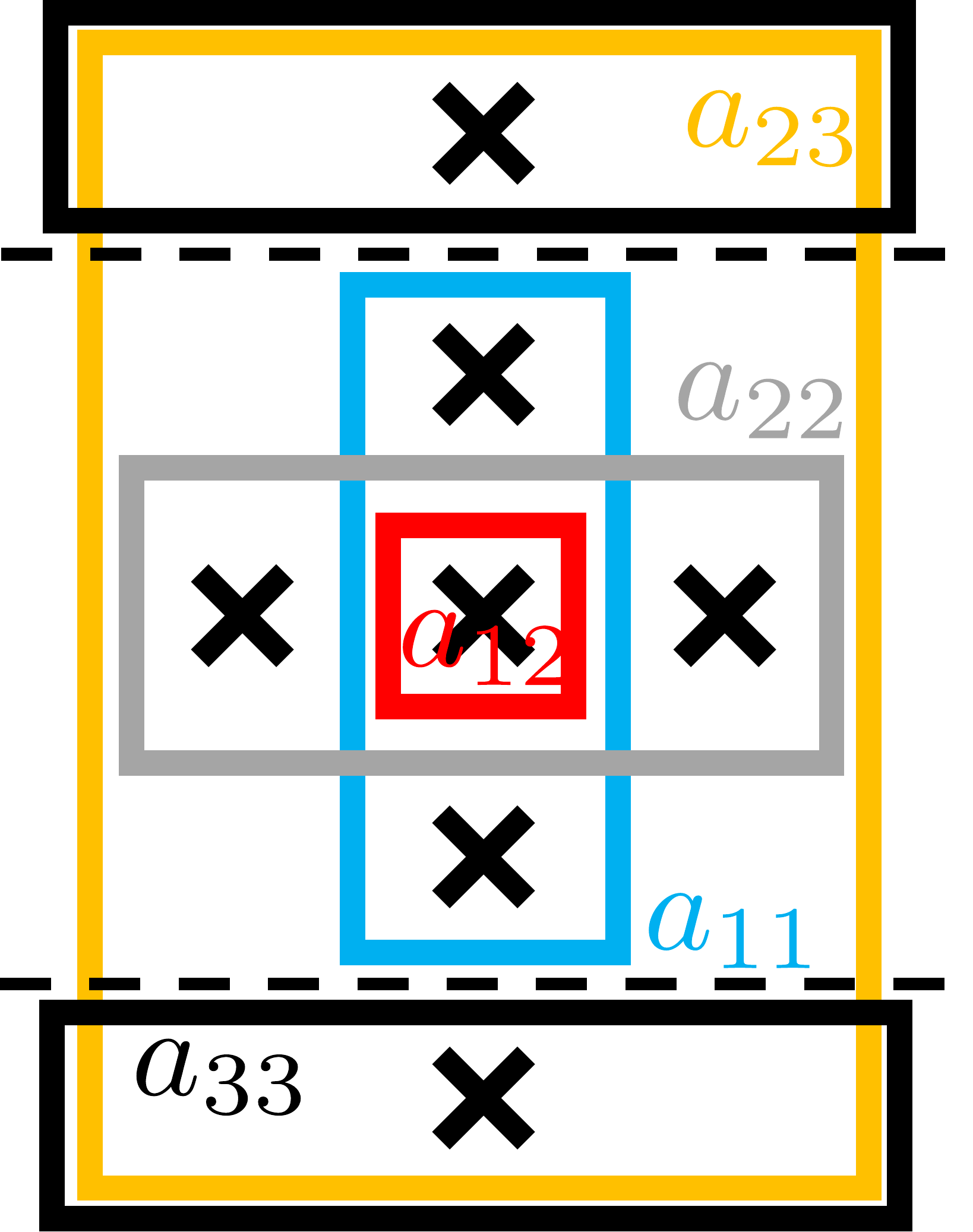}} \hspace{0.1cm} = \Big(\frac{ \mathcal{D}}{d_{33}} \Big)^2 \delta_{a_{31},a_{33}} \hspace{0.3cm}\adjustbox{valign = c}{\includegraphics[height = 3 cm]{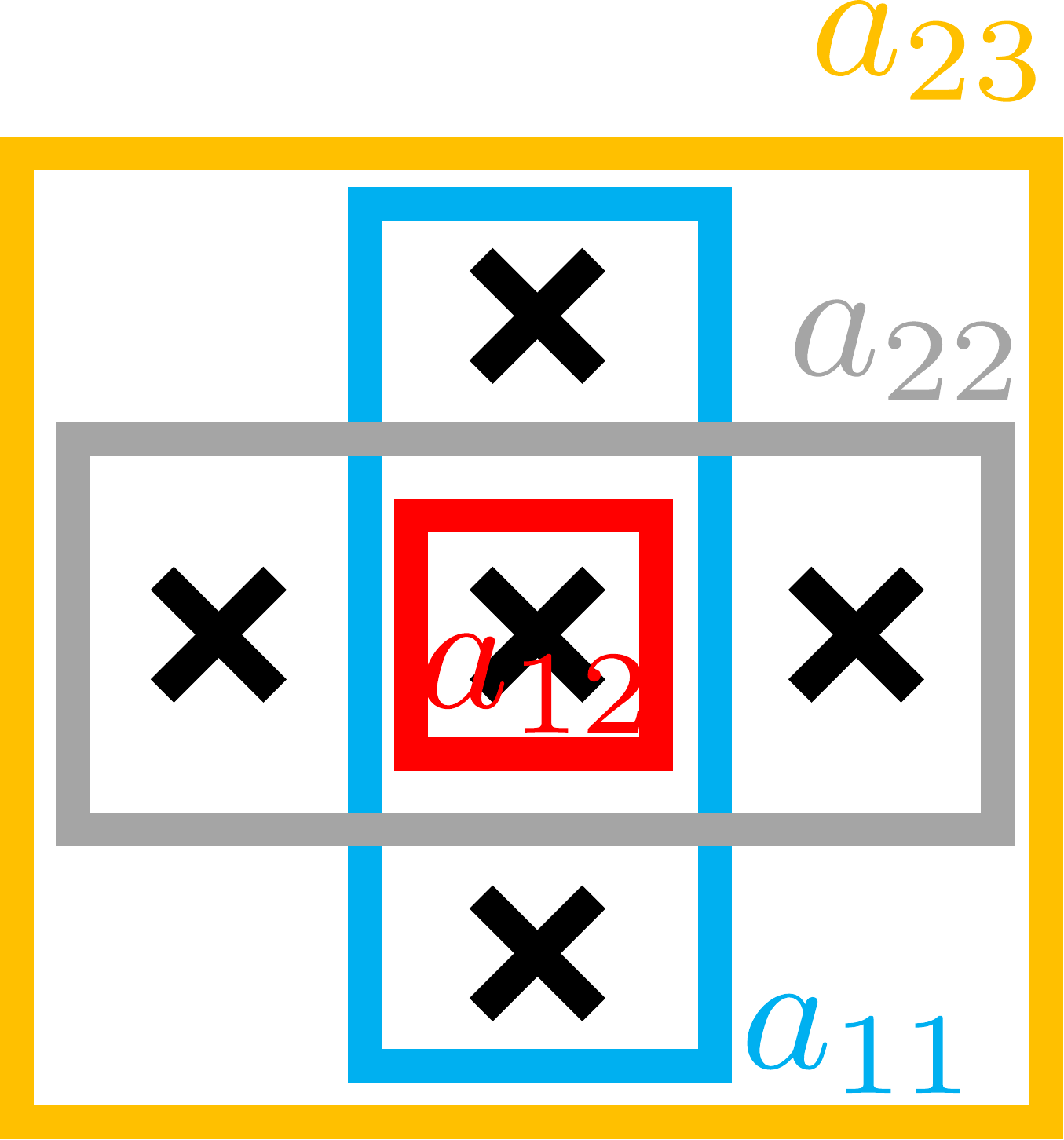}}.
\end{equation}
Finally, we rename the row indices to achieve a configuration equivalent to $\mathrm{Tr} (\rho_A)^2$. 
Therefore, from Eq.~(\ref{Eq:SurgeryLeft}), Eq.~(\ref{Eq:SurgeryRight}) and Eq.~(\ref{Eq:SurgeryUD}), we have
\begin{equation}
    \adjustbox{valign = c}{\includegraphics[height = 3 cm]{Figure/Cross3Tr3.pdf}} \hspace{0.1cm} = \Big(\frac{ \mathcal{D}}{d_{12}} \Big)^4 \delta_{a_{11},a_{12}} \delta_{a_{12},a_{23}} \delta_{a_{22},a_{23}}.
\end{equation}
Combining these results, we arrive at the expression:
\begin{equation}
    \mathrm{Tr} (\rho_A)^3 = \sum_{a} |\psi_a|^6 \Big(\frac{ \mathcal{D}}{d_a} \Big)^{8}.
\end{equation}

In general, for $\mathrm{Tr} (\rho_A)^n$, we have $n-1$ layers. Each layer consists of $4$ holes positioned on the left, right, top, and bottom. Including the hole in the center, this gives a total of $4(n-1)+1$ holes, as discussed in Sec.~\ref{SubSec:Vacuum}
\begin{equation}
    \mathrm{Tr} (\rho_A)^n = \adjustbox{valign = c}{\includegraphics[height = 4 cm]{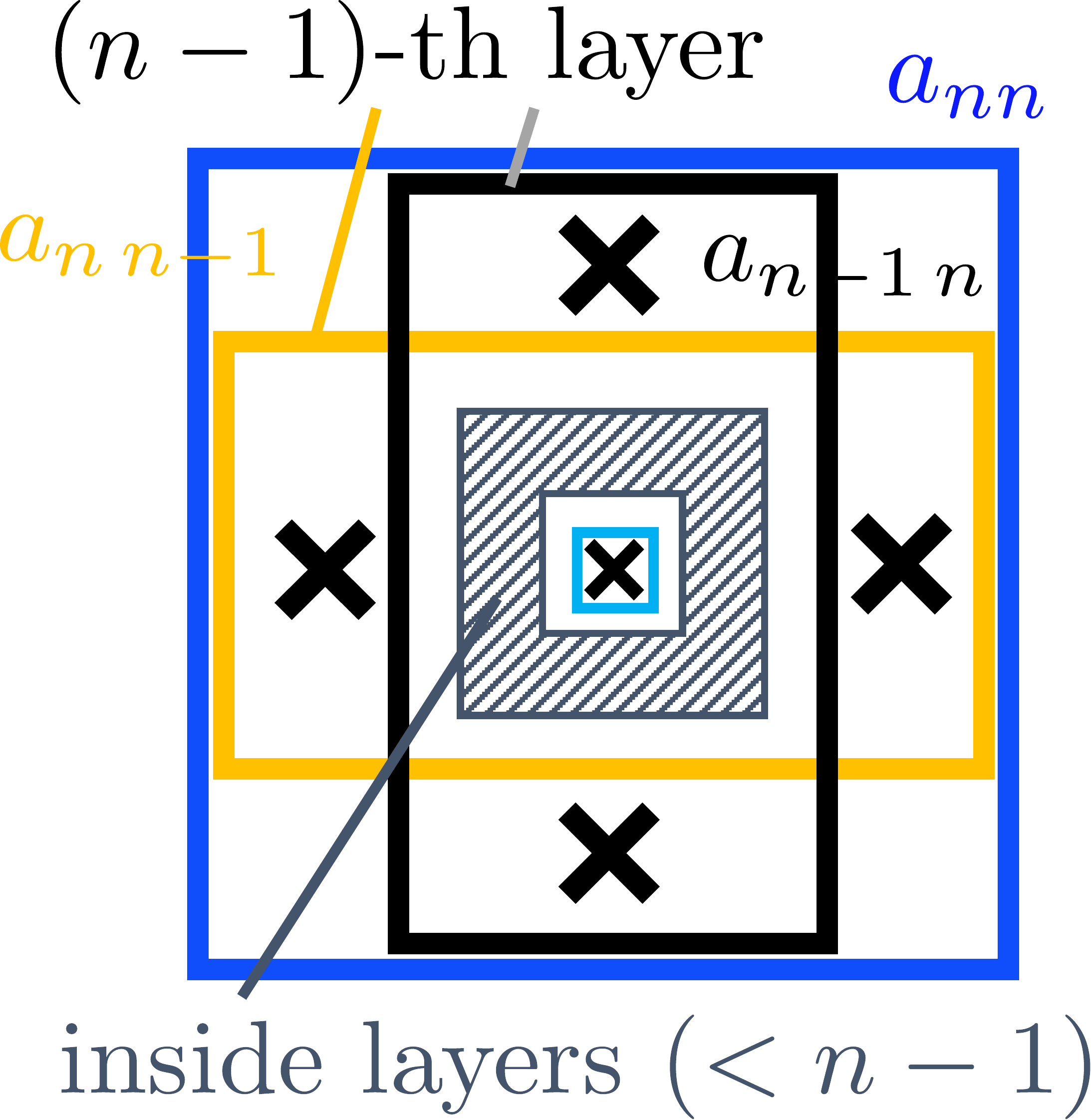}}
\end{equation}
By applying the surgery method to the outermost layer of holes, we obtain the configuration for $\mathrm{Tr} (\rho_A)^{n-1}$
\begin{align}
   & \adjustbox{valign = c}{\includegraphics[height = 4 cm]{Figure/CrossTrn.pdf}}  \notag\\
    = &\Big( \frac{D}{d_{\color{blue}\bullet}} \Big)^4 \delta_{{\color{black}\bullet}{\color{blue}\bullet}} \delta_{{\color{orange}\bullet}{\color{blue}\bullet}}\adjustbox{valign = c}{\includegraphics[height = 2 cm]{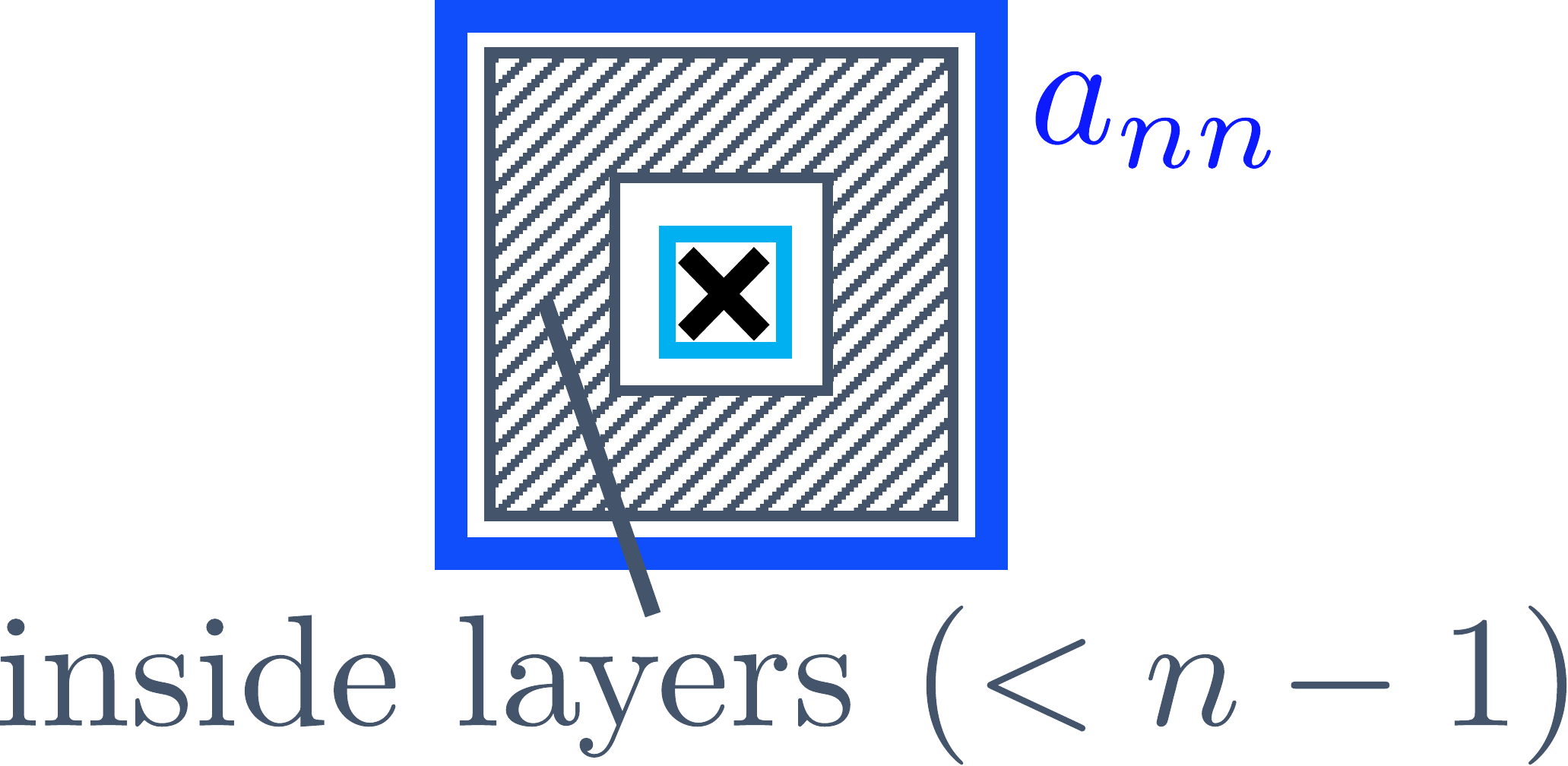}},
\end{align}
where we use the colored dots to label the Wilson lines and the delta functions constrain all the outer layer Wilson lines to be identical.
By induction, we perform surgery for a total of $n-1$ layers, resulting in a cumulative contribution of $-4(n-1)$th power of $S_{0a}$.
Thus, the overall result is:
\begin{equation}
    \mathrm{Tr} (\rho_A)^n = \sum_{a} |\psi_a|^{2n} \Big(\frac{ \mathcal{D}}{d_a} \Big)^{4(n-1)}.
\end{equation}
The TEE is then given by:
\begin{align}
    S_\text{TEE} &= \lim_{n\rightarrow 1} \frac{1}{1-n} \ln \mathrm{Tr} (\rho_A)^n \notag\\
    &= \sum_a 2|\psi_a|^2 (2\ln d_a - \ln |\psi_a|) - 4 \ln  \mathcal{D}.
\end{align}

To find the minimum under the constraints $0 \leq |\psi_a| \leq 1 $ and $\sum_a |\psi_a|^2 = 1$, we consider that the minimum must occur at a local extremum or on the boundary of the feasible region. Utilizing the method of Lagrange multipliers, as discussed in~\cite{KnotTEE}, we determine that the global minimum is attained at the boundary when $|\psi_a| = \delta_{0a}$.

This minimum corresponds to the scenario where no Wilson lines are inserted into the bulk. This aligns with our intuition that Wilson lines act as entangled pairs of quasi-particles across subregions, thereby increasing the overall entanglement.

Consequently, we complete the proof for the special case where $m=2$ in the previous section, showing that the intrinsic topological entanglement entropy for a canonical $R_2$ bipartition is given by $-4 \ln  \mathcal{D}$.

\subsection{Generic state on \texorpdfstring{$R_3$}{R\_3} and \texorpdfstring{$R_m$}{R\_m} bipartition}

Before generalizing the result to arbitrary rings, we will first examine the $R_3$ bipartition.
Consider a generic state $\ket{\psi} = \sum_a \psi_a \ket{a}$ with the $R_3$ bipartition.
The bulk configuration for $\mathrm{Tr} (\rho_A)^2$ now consists of six nodes arranged in a circle, with each neighboring pair connected by two lines.
Using similar pictorial notation as in the previous section, we have
\begin{align}
    & \mathrm{Tr} (\rho_A)^2   \notag\\
     = &\sum_{a_{11},a_{12},a_{22},a_{21}} \psi_{11} \Bar{\psi_{12}} \psi_{22} \Bar{\psi_{21}} \hspace{0.3cm} \adjustbox{valign = c}{\includegraphics[height = 3 cm]{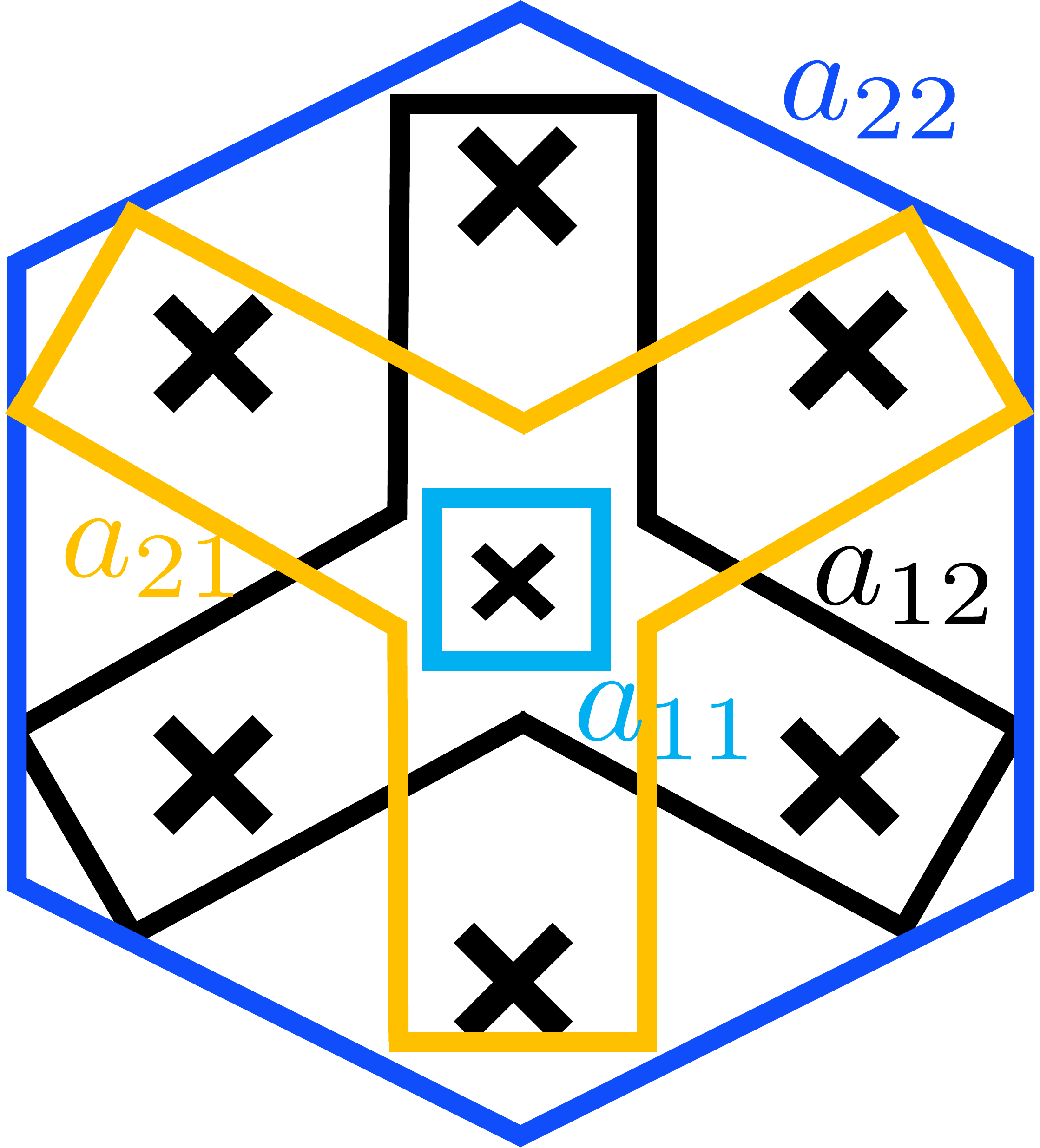}}.
\end{align}
In the figure, there are six rows of holes oriented at an angle of 
$\frac{n\pi}{3}$, where $n =0,1, \dots 5$. We separate these rows into two groups based on whether $n$ is even or odd. We denote a Wilson loop $a_{ij}$ if it threads through the $i$-th layer on the even rows and $j$-th layer on the odd rows.
We apply the surgery method to the outermost holes of the even rows, resulting in
\begin{align}
    &\adjustbox{valign = c}{\includegraphics[height = 3 cm]{Figure/Star.pdf}} \hspace{0.3cm}  \notag\\
    =& \Big( \frac{ \mathcal{D}}{d_{22}} \Big)^3 \delta_{a_{22},a_{12}} \hspace{0.3cm} \adjustbox{valign = c}{\includegraphics[height = 3 cm]{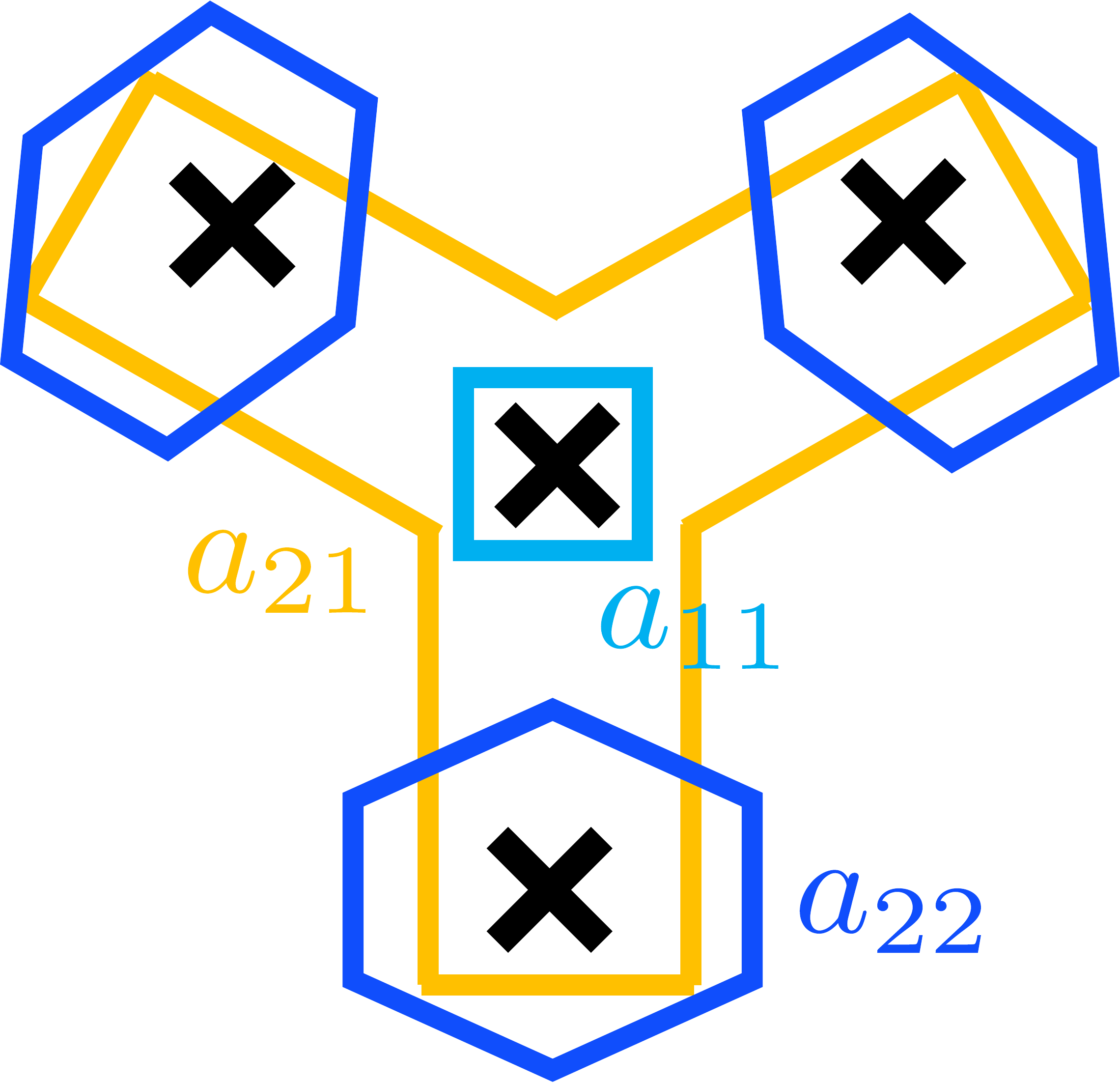}}.
\end{align}
Next, we perform the surgery on the outermost holes of the odd rows:
\begin{equation}
    \adjustbox{valign = c}{\includegraphics[height = 3 cm]{Figure/Star2.pdf}} \hspace{0.3cm} = \Big( \frac{ \mathcal{D}}{d_{22}} \Big)^3 \delta_{a_{22},a_{21}}\delta_{a_{22},a_{11}}.
\end{equation}
Combining the results, we obtain
\begin{equation}
    \mathrm{Tr} (\rho_A)^2 = \sum_{a} |\psi_a|^4 \Big( \frac{ \mathcal{D}}{d_a} \Big)^6.
\end{equation}

For $ \mathrm{Tr} (\rho_A)^n$, the bulk configuration consists of six nodes with $n$ lines connecting neighboring nodes.
This configuration translates to a figure with $n-1$ layers of holes, where each layer consists of six holes.
We then repeat the surgery method on the even and odd rows, layer by layer, obtaining
\begin{equation}
    \mathrm{Tr} (\rho_A)^n = \sum_{a} |\psi_a|^{2n} \Big( \frac{ \mathcal{D}}{d_a} \Big)^{6(n-1)}.
\end{equation}
The TEE is therefore given by
\begin{equation}
    S_\text{TEE} = \sum_a 2|\psi_a|^2 (3\ln d_a - \ln |\psi_a|) - 6 \ln  \mathcal{D},
\end{equation}
which reaches its minimum value of $- 6 \ln  \mathcal{D}$ when there are no Wilson lines present.

Similar arguments can be generalized to the $R_m$ bipartition. In this case, the configuration will consist of $2m$ rows of holes, which can be separated into even and odd rows.
Surgery method is then applied layer by layer to the $n-1$ layers of holes.
For each layer, we perform the surgery on the even and odd rows sequentially to remove the $2m$ holes:
\begin{equation}
\begin{aligned}
    &\adjustbox{valign = c}{\includegraphics[height = 4 cm]{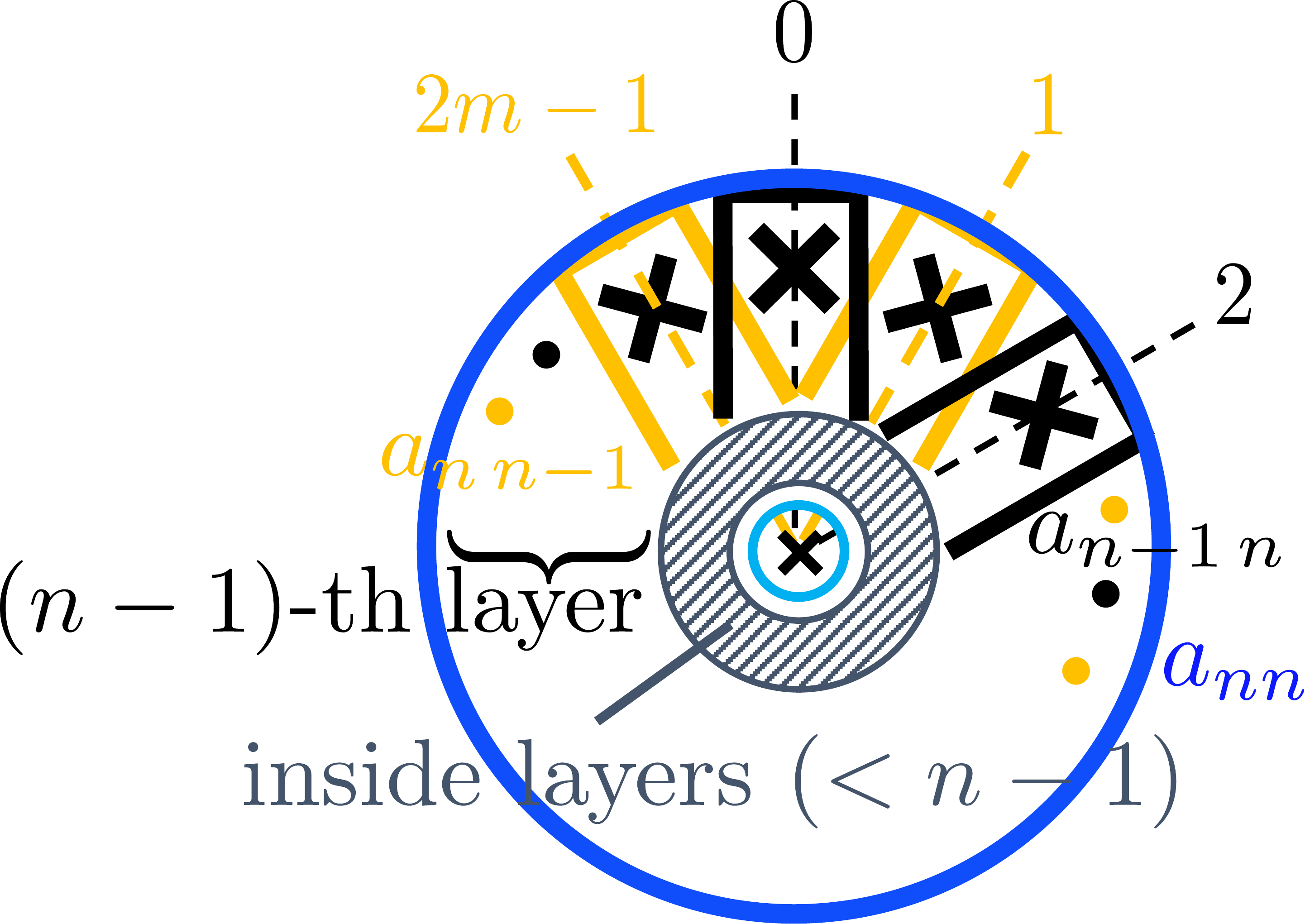}} \hspace{0.3cm} \notag \\
   = &\Big( \frac{ \mathcal{D}}{d_{\color{blue}\bullet}} \Big)^m \delta_{{\color{black}\bullet}{\color{blue}\bullet}} \hspace{0.3cm} 
    \adjustbox{valign = c}{\includegraphics[height = 3 cm]{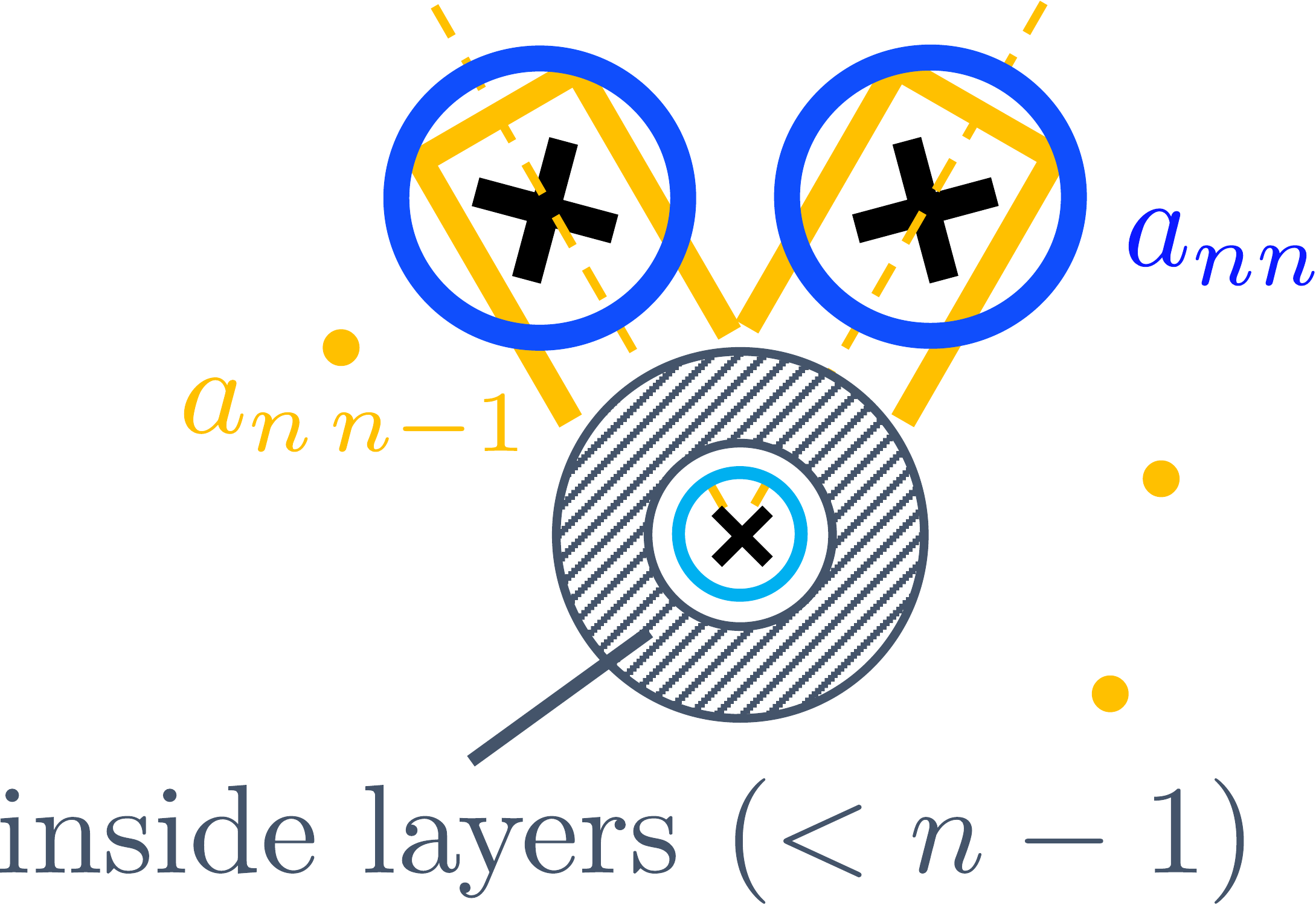}} \hspace{0.3cm}\\
    = &\Big( \frac{ \mathcal{D}}{d_{\color{blue}\bullet}} \Big)^{2m}  \delta_{{\color{blue}\bullet}{\color{orange}\bullet}} \delta_{{\color{black}\bullet}{\color{blue}\bullet}} \hspace{0.3cm}
    \adjustbox{valign = c}{\includegraphics[height = 2 cm]{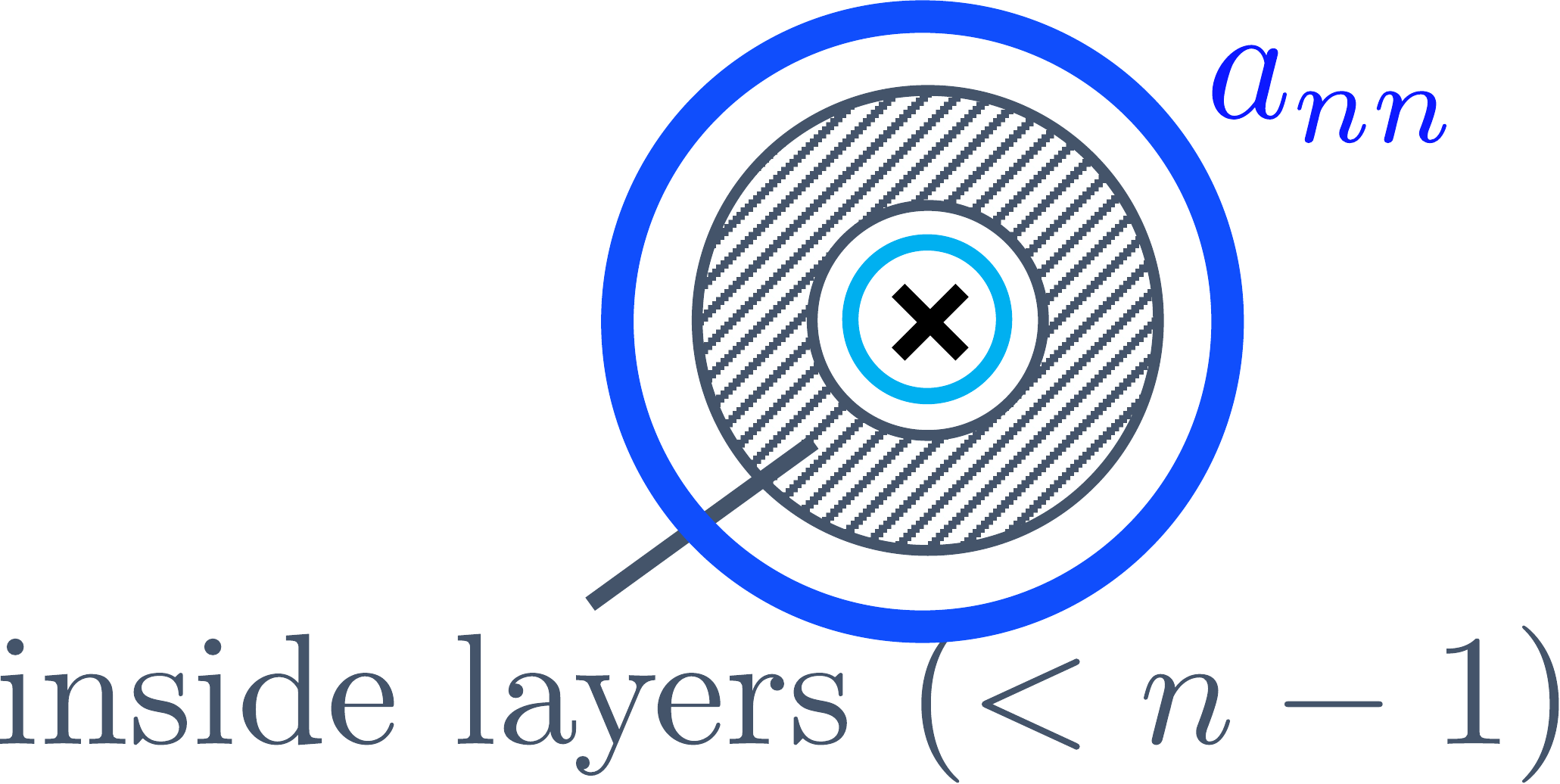}} \hspace{0.3cm}.
\end{aligned}
\end{equation}

By induction, after peeling off all $n-1$ layers, we obtain
\begin{equation}
    \mathrm{Tr} (\rho_A)^n = \sum_{a} |\psi_a|^{2n} \Big( \frac{ \mathcal{D}}{d_a} \Big)^{2m(n-1)}.
\end{equation}
The TEE is then given by
\begin{equation}\label{Eq:GeneralTEE}
    S_\text{TEE} = \sum_a p_a (m \ln d_a - \ln p_a )+ -2m \ln  \mathcal{D},
\end{equation}
where $p_a = |\psi_a|^2$ plays the role of classical probability.

The TEE can be separated into three distinct components:
\begin{enumerate}
    \item \textbf{Intrinsic TEE} $S_{\text{iTEE}} = -2m \ln  \mathcal{D}$: This is a non-positive contribution that depends solely on the number of interfaces between the two subregions of the system. It does not depend on any specific ground state configuration.
    \item \textbf{Wilson line} contribution $S_{\mathrm{Wil}} = 2m \sum_a p_a \ln d_a$: This term represents a non-negative contribution to the entanglement due to the presence of Wilson lines linking the two subregions. The linkage between subregions is proportional to the number of non-contractible interfaces and contributes to the total entanglement through the quantum dimensions $d_a$ of different quasi-particle types.
    \item \textbf{Classical} contribution $S_{\mathrm{cl}} = -2 \sum_a p_a \ln p_a$: This non-negative contribution arises from the classical superposition of degenerate ground states. The term is directly related to the Shannon entropy of the ground state probabilities. Importantly, this classical contribution only manifests when the bipartition of the system encloses a non-contractible loop, meaning that the system's topology plays a critical role in its appearance.
\end{enumerate}
The latter two contributions, $S_{\mathrm{Wil}}$ and $S_{\mathrm{cl}}$ both arise due to the degeneracy of the ground state in a topologically ordered system. These contributions reflect entanglement properties that depend on the specific ground state configuration.

We refer to the combination of these two contributions as the ground state TEE:
\begin{equation}\label{Eq:GSTEE}
    S_{\text{gs}} = S_{\mathrm{Wil}} + S_{\mathrm{cl}} = \sum_a p_a (m \ln d_a - \ln p_a ).
\end{equation}
This ground state TEE is always non-negative and only appears when $m>1$, meaning that the bipartition of the system winds around a non-contractible cycle on the torus.

In summary, given a general bipartition of a torus with boundary $\partial A = (\amalg_{i=1}^n C_{i}) \amalg (\amalg_{j=1}^{2m} K_j(p,q))$, the bipartition can be simplified to the canonical form $R_m$. This is done by removing the contractible interfaces, which introduces a correction term $-n \ln \mathcal{D}$, and applying a coordinate transformation that maps the original state $\ket{\psi}$ to an effective state $\ket{\mathcal{O}\psi}$. The resulting TEE is:
\begin{equation}
    S_{\text{TEE}}(A, \ket{\psi}) = S_{\text{TEE}}(R_m, \ket{\mathcal{O}\psi}) - n\ln  \mathcal{D}.
\end{equation}
The TEE for the $R_m$ bipartition can be further decomposed as:
\begin{equation}
    S_{\text{TEE}}(R_m, \ket{\psi}) = - 2m\ln  \mathcal{D} + S_{\mathrm{gs}}(R_m, \ket{\psi}).
\end{equation}
The intrinsic TEE of $A$ is given by
\begin{equation}
    S_{\text{iTEE}}(A) = -\pi_{\partial A} \ln  \mathcal{D}.
\end{equation}
where $\pi_{\partial A} = 2m+n$ represents the number of interface components (contractible and non-contractible). The intrinsic TEE also serves as the lower bound for the TEE.

\bibliography{references}

\begin{thebibliography}{36}%
\makeatletter
\providecommand \@ifxundefined [1]{%
 \@ifx{#1\undefined}
}%
\providecommand \@ifnum [1]{%
 \ifnum #1\expandafter \@firstoftwo
 \else \expandafter \@secondoftwo
 \fi
}%
\providecommand \@ifx [1]{%
 \ifx #1\expandafter \@firstoftwo
 \else \expandafter \@secondoftwo
 \fi
}%
\providecommand \natexlab [1]{#1}%
\providecommand \enquote  [1]{``#1''}%
\providecommand \bibnamefont  [1]{#1}%
\providecommand \bibfnamefont [1]{#1}%
\providecommand \citenamefont [1]{#1}%
\providecommand \href@noop [0]{\@secondoftwo}%
\providecommand \href [0]{\begingroup \@sanitize@url \@href}%
\providecommand \@href[1]{\@@startlink{#1}\@@href}%
\providecommand \@@href[1]{\endgroup#1\@@endlink}%
\providecommand \@sanitize@url [0]{\catcode `\\12\catcode `\$12\catcode `\&12\catcode `\#12\catcode `\^12\catcode `\_12\catcode `\%12\relax}%
\providecommand \@@startlink[1]{}%
\providecommand \@@endlink[0]{}%
\providecommand \url  [0]{\begingroup\@sanitize@url \@url }%
\providecommand \@url [1]{\endgroup\@href {#1}{\urlprefix }}%
\providecommand \urlprefix  [0]{URL }%
\providecommand \Eprint [0]{\href }%
\providecommand \doibase [0]{https://doi.org/}%
\providecommand \selectlanguage [0]{\@gobble}%
\providecommand \bibinfo  [0]{\@secondoftwo}%
\providecommand \bibfield  [0]{\@secondoftwo}%
\providecommand \translation [1]{[#1]}%
\providecommand \BibitemOpen [0]{}%
\providecommand \bibitemStop [0]{}%
\providecommand \bibitemNoStop [0]{.\EOS\space}%
\providecommand \EOS [0]{\spacefactor3000\relax}%
\providecommand \BibitemShut  [1]{\csname bibitem#1\endcsname}%
\let\auto@bib@innerbib\@empty
\bibitem [{\citenamefont {WEN}(1990)}]{TO}%
  \BibitemOpen
  \bibfield  {author} {\bibinfo {author} {\bibfnamefont {X.~G.}\ \bibnamefont {WEN}},\ }\href {https://doi.org/10.1142/S0217979290000139} {\bibfield  {journal} {\bibinfo  {journal} {International Journal of Modern Physics B}\ }\textbf {\bibinfo {volume} {04}},\ \bibinfo {pages} {239} (\bibinfo {year} {1990})},\ \Eprint {https://arxiv.org/abs/https://doi.org/10.1142/S0217979290000139} {https://doi.org/10.1142/S0217979290000139} \BibitemShut {NoStop}%
\bibitem [{\citenamefont {Tsui}\ \emph {et~al.}(1982)\citenamefont {Tsui}, \citenamefont {Stormer},\ and\ \citenamefont {Gossard}}]{FQH1}%
  \BibitemOpen
  \bibfield  {author} {\bibinfo {author} {\bibfnamefont {D.~C.}\ \bibnamefont {Tsui}}, \bibinfo {author} {\bibfnamefont {H.~L.}\ \bibnamefont {Stormer}},\ and\ \bibinfo {author} {\bibfnamefont {A.~C.}\ \bibnamefont {Gossard}},\ }\href {https://doi.org/10.1103/PhysRevLett.48.1559} {\bibfield  {journal} {\bibinfo  {journal} {Phys. Rev. Lett.}\ }\textbf {\bibinfo {volume} {48}},\ \bibinfo {pages} {1559} (\bibinfo {year} {1982})}\BibitemShut {NoStop}%
\bibitem [{\citenamefont {Laughlin}(1983)}]{FQH2}%
  \BibitemOpen
  \bibfield  {author} {\bibinfo {author} {\bibfnamefont {R.~B.}\ \bibnamefont {Laughlin}},\ }\href {https://doi.org/10.1103/PhysRevLett.50.1395} {\bibfield  {journal} {\bibinfo  {journal} {Phys. Rev. Lett.}\ }\textbf {\bibinfo {volume} {50}},\ \bibinfo {pages} {1395} (\bibinfo {year} {1983})}\BibitemShut {NoStop}%
\bibitem [{\citenamefont {Wen}(2002)}]{SpinLiquid}%
  \BibitemOpen
  \bibfield  {author} {\bibinfo {author} {\bibfnamefont {X.-G.}\ \bibnamefont {Wen}},\ }\href {https://doi.org/10.1103/PhysRevB.65.165113} {\bibfield  {journal} {\bibinfo  {journal} {Phys. Rev. B}\ }\textbf {\bibinfo {volume} {65}},\ \bibinfo {pages} {165113} (\bibinfo {year} {2002})}\BibitemShut {NoStop}%
\bibitem [{\citenamefont {Stone}\ and\ \citenamefont {Chung}(2006)}]{px+ipy}%
  \BibitemOpen
  \bibfield  {author} {\bibinfo {author} {\bibfnamefont {M.}~\bibnamefont {Stone}}\ and\ \bibinfo {author} {\bibfnamefont {S.-B.}\ \bibnamefont {Chung}},\ }\href {https://doi.org/10.1103/PhysRevB.73.014505} {\bibfield  {journal} {\bibinfo  {journal} {Phys. Rev. B}\ }\textbf {\bibinfo {volume} {73}},\ \bibinfo {pages} {014505} (\bibinfo {year} {2006})}\BibitemShut {NoStop}%
\bibitem [{\citenamefont {Kitaev}\ and\ \citenamefont {Preskill}(2006)}]{TEE}%
  \BibitemOpen
  \bibfield  {author} {\bibinfo {author} {\bibfnamefont {A.}~\bibnamefont {Kitaev}}\ and\ \bibinfo {author} {\bibfnamefont {J.}~\bibnamefont {Preskill}},\ }\href {https://doi.org/10.1103/PhysRevLett.96.110404} {\bibfield  {journal} {\bibinfo  {journal} {Phys. Rev. Lett.}\ }\textbf {\bibinfo {volume} {96}},\ \bibinfo {pages} {110404} (\bibinfo {year} {2006})}\BibitemShut {NoStop}%
\bibitem [{\citenamefont {Levin}\ and\ \citenamefont {Wen}(2006)}]{TEE2}%
  \BibitemOpen
  \bibfield  {author} {\bibinfo {author} {\bibfnamefont {M.}~\bibnamefont {Levin}}\ and\ \bibinfo {author} {\bibfnamefont {X.-G.}\ \bibnamefont {Wen}},\ }\href {https://doi.org/10.1103/PhysRevLett.96.110405} {\bibfield  {journal} {\bibinfo  {journal} {Phys. Rev. Lett.}\ }\textbf {\bibinfo {volume} {96}},\ \bibinfo {pages} {110405} (\bibinfo {year} {2006})}\BibitemShut {NoStop}%
\bibitem [{\citenamefont {Wen}\ \emph {et~al.}(2016)\citenamefont {Wen}, \citenamefont {Matsuura},\ and\ \citenamefont {Ryu}}]{EdgeTEE}%
  \BibitemOpen
  \bibfield  {author} {\bibinfo {author} {\bibfnamefont {X.}~\bibnamefont {Wen}}, \bibinfo {author} {\bibfnamefont {S.}~\bibnamefont {Matsuura}},\ and\ \bibinfo {author} {\bibfnamefont {S.}~\bibnamefont {Ryu}},\ }\href {https://doi.org/10.1103/PhysRevB.93.245140} {\bibfield  {journal} {\bibinfo  {journal} {Phys. Rev. B}\ }\textbf {\bibinfo {volume} {93}},\ \bibinfo {pages} {245140} (\bibinfo {year} {2016})}\BibitemShut {NoStop}%
\bibitem [{\citenamefont {Cardy}(2006)}]{CARDY2006333}%
  \BibitemOpen
  \bibfield  {author} {\bibinfo {author} {\bibfnamefont {J.}~\bibnamefont {Cardy}},\ }in\ \href {https://doi.org/https://doi.org/10.1016/B0-12-512666-2/00398-9} {\emph {\bibinfo {booktitle} {Encyclopedia of Mathematical Physics}}},\ \bibinfo {editor} {edited by\ \bibinfo {editor} {\bibfnamefont {J.-P.}\ \bibnamefont {FranÃ§oise}}, \bibinfo {editor} {\bibfnamefont {G.~L.}\ \bibnamefont {Naber}},\ and\ \bibinfo {editor} {\bibfnamefont {T.~S.}\ \bibnamefont {Tsun}}}\ (\bibinfo  {publisher} {Academic Press},\ \bibinfo {address} {Oxford},\ \bibinfo {year} {2006})\ pp.\ \bibinfo {pages} {333--340}\BibitemShut {NoStop}%
\bibitem [{\citenamefont {Qi}\ \emph {et~al.}(2012)\citenamefont {Qi}, \citenamefont {Katsura},\ and\ \citenamefont {Ludwig}}]{BCFT1}%
  \BibitemOpen
  \bibfield  {author} {\bibinfo {author} {\bibfnamefont {X.-L.}\ \bibnamefont {Qi}}, \bibinfo {author} {\bibfnamefont {H.}~\bibnamefont {Katsura}},\ and\ \bibinfo {author} {\bibfnamefont {A.~W.~W.}\ \bibnamefont {Ludwig}},\ }\href {https://doi.org/10.1103/PhysRevLett.108.196402} {\bibfield  {journal} {\bibinfo  {journal} {Phys. Rev. Lett.}\ }\textbf {\bibinfo {volume} {108}},\ \bibinfo {pages} {196402} (\bibinfo {year} {2012})}\BibitemShut {NoStop}%
\bibitem [{\citenamefont {Das}\ and\ \citenamefont {Datta}(2015)}]{BCFT2}%
  \BibitemOpen
  \bibfield  {author} {\bibinfo {author} {\bibfnamefont {D.}~\bibnamefont {Das}}\ and\ \bibinfo {author} {\bibfnamefont {S.}~\bibnamefont {Datta}},\ }\href {https://doi.org/10.1103/PhysRevLett.115.131602} {\bibfield  {journal} {\bibinfo  {journal} {Phys. Rev. Lett.}\ }\textbf {\bibinfo {volume} {115}},\ \bibinfo {pages} {131602} (\bibinfo {year} {2015})}\BibitemShut {NoStop}%
\bibitem [{\citenamefont {Haque}\ \emph {et~al.}(2007)\citenamefont {Haque}, \citenamefont {Zozulya},\ and\ \citenamefont {Schoutens}}]{FQHTO1}%
  \BibitemOpen
  \bibfield  {author} {\bibinfo {author} {\bibfnamefont {M.}~\bibnamefont {Haque}}, \bibinfo {author} {\bibfnamefont {O.}~\bibnamefont {Zozulya}},\ and\ \bibinfo {author} {\bibfnamefont {K.}~\bibnamefont {Schoutens}},\ }\href {https://doi.org/10.1103/PhysRevLett.98.060401} {\bibfield  {journal} {\bibinfo  {journal} {Phys. Rev. Lett.}\ }\textbf {\bibinfo {volume} {98}},\ \bibinfo {pages} {060401} (\bibinfo {year} {2007})}\BibitemShut {NoStop}%
\bibitem [{\citenamefont {Zozulya}\ \emph {et~al.}(2007)\citenamefont {Zozulya}, \citenamefont {Haque}, \citenamefont {Schoutens},\ and\ \citenamefont {Rezayi}}]{FQHTO2}%
  \BibitemOpen
  \bibfield  {author} {\bibinfo {author} {\bibfnamefont {O.~S.}\ \bibnamefont {Zozulya}}, \bibinfo {author} {\bibfnamefont {M.}~\bibnamefont {Haque}}, \bibinfo {author} {\bibfnamefont {K.}~\bibnamefont {Schoutens}},\ and\ \bibinfo {author} {\bibfnamefont {E.~H.}\ \bibnamefont {Rezayi}},\ }\href {https://doi.org/10.1103/PhysRevB.76.125310} {\bibfield  {journal} {\bibinfo  {journal} {Phys. Rev. B}\ }\textbf {\bibinfo {volume} {76}},\ \bibinfo {pages} {125310} (\bibinfo {year} {2007})}\BibitemShut {NoStop}%
\bibitem [{\citenamefont {Friedman}\ and\ \citenamefont {Levine}(2008)}]{FQHTO3}%
  \BibitemOpen
  \bibfield  {author} {\bibinfo {author} {\bibfnamefont {B.~A.}\ \bibnamefont {Friedman}}\ and\ \bibinfo {author} {\bibfnamefont {G.~C.}\ \bibnamefont {Levine}},\ }\href {https://doi.org/10.1103/PhysRevB.78.035320} {\bibfield  {journal} {\bibinfo  {journal} {Phys. Rev. B}\ }\textbf {\bibinfo {volume} {78}},\ \bibinfo {pages} {035320} (\bibinfo {year} {2008})}\BibitemShut {NoStop}%
\bibitem [{\citenamefont {Isakov}\ \emph {et~al.}(2011)\citenamefont {Isakov}, \citenamefont {Hastings},\ and\ \citenamefont {Melko}}]{SpinLiquidTO1}%
  \BibitemOpen
  \bibfield  {author} {\bibinfo {author} {\bibfnamefont {S.~V.}\ \bibnamefont {Isakov}}, \bibinfo {author} {\bibfnamefont {M.~B.}\ \bibnamefont {Hastings}},\ and\ \bibinfo {author} {\bibfnamefont {R.~G.}\ \bibnamefont {Melko}},\ }\href {https://doi.org/10.1038/nphys2036} {\bibfield  {journal} {\bibinfo  {journal} {Nature Physics}\ }\textbf {\bibinfo {volume} {7}},\ \bibinfo {pages} {772} (\bibinfo {year} {2011})}\BibitemShut {NoStop}%
\bibitem [{\citenamefont {Jiang}\ \emph {et~al.}(2012)\citenamefont {Jiang}, \citenamefont {Wang},\ and\ \citenamefont {Balents}}]{SpinLiquidTO2}%
  \BibitemOpen
  \bibfield  {author} {\bibinfo {author} {\bibfnamefont {H.-C.}\ \bibnamefont {Jiang}}, \bibinfo {author} {\bibfnamefont {Z.}~\bibnamefont {Wang}},\ and\ \bibinfo {author} {\bibfnamefont {L.}~\bibnamefont {Balents}},\ }\href {https://doi.org/10.1038/nphys2465} {\bibfield  {journal} {\bibinfo  {journal} {Nature Physics}\ }\textbf {\bibinfo {volume} {8}},\ \bibinfo {pages} {902} (\bibinfo {year} {2012})}\BibitemShut {NoStop}%
\bibitem [{\citenamefont {Castelnovo}\ and\ \citenamefont {Chamon}(2007)}]{ToricCodeTO}%
  \BibitemOpen
  \bibfield  {author} {\bibinfo {author} {\bibfnamefont {C.}~\bibnamefont {Castelnovo}}\ and\ \bibinfo {author} {\bibfnamefont {C.}~\bibnamefont {Chamon}},\ }\href {https://doi.org/10.1103/PhysRevB.76.184442} {\bibfield  {journal} {\bibinfo  {journal} {Phys. Rev. B}\ }\textbf {\bibinfo {volume} {76}},\ \bibinfo {pages} {184442} (\bibinfo {year} {2007})}\BibitemShut {NoStop}%
\bibitem [{\citenamefont {Furukawa}\ and\ \citenamefont {Misguich}(2007)}]{DimmerTO}%
  \BibitemOpen
  \bibfield  {author} {\bibinfo {author} {\bibfnamefont {S.}~\bibnamefont {Furukawa}}\ and\ \bibinfo {author} {\bibfnamefont {G.}~\bibnamefont {Misguich}},\ }\href {https://doi.org/10.1103/PhysRevB.75.214407} {\bibfield  {journal} {\bibinfo  {journal} {Phys. Rev. B}\ }\textbf {\bibinfo {volume} {75}},\ \bibinfo {pages} {214407} (\bibinfo {year} {2007})}\BibitemShut {NoStop}%
\bibitem [{\citenamefont {Zou}\ and\ \citenamefont {Haah}(2016)}]{Spurious1}%
  \BibitemOpen
  \bibfield  {author} {\bibinfo {author} {\bibfnamefont {L.}~\bibnamefont {Zou}}\ and\ \bibinfo {author} {\bibfnamefont {J.}~\bibnamefont {Haah}},\ }\href {https://doi.org/10.1103/PhysRevB.94.075151} {\bibfield  {journal} {\bibinfo  {journal} {Phys. Rev. B}\ }\textbf {\bibinfo {volume} {94}},\ \bibinfo {pages} {075151} (\bibinfo {year} {2016})}\BibitemShut {NoStop}%
\bibitem [{\citenamefont {Williamson}\ \emph {et~al.}(2019)\citenamefont {Williamson}, \citenamefont {Dua},\ and\ \citenamefont {Cheng}}]{Spurious2}%
  \BibitemOpen
  \bibfield  {author} {\bibinfo {author} {\bibfnamefont {D.~J.}\ \bibnamefont {Williamson}}, \bibinfo {author} {\bibfnamefont {A.}~\bibnamefont {Dua}},\ and\ \bibinfo {author} {\bibfnamefont {M.}~\bibnamefont {Cheng}},\ }\href {https://doi.org/10.1103/PhysRevLett.122.140506} {\bibfield  {journal} {\bibinfo  {journal} {Phys. Rev. Lett.}\ }\textbf {\bibinfo {volume} {122}},\ \bibinfo {pages} {140506} (\bibinfo {year} {2019})}\BibitemShut {NoStop}%
\bibitem [{\citenamefont {Fliss}\ \emph {et~al.}(2017)\citenamefont {Fliss}, \citenamefont {Wen}, \citenamefont {Parrikar}, \citenamefont {Hsieh}, \citenamefont {Han}, \citenamefont {Hughes},\ and\ \citenamefont {Leigh}}]{Spurious3}%
  \BibitemOpen
  \bibfield  {author} {\bibinfo {author} {\bibfnamefont {J.~R.}\ \bibnamefont {Fliss}}, \bibinfo {author} {\bibfnamefont {X.}~\bibnamefont {Wen}}, \bibinfo {author} {\bibfnamefont {O.}~\bibnamefont {Parrikar}}, \bibinfo {author} {\bibfnamefont {C.-T.}\ \bibnamefont {Hsieh}}, \bibinfo {author} {\bibfnamefont {B.}~\bibnamefont {Han}}, \bibinfo {author} {\bibfnamefont {T.~L.}\ \bibnamefont {Hughes}},\ and\ \bibinfo {author} {\bibfnamefont {R.~G.}\ \bibnamefont {Leigh}},\ }\href {https://doi.org/10.1007/JHEP09(2017)056} {\bibfield  {journal} {\bibinfo  {journal} {JHEP 09 (2017) 056}\ } (\bibinfo {year} {2017})}\BibitemShut {NoStop}%
\bibitem [{\citenamefont {Santos}\ \emph {et~al.}(2018)\citenamefont {Santos}, \citenamefont {Cano}, \citenamefont {Mulligan},\ and\ \citenamefont {Hughes}}]{Spurious4}%
  \BibitemOpen
  \bibfield  {author} {\bibinfo {author} {\bibfnamefont {L.~H.}\ \bibnamefont {Santos}}, \bibinfo {author} {\bibfnamefont {J.}~\bibnamefont {Cano}}, \bibinfo {author} {\bibfnamefont {M.}~\bibnamefont {Mulligan}},\ and\ \bibinfo {author} {\bibfnamefont {T.~L.}\ \bibnamefont {Hughes}},\ }\href {https://doi.org/10.1103/PhysRevB.98.075131} {\bibfield  {journal} {\bibinfo  {journal} {Phys. Rev. B}\ }\textbf {\bibinfo {volume} {98}},\ \bibinfo {pages} {075131} (\bibinfo {year} {2018})}\BibitemShut {NoStop}%
\bibitem [{\citenamefont {Stephen}\ \emph {et~al.}(2019)\citenamefont {Stephen}, \citenamefont {Dreyer}, \citenamefont {Iqbal},\ and\ \citenamefont {Schuch}}]{Spurious5}%
  \BibitemOpen
  \bibfield  {author} {\bibinfo {author} {\bibfnamefont {D.~T.}\ \bibnamefont {Stephen}}, \bibinfo {author} {\bibfnamefont {H.}~\bibnamefont {Dreyer}}, \bibinfo {author} {\bibfnamefont {M.}~\bibnamefont {Iqbal}},\ and\ \bibinfo {author} {\bibfnamefont {N.}~\bibnamefont {Schuch}},\ }\href {https://doi.org/10.1103/PhysRevB.100.115112} {\bibfield  {journal} {\bibinfo  {journal} {Phys. Rev. B}\ }\textbf {\bibinfo {volume} {100}},\ \bibinfo {pages} {115112} (\bibinfo {year} {2019})}\BibitemShut {NoStop}%
\bibitem [{\citenamefont {Nielsen}\ and\ \citenamefont {Petz}(2005)}]{Nielsen}%
  \BibitemOpen
  \bibfield  {author} {\bibinfo {author} {\bibfnamefont {M.~A.}\ \bibnamefont {Nielsen}}\ and\ \bibinfo {author} {\bibfnamefont {D.}~\bibnamefont {Petz}},\ }\href {https://arxiv.org/abs/quant-ph/0408130} {\bibinfo {title} {A simple proof of the strong subadditivity inequality}} (\bibinfo {year} {2005}),\ \Eprint {https://arxiv.org/abs/quant-ph/0408130} {arXiv:quant-ph/0408130 [quant-ph]} \BibitemShut {NoStop}%
\bibitem [{Note1()}]{Note1}%
  \BibitemOpen
  \bibinfo {note} {The numerical data for checking the validity of the inequality for UMTCs up to rank 11 is available at https://github.com/chihyulo/SSA-of-TEE-for-UMTC-up-to-rank-11}\BibitemShut {NoStop}%
\bibitem [{Note2()}]{Note2}%
  \BibitemOpen
  \bibinfo {note} {For simplicity, we will discuss the case without Wilson line insertions. However, the argument remains valid in the presence of Wilson lines.}\BibitemShut {Stop}%
\bibitem [{Note3()}]{Note3}%
  \BibitemOpen
  \bibinfo {note} {Any Wilson lines should be avoided in this consideration.}\BibitemShut {Stop}%
\bibitem [{\citenamefont {Lo}\ and\ \citenamefont {Chang}(2024)}]{KnotTEE}%
  \BibitemOpen
  \bibfield  {author} {\bibinfo {author} {\bibfnamefont {C.-Y.}\ \bibnamefont {Lo}}\ and\ \bibinfo {author} {\bibfnamefont {P.-Y.}\ \bibnamefont {Chang}},\ }\href {https://doi.org/10.1007/JHEP02(2024)117} {\bibfield  {journal} {\bibinfo  {journal} {Journal of High Energy Physics}\ }\textbf {\bibinfo {volume} {2024}},\ \bibinfo {pages} {117} (\bibinfo {year} {2024})}\BibitemShut {NoStop}%
\bibitem [{\citenamefont {Berthiere}\ \emph {et~al.}(2021)\citenamefont {Berthiere}, \citenamefont {Chen}, \citenamefont {Liu},\ and\ \citenamefont {Chen}}]{Reflect}%
  \BibitemOpen
  \bibfield  {author} {\bibinfo {author} {\bibfnamefont {C.}~\bibnamefont {Berthiere}}, \bibinfo {author} {\bibfnamefont {H.}~\bibnamefont {Chen}}, \bibinfo {author} {\bibfnamefont {Y.}~\bibnamefont {Liu}},\ and\ \bibinfo {author} {\bibfnamefont {B.}~\bibnamefont {Chen}},\ }\href {https://doi.org/10.1103/PhysRevB.103.035149} {\bibfield  {journal} {\bibinfo  {journal} {Phys. Rev. B}\ }\textbf {\bibinfo {volume} {103}},\ \bibinfo {pages} {035149} (\bibinfo {year} {2021})}\BibitemShut {NoStop}%
\bibitem [{Note4()}]{Note4}%
  \BibitemOpen
  \bibinfo {note} {In general, one can also consider extra punctures in the connected components. However, as we will show in later sections, these punctures will not affect the SSA properties.}\BibitemShut {Stop}%
\bibitem [{Note5()}]{Note5}%
  \BibitemOpen
  \bibinfo {note} {Contractible is essential here. For the case where $A$ and $B$ are disjoint $R_1$ bipartitions, for which $\protect \mathcal {I}_\protect \text {TEE}(A:B) = S_\protect \text {cl} \geq 0$. If we introduce an intersection between $A$ and $B$ at a $R_1$ region, we obtain $\protect \mathcal {I}_\protect \text {TEE}(A:B) = 0$, meaning that the topological CMI actually decreases when a non-contractible intersection is introduced.}\BibitemShut {Stop}%
\bibitem [{\citenamefont {Ng}\ \emph {et~al.}(2023)\citenamefont {Ng}, \citenamefont {Rowell},\ and\ \citenamefont {Wen}}]{ModularData}%
  \BibitemOpen
  \bibfield  {author} {\bibinfo {author} {\bibfnamefont {S.-H.}\ \bibnamefont {Ng}}, \bibinfo {author} {\bibfnamefont {E.~C.}\ \bibnamefont {Rowell}},\ and\ \bibinfo {author} {\bibfnamefont {X.-G.}\ \bibnamefont {Wen}},\ }\href {https://arxiv.org/abs/2308.09670} {\bibinfo {title} {Classification of modular data up to rank 11}} (\bibinfo {year} {2023}),\ \Eprint {https://arxiv.org/abs/2308.09670} {arXiv:2308.09670 [math.QA]} \BibitemShut {NoStop}%
\bibitem [{\citenamefont {Witten}(1989)}]{Witten1989}%
  \BibitemOpen
  \bibfield  {author} {\bibinfo {author} {\bibfnamefont {E.}~\bibnamefont {Witten}},\ }\href {https://doi.org/10.1007/BF01217730} {\bibfield  {journal} {\bibinfo  {journal} {Communications in Mathematical Physics}\ }\textbf {\bibinfo {volume} {121}},\ \bibinfo {pages} {351} (\bibinfo {year} {1989})}\BibitemShut {NoStop}%
\bibitem [{Note6()}]{Note6}%
  \BibitemOpen
  \bibinfo {note} {A simple way to compute the number of holes $F$ is to apply the Euler characteristic formula with $\chi = 1$, $E = 4n$ and $V=4$.}\BibitemShut {Stop}%
\bibitem [{Note7()}]{Note7}%
  \BibitemOpen
  \bibinfo {note} {The gray line is actually continuous; the diagram is simplified for ease of visualization.}\BibitemShut {Stop}%
\bibitem [{\citenamefont {Bonderson}\ \emph {et~al.}(2008)\citenamefont {Bonderson}, \citenamefont {Shtengel},\ and\ \citenamefont {Slingerland}}]{Bonderson}%
  \BibitemOpen
  \bibfield  {author} {\bibinfo {author} {\bibfnamefont {P.}~\bibnamefont {Bonderson}}, \bibinfo {author} {\bibfnamefont {K.}~\bibnamefont {Shtengel}},\ and\ \bibinfo {author} {\bibfnamefont {J.}~\bibnamefont {Slingerland}},\ }\href {https://doi.org/https://doi.org/10.1016/j.aop.2008.01.012} {\bibfield  {journal} {\bibinfo  {journal} {Annals of Physics}\ }\textbf {\bibinfo {volume} {323}},\ \bibinfo {pages} {2709} (\bibinfo {year} {2008})}\BibitemShut {NoStop}%
\end{thebibliography}%

\end{document}